\documentclass[12pt]{article}
\pdfoutput=1
\usepackage{amssymb,amsmath,dsfont,mathrsfs}
\usepackage{mathtools}
\usepackage{array}
\usepackage{physics}
\usepackage{slashed}
\usepackage[normalem]{ulem}
\usepackage{epsfig}
\usepackage{epstopdf}
\usepackage{latexsym}
\usepackage{tikz}
\usepackage{tikz-cd, graphicx}
\usetikzlibrary{calc, shapes.geometric, positioning, arrows.meta, decorations.pathreplacing,tikzmark}
\usepackage{booktabs}
\usepackage{bbm}
\usepackage{xspace}
\usepackage{cancel}
\usepackage[utf8]{inputenc}
\usepackage{csquotes}
\usepackage{enumitem}
\usepackage{braket}
\usepackage[numbers,compress]{natbib}
\usepackage[T1]{fontenc}
\usepackage[margin=20pt,small]{caption}
\usepackage[toc]{appendix}
\usepackage{color}
\usepackage{float}
\usepackage{yfonts}
\usepackage{datetime}
\usepackage[
colorlinks=false,
linkcolor=darkblue,  
urlcolor=blue,    
filecolor=blue,     
citecolor=red,
linktocpage=true,
pdfstartview=FitV,
bookmarksopen=true,
pdfencoding=auto,
pagebackref]{hyperref}
\numberwithin{equation}{section}
\usepackage{anyfontsize}
\usepackage{subcaption}
\usepackage{diagbox}

\usepackage[scr=dutchcal]{mathalpha}

\definecolor{cardinal}{rgb}{0.6,0,0}
\definecolor{darkgreen}{rgb}{0,0.5,0}
\definecolor{golden}{rgb}{0.92, 0.7, 0}
\definecolor{midnight}{rgb}{0, 0, 0.5}
\definecolor{darkblue}{rgb}{0.2, 0, 0.8}

\usepackage[most]{tcolorbox}

\begin{document}  

	\begin{titlepage}
	\begin{flushright}
	SISSA  11/2026/FISI
	\end{flushright}
		\bigskip
		\begin{center} 
			{\fontsize{22pt}{0pt} \bf Symmetry extension by condensation defects II:
            \vskip 15pt
            general dimensions and higher-groups}

			\vskip30pt

			{\large Matteo Bertolini$^{a,b}$, Lorenzo Di Pietro$^{b,c}$,
            \\
            \vskip 2mm
            Stefano C. Lanza$^{a,b}$,  Antonio Santaniello$^{b,c}$\\}
			\bigskip
                    ${}^{a}$ SISSA, Via Bonomea 265, 34136 Trieste, Italy \\
                    \vskip 2mm
                    ${}^{b}$
            	INFN, Sezione di Trieste, Via Valerio 2, 34127 Trieste, Italy\\
                \vskip 2mm
           		${}^{c}$
            	Dipartimento di Fisica, Universit\`{a} di Trieste, Strada Costiera 11, 34151 Trieste, Italy\\
                    \vskip 5mm

			\texttt{bertmat@sissa.it,~ldipietro@units.it,\\
            slanza@sissa.it,~antonio.santaniello@phd.units.it}\\
		\end{center}
		
		\bigskip
		\bigskip

        
		\begin{abstract}
\noindent 
We discuss a class of symmetry structures in general spacetime dimension that arise when gauging symmetries in theories with a cubic 't Hooft anomaly. The anomaly mixes two Abelian discrete symmetries $A$ and $B$ with a characteristic class for an additional symmetry $C$, and we gauge $A\times B$. The novelty of this construction is that the resulting symmetry structure involves certain condensation defects of the dual symmetry $\widehat{A}\times\widehat{B}$. Specifically, these defects generate an invertible symmetry that extends $C$, either as an ordinary group extension or as a higher-group, depending on the characteristic class. We describe the corresponding charged operators and states, and illustrate the mechanism in several examples.

		\end{abstract}

	\end{titlepage}
	
	
	\setcounter{tocdepth}{2}	

\tableofcontents

\section{Introduction}

In this work we describe a novel symmetry structure that can arise in a QFT upon gauging anomaly-free subgroups of anomalous symmetries. Our starting point is the cubic mixed anomaly that we studied in \cite{Bertolini:2025wyj} between the $\mathbb{Z}^{(1)}_N$ one-form symmetry and the $U(1)^{(0)}_I$ instantonic 0-form symmetry of five-dimensional $SU(N)$ Yang-Mills (YM) theory, captured by the inflow action \cite{BenettiGenolini:2020doj, Gukov:2020btk, Genolini:2022mpi}
\begin{equation}\label{eq:BGTanointro}
S_{\mathrm{inflow}} =
\frac{2\pi i}{N} \int_{\mathcal{Y}_{6}}
\frac{\mathcal{P}(\mathcal{B}_2)}{2} \cup \left[\frac{dA_1}{2\pi}\right]_N \ .
\end{equation}
In particular, we considered gauging a subgroup of the one-form symmetry, to obtain gauge theories with a general global form $SU(N)/\mathbb{Z}_k$ for the gauge group, with $k$ a divisor of $N$. The imprint of the anomaly is a particular symmetry structure in the gauged theory, namely an extension of the instantonic symmetry by a condensation defect of the ``dual'' magnetic two-form symmetry. This phenomenon can be equivalently rephrased in terms of charged objects, as the statement that the configuration of two 't Hooft surfaces that link with each other carries a fractional instantonic charge. In \cite{Bertolini:2025wyj} we applied these observations to the interacting UV fixed points of 5d $\mathcal{N}=1$ supersymmetric YM (SYM), finding, among other things, that the UV fixed point of  $SO(3)$ SYM has an instantonic global symmetry with group $SU(2)_I$.

It is natural to wonder whether these effects due to the anomaly \eqref{eq:BGTanointro} are a special case of a larger framework, of more general applicability. The important ingredients in \eqref{eq:BGTanointro} are: 
\begin{itemize}
\item{the background gauge field $\mathcal{B}_2$ that we promote to be dynamical appears quadratically: this is the reason why the condensation defect of the dual symmetry enters the symmetry structure of the gauged theory;}
\item{the background gauge field $A_1$ of the symmetry that remains global appears through a certain characteristic class, in this case simply the (mod $N$ reduction of the) first Chern class. In the gauged theory, this class determines the precise relation between this symmetry and the condensation defect, in this case a symmetry extension.}
\end{itemize}

In order to keep both of these features in a more general setup, in this paper we consider a theory $\mathcal{T}$ with a cubic anomaly of the form
\begin{equation}\label{eq:anomalyintro}  
\mathcal{T}\,:\quad S_{\mathrm{inflow}}=2\pi i\int_{\mathcal{Y}_{d+1}} (\mathcal{A}_{p+1}\cup \mathcal{B}_{q+1})\cup t_{n+r+1}(C_{r+1})\ ,
\end{equation}
with $p+q+n+r+3 = d+1$. Here $\mathcal{A}_{p+1}$ and $\mathcal{B}_{q+1}$ are background gauge fields for discrete $p$- and $q$-form symmetries $A^{(p)}$ and $B^{(q)}$. We also consider the case in which there is a single $p$-form symmetry $A^{(p)}$ and  $\mathcal{A}_{p+1}\cup \mathcal{B}_{q+1}$ is replaced by an appropriate square of $\mathcal{A}_{p+1}$. Finally, $t_{n+r+1}(C_{r+1})$ is a degree-$(n+r+1)$ characteristic class associated with an $r$-form symmetry $C^{(r)}$, which could be either continuous or discrete, with background gauge field $C_{r+1}$. The class $t_{n+r+1}(C_{r+1})$ is assumed to take values in a discrete Abelian group $D$. Correspondingly, the product $\mathcal{A}_{p+1}\cup \mathcal{B}_{q+1}$ is defined to take values in the Pontryagin dual group $\widehat{D}$. This ensures that the expression \eqref{eq:anomalyintro} is a well-defined anomaly inflow. 

Our goal is to study the symmetry structure of the theory 
\begin{equation}
\widehat{\mathcal{T}} =  \mathcal{T}/(A^{(p)}\times B^{(q)})~,
\end{equation}
obtained from $\mathcal{T}$ upon gauging $A^{(p)}\times B^{(q)}$. As is familiar, this theory features ``dual'' $\widehat{A}^{(d-p-2)}\times \widehat{B}^{(d-q-2)}$ symmetries. We wish to determine the precise relation between these symmetries and $C^{(r)}$ that reflects the existence of the anomaly in the parent theory. Questions of this type have already received considerable attention in the literature, some foundational references being \cite{Kapustin:2013uxa,Bhardwaj:2017xup,Tachikawa:2017gyf, Kaidi:2021xfk, Choi:2021kmx}, while for some recent works see e.g. \cite{Moradi:2023dan,Su:2024vrk,Villa:2026jmd,Bason:2026njw}. However, a general answer for the setup with the cubic anomaly as in \eqref{eq:anomalyintro} has not been derived so far. Typically, an 't Hooft anomaly gives rise, in the gauged theory, to an extension/higher-group structure \cite{Sharpe:2015mja,Cordova:2018cvg,Benini:2018reh}, or to the presence of non-invertible fusion rules \cite{Petkova:2000ip,Frohlich:2004ef,Frohlich:2006ch,Chang:2018iay,Thorngren:2019iar}. The novelty here is that, since the gauged symmetry $\widehat{A}^{(d-p-2)}\times \widehat{B}^{(d-q-2)}$ appears quadratically in the anomaly, the resulting structure involves \textit{condensation defects} \cite{Roumpedakis:2022aik, Choi:2022zal} of the ``dual'' symmetries $\widehat{A}^{(d-p-2)}\times \widehat{B}^{(d-q-2)}$. These defects are defined by a ``higher-gauging'' procedure, i.e. gauging $\widehat{A}^{(d-p-2)}\times \widehat{B}^{(d-q-2)}$ only on a submanifold of dimension $p+q+2 = d-n-r$. Thanks to the inclusion of a certain discrete torsion term, this condensation defect has invertible fusion rules described by the group $D$, i.e. it gives a $D^{(n+r-1)}$ symmetry in the theory $\widehat{\mathcal{T}}$. Our main result is to show that the symmetry structure of $\widehat{\mathcal{T}}$ is a certain extension of $C^{(r)}$ by $D^{(n+r-1)}$. The precise nature of the extension depends on the number $n$ associated with the characteristic class $t_{n+r+1}$: for $n=1$ it is a standard group extension of $C^{(r)}$ by $D^{(r)}$, while, for $n\geq 2$, $C^{(r)}$ and $D^{(n+r-1)}$ form a higher-group structure, specifically a $(n+r)$-group. 

Besides describing the symmetry structure, i.e. the relation between the topological charge defects, we also discuss the associated representations, i.e. the charged operators. For this purpose we focus our attention on the cases $n=1$ and $n=2$. First, we explain that, due to the special nature of the charge defects, the configurations charged under $D^{(n+r-1)}$ are not extended operators with $(n+r-1)$-dimensional support, as one might expect for an $(n+r-1)$-symmetry. Rather, they are obtained by inserting two operators, charged respectively under $\widehat{A}^{(d-p-2)}$ and $\widehat{B}^{(d-q-2)}$, in a configuration that can have nontrivial {\it triple-linking} \cite{Putrov:2016qdo,Kaidi:2022cpf} with the $D^{(n+r-1)}$ charge. Then, we discuss what this implies for the action of the structure in which $D^{(n+r-1)}$ and $C^{(r)}$ are embedded. The result can be summarized by saying that the triple-linking configuration, in addition to the $D^{(n+r-1)}$ charges, also gives rise to projective representations of $C^{(r)}$. The charges under $\widehat{A}^{(d-p-2)}$ and $\widehat{B}^{(d-q-2)}$, together with the projective representation of $C^{(r)}$, fully characterize the action of the symmetry structure of $\widehat{\mathcal{T}}$ on charged operators. We also explain which Hilbert spaces for the theory $\widehat{\mathcal{T}}$ carry non-trivial representations of the extension ($n=1$) or of the $(r+2)$-group ($n=2$).
  
We then discuss several examples where this structure arises:
\begin{itemize}
\item{In three-dimensional quantum electrodynamics (QED$_3$) with matter content given by $N$ complex scalar fields, a discrete subgroup $A^{(0)}\times B^{(0)}=\mathbb{Z}^{(0)}_N\times \mathbb{Z}^{(0)}_N$ of the flavor symmetry has an anomaly of the form \eqref{eq:anomalyintro} with the $U(1)^{(0)}_M$ magnetic symmetry. The latter enters the anomaly through the mod $N$ reduction of the first Chern class. Applying the general result to this case, we obtain that, upon gauging $A^{(0)}\times B^{(0)}$, $U(1)^{(0)}_M$ is extended by the $\mathbb{Z}^{(0)}_N$ symmetry generated by a condensation defect of $\widehat{A}^{(1)}\times \widehat{B}^{(1)}$. The operators charged under the extension are links of twist operators for $A^{(0)}$ and $B^{(0)}$, that become genuine line operators in the gauged theory.}
\item{In the $\mathcal{N}=4$ supersymmetric version of QED$_3$, coupled to $N=2$ matter multiplets, the anomaly involving the flavor and magnetic symmetries is analogous to the non supersymmetric case. The theory flows to an interacting superconformal field theory (SCFT) in the IR, where the magnetic symmetry is enhanced to a non-Abelian $SO(3)^{(0)}_M$ magnetic symmetry. We show that, upon gauging again an $A^{(0)}\times B^{(0)}=\mathbb{Z}^{(0)}_2\times \mathbb{Z}^{(0)}_2$ subgroup of the flavor symmetry, the symmetry of the IR fixed point gets extended to $SU(2)^{(0)}_M$ by a $\mathbb{Z}_2^{(0)}$ group generated by a condensation defect. The IR fixed point of supersymmetric QED$_3$ with $N=2$ flavors can be also viewed as the $T(SU(2))$ theory of Gaiotto and Witten \cite{Gaiotto:2008ak}. We show that a similar phenomenon also applies to the more general $T(SU(N))$ SCFT with $N>2$, where one $PSU(N)$ factor of the global symmetry gets extended to $SU(N)$ upon gauging a $\mathbb{Z}^{(0)}_N\times \mathbb{Z}^{(0)}_N$ subgroup of the other $PSU(N)$ factor.}
\item{In the example of five-dimensional $\mathcal{N}=1$ supersymmetric Yang-Mills (SYM) theory, with the anomaly \eqref{eq:BGTanointro} already studied in \cite{Bertolini:2025wyj}, we provide additional strong evidence for the extension of the instantonic symmetry $U(1)_I^{(0)}$ using the supersymmetric partition function. Following the general discussion about states charged under the extension, in this case we focus on the partition function on the background $S^1 \times S^2 \times S^2$, because the Hilbert space on $S^2\times S^2$ (unlike $S^4$) can carry the expected fractional instantonic charges. Building on the recent result \cite{Kim:2025fpz} for these partition functions, we show explicitly that in the theory with gauge group $SO(3)$ the fugacity for $U(1)_I^{(0)}$ enters with a doubled periodicity compared to the theory with gauge group $SU(2)$, signaling the $\mathbb{Z}_2$ extension.}
\item{In the $E_1$ SCFT, the UV fixed point of five-dimensional $\mathcal{N}=1$ SYM with gauge group $SU(2)$ \cite{Seiberg:1996bd}, the instantonic symmetry has a non-Abelian enhancement to $SO(3)^{(0)}_I$, and the 't Hooft anomaly involving $SO(3)^{(0)}_I$ and the $\mathbb{Z}_2^{(1)}$ one-form symmetry is determined by matching with \eqref{eq:BGTanointro}, as discussed in \cite{BenettiGenolini:2020doj, Genolini:2022mpi}. We show that an $A^{(0)}\times B^{(0)}=\mathbb{Z}^{(0)}_2\times \mathbb{Z}^{(0)}_2$ subgroup of $SO(3)^{(0)}_I$ has a mixed anomaly with $\mathbb{Z}_2^{(1)}$ precisely of the form \eqref{eq:anomalyintro} with $n=2$. As a result, gauging this $\mathbb{Z}^{(0)}_2\times \mathbb{Z}^{(0)}_2$ symmetry of $E_1$, we obtain a theory with a 3-group involving $\mathbb{Z}_2^{(1)}$ and a two-form symmetry $\mathbb{Z}_2^{(2)}$ generated by a condensation defect.}
\end{itemize}

The general analysis of the symmetry structure is discussed in section \ref{sec:CUBIC}, while in section \ref{SEC:examples} we present the aforementioned examples.
Several appendices contain derivations of some technical results needed in the main text, as well as a detailed pedagogical exposition of charged operators and states in the case of extensions and higher-groups of standard type (i.e. not associated with a condensation defect).

\section{Dual of a cubic anomaly}
\label{sec:CUBIC}
In this section we discuss the symmetry structure that reflects the anomaly \eqref{eq:anomalyintro} after gauging, first at the level of background fields and topological operators, and then at the level of its representations on appropriate Hilbert spaces.

\subsection{Symmetry structure}
\paragraph{The anomalous theory $\mathcal{T}$} Let us consider a $d$-dimensional theory $\mathcal{T}$ with a symmetry $A^{(p)}\times B^{(q)}\times C^{(r)}$, where the superscripts denote that the factors are $p$-,$q$- and $r$-form symmetries. The groups $A$ and $B$ are assumed finite and Abelian, while $C$ can be either finite or a compact Lie group and, for $r=0$, also non-Abelian. We are interested in the case where these symmetries have a mixed 't Hooft anomaly described by the anomaly inflow\footnote{In the case $p=q$ we will also allow the possibility that the groups $A$ and $B$ coincide, i.e.~the symmetry is just $A^{(p)}\times C^{(r)}$ and the cubic anomaly involves a term quadratic in the $A^{(p)}$ gauge field. It is mostly straightforward to adapt the following equations to this case, we will explicitly comment on it when more care is needed.}
\begin{equation}\label{eq:anomaly}  S_{\mathrm{inflow}}=2\pi i \int_{\mathcal{Y}_{d+1}} (\mathcal{A}_{p+1}\cup_\mathbb{Z} \mathcal{B}_{q+1})\cup t_{n+r+1}(C_{r+1}) \ .
\end{equation}
We denote with uppercase calligraphic latin letters the discrete background gauge fields 
\begin{equation}
\begin{split}
 \mathcal{A}_{p+1}&\in H^{p+1}(\mathcal{Y}_{d+1},A) \ ,\\
   \mathcal{B}_{q+1}&\in H^{q+1}(\mathcal{Y}_{d+1},B) \ ,
\end{split}
\end{equation}
for the respective symmetries, while $C_{r+1}$ is an $(r+1)$-form background gauge field for the $C^{(r)}$ symmetry. We further assume that $C^{(r)}$ bundles admit a characteristic class with values in a discrete Abelian group $D$, $ t_{n+r+1}\in H^{n+r+1}(B^{r+1}C,D)$, where we denote with $B^{r+1}C$ the classifying space of $C^{(r)}$ bundles, see e.g.~Appendix A of \cite{Wan:2024kaf} for a review for physicists.\footnote{
If $C$ is discrete, $B^{r+1}C$ is homotopically equivalent to the Eilenberg-MacLane space $K(C,r+1)$, i.e.~a topological space whose $(r+1)^{\text{th}}$ homotopy group is $C$ and all the others are trivial. If $C=U(1)$ (with standard topology), then $B^{r+1}C$ is homotopically equivalent to the Eilenberg-MacLane space $K(\mathbb{Z},r+2)$ \cite{Davis:2001woh}.\label{EML}} The characteristic class pulled back on the inflow manifold through the gauge field $C_{r+1}$ gives
\begin{equation}\label{eq:pullback}
    t_{n+r+1}(C_{r+1})\in H^{n+r+1}(\mathcal{Y}_{d+1},D) \ .
\end{equation}
The integer $n$ characterizes the class, and it must satisfy $d+1=n+p+q+r+3$.
A natural choice for the image of $t_{n+r+1}$ is the group $D=\widehat{A\otimes_\mathbb{Z} B}$, where $A\otimes_\mathbb{Z} B$ is the natural tensor product between finite Abelian groups, viewed as $\mathbb{Z}$ modules,\footnote{For example, for $A=\mathbb{Z}_N$ and $B=\mathbb{Z}_M$, $A\otimes_\mathbb{Z}B=\mathbb{Z}_{\text{gcd}(N,M)}$. This product is explicitly constructed as follows: given a pair $n\in \mathbb{Z}_N $ and $m \in \mathbb{Z}_M$, the associated element $n\otimes_\mathbb{Z}m \in \mathbb{Z}_{\text{gcd}(N,M)}$ is given by lifting $n,m$ to $\mathbb{Z}$, computing the product in $\mathbb{Z}$ and reducing it modulo gcd$(N,M)$, which ensures the independence of the $\mathbb{Z}$ lifts. This generalizes to arbitrary $A=\prod_i \mathbb{Z}_{N_i}$ and $B=\prod_j \mathbb{Z}_{M_j}$ as $A\otimes_\mathbb{Z}B=\prod_{i,j}\mathbb{Z}_{\text{gcd}(N_i,M_j)}$. It also follows that $\mathbb{Z}_N\otimes_\mathbb{Z}\mathbb{Z} = \mathbb{Z}_N$. \label{footnote:Ztensor}} and the hat denotes the Pontryagin dual group $\widehat{G}=\text{Hom}(G,U(1))$ for any Abelian group $G$.
With this choice, the first cup product
\begin{equation}
    \cup_\mathbb{Z} : H^{p+1}(\mathcal{Y}_{d+1},A)\times H^{q+1}(\mathcal{Y}_{d+1},B) \rightarrow H^{p+q+2}(\mathcal{Y}_{d+1},\widehat{D}) \ ,
\end{equation}
  is associated with the canonical bi-homomorphism $\otimes_\mathbb{Z}: A \times B \rightarrow \widehat{D}=A\otimes_\mathbb{Z}B$, as in footnote \ref{footnote:Ztensor}. The second cup product  
\begin{equation}
   \cup: H^{p+q+2}(\mathcal{Y}_{d+1},\widehat{D})\times  H^{n+r+1}(\mathcal{Y}_{d+1},D)\rightarrow H^{d+1}(\mathcal{Y}_{d+1},U(1)) \ ,
\end{equation}
is associated with the natural pairing $\cdot :\widehat{D}\times D \rightarrow U(1)$\footnote{In the specific case of ${D}=\widehat{D}=\mathbb{Z}_N$, a common notation for the pairing, that we will use in examples, is $d\cdot\widehat{d}=\frac{d\,\widehat{d}}{N}\text{ mod }1$, where $d$ and $\widehat{d}$ in the right-hand side denote $\mathbb{Z}$ lifts.\label{foot:normalization}}. Finally, the integral defines a pairing between chains and cochains, with the  convention that $S_{\mathrm{inflow}}\sim  S_{\mathrm{inflow}} + 2\pi i$ gives an element of $U(1)$ in additive notation, or equivalently $e^{-S_{\mathrm{inflow}}}$ gives an element of $U(1)$ in multiplicative notation. 
Under a generic background gauge transformation we have 
\begin{equation}
\label{eq:gauge_tr_1}
    \begin{split}
        \mathcal{A}_{p+1}&\mapsto \mathcal{A}_{p+1} + \delta \alpha_p \ ,\\
        \mathcal{B}_{q+1}&\mapsto \mathcal{B}_{q+1} + \delta \beta_q \ ,\\
        C_{r+1}&\mapsto 
        \begin{cases}
           C_{r+1} + d\gamma_r\ , &  r\neq 0\\
           U_{\gamma_0}^{-1}C_1U_{\gamma_0} + U_{\gamma_0}^{-1}dU_{\gamma_0}\ , & r=0
        \end{cases}\\
        t_{n+r+1}(C_{r+1})&\mapsto t_{n+r+1}(C_{r+1})+\delta \tau_{n+r}(C_{r+1},\gamma_r)\ .
    \end{split}
\end{equation}
The inflow action \eqref{eq:anomaly} is defined by the property that the phase picked by the partition function $Z_\mathcal{T}$ in the presence of background gauge fields, under the most generic background gauge transformation, equals the phase picked by $e^{-S_{\mathrm{inflow}}}$. In the case at hand, at linear order in the parameters, this phase is
\begin{align}\label{Ano-phase}
\begin{split}
\exp \left(-2\pi i\int_{\mathcal{M}_d}
(\alpha_p \cup_\mathbb{Z} \mathcal{B}_{q+1}
 \right. &+(-1)^{p+1}\mathcal{A}_{p+1}\cup_\mathbb{Z}\beta_q
)\cup t_{n+r+1} \\
& ~\left.\phantom{\int} +(-1)^{p+q}\mathcal{A}_{p+1}\cup_\mathbb{Z} \mathcal{B}_{q+1}\cup \tau_{n+r}\right) \ .
\end{split}
\end{align}

\paragraph{Symmetries of $\widehat{\mathcal{T}}$} We want to examine the symmetry structure in the theory $\widehat{\mathcal{T}}=\mathcal{T}/(A^{(p)}\times B^{(q)})$ obtained from $\mathcal{T}$ by gauging the symmetry $A^{(p)}\times B^{(q)}$, i.e.~summing over the dynamical gauge fields $\mathscr{a}_{p+1}, \mathscr{b}_{q+1}$, denoted in lowercase calligraphic latin letters. We focus on the interplay between the topological defects associated with the $C^{(r)}$ symmetry of $\mathcal{T}$, and the dual symmetry $\widehat{A}^{(d-p-2)}\times \widehat{B}^{(d-q-2)}$, generated respectively by the operators
\begin{equation}
    U_{\widehat{a}}(\Sigma_{p+1})=\exp(2\pi i \, \widehat{a} \cdot \int_{\Sigma_{p+1}} \mathscr{a}_{p+1}) \ , \hspace{.7cm} U_{\widehat{b}}(\Sigma_{q+1})=\exp(2\pi i \, \widehat{b} \cdot \int_{\Sigma_{q+1}} \mathscr{b}_{q+1}) \ ,
\end{equation}
where $\widehat{a}$ and $\widehat{b}$ are group elements of $\widehat{A}$ and $\widehat{B}$ labeling the operators. Here we focus on a specific class of topological defects that are present in the theory after gauging the symmetry $A^{(p)}\times B^{(q)}$, namely the operators
\begin{equation}\label{aub}
U_d[\Sigma_{p+q+2}]=\exp\left(2\pi i \, d\cdot \int_{\Sigma_{p+q+2}}\mathscr{a}_{p+1} \cup_\mathbb{Z} \mathscr{b}_{q+1}\right) \ ,
\end{equation}
 labeled by an element $d \in D$, where again the dot $\cdot$ denotes the natural pairing with $\widehat{D}$. These operators generate a $D^{(n+r-1)}$ $(n+r-1)$-form symmetry. They can be seen as condensation defects of the dual symmetry $\widehat{A}^{(d-p-2)}\times \widehat{B}^{(d-q-2)}$, i.e.~topological defects obtained by gauging this symmetry on a submanifold $\Sigma_{p+q+2}$. The specific gauging that produces $U_d[\Sigma_{p+q+2}]$ requires the inclusion of additional dynamical gauge fields with a discrete torsion term. The resulting expression reads
\begin{equation}\label{cond-def}
\begin{split}
 U_d[\Sigma_{p+q+2}]\propto
 \sum_{\substack{
 \widehat{\mathscr{a}}_{q+1},\ \widehat{\mathscr{b}}_{p+1}\\
 \mathscr{a}'_{p+1},\ \mathscr{b}'_{q+1}}}
 &
 \exp(2\pi i\int_{\Sigma_{p+q+2}} 
 (\mathscr{a}_{p+1}-\mathscr{a}'_{p+1})\cup \widehat{\mathscr{a}}_{q+1}
 +
 (\mathscr{b}_{q+1}-\mathscr{b}'_{q+1})\cup \widehat{\mathscr{b}}_{p+1}) 
 \times \\
 & \exp(2\pi i \, d\cdot\int_{\Sigma_{p+q+2}}{\mathscr{a}}'_{p+1}\cup_\mathbb{Z} {\mathscr{b}}'_{q+1}) \ ,
\end{split}
\end{equation}
where $\widehat{\mathscr{a}}_{q+1}=\widehat{\mathscr{a}}_{d-p-1-(d-p-q-2)}$ and $\widehat{\mathscr{b}}_{p+1}=\widehat{\mathscr{b}}_{d-q-1-(d-p-q-2)}$ are the gauge fields of the dual $\widehat{A}^{(d-p-2)}\times \widehat{B}^{(d-q-2)}$ symmetry {\it restricted} to the defect $\Sigma_{p+q+2}$, and $\mathscr{a}'_{p+1}, \mathscr{b}'_{q+1}$ are the additional dynamical $A$ and $B$ gauge fields, respectively. In \eqref{cond-def} we are neglecting the normalization factors. For their precise form in an example see e.g.~\cite{Bertolini:2025wyj}. We introduce a background gauge field $\mathcal{D}_{n+r}$ for the $D^{(n+r-1)}$ symmetry, with gauge transformation
\begin{equation}
 \mathcal{D}_{n+r} \mapsto \mathcal{D}_{n+r} + \Lambda_{n+r} \ ,
\end{equation}
where $\Lambda_{n+r}$ is yet to be determined. Note that we will need to allow $\mathcal{D}_{n+r}$ to be not closed, i.e.~a generic $D$-valued $(n+r)$-cochain. 

The interplay between the symmetries $C^{(r)}$ and $D^{(n+r-1)}$ can be derived from the partition function of $\widehat{\mathcal{T}}$ coupled to the background gauge fields
\begin{equation}\label{Zthat}
    Z_{\widehat{\mathcal{T}}}[C_{r+1},\mathcal{D}_{n+r}]=\sum_{\mathscr{a}_{p+1}, \mathscr{b}_{q+1}}\exp(2\pi i\int_{\mathcal{M}_d} \mathscr{a}_{p+1}\cup_\mathbb{Z} \mathscr{b}_{q+1} \cup \mathcal{D}_{n+r})Z_\mathcal{T}[\mathscr{a}_{p+1},\mathscr{b}_{q+1},C_{r+1}]\ ,
\end{equation}
by requiring that the summand is both invariant under the gauge transformations of the dynamical fields, and that it does not produce operator-valued phases under background gauge transformations. The gauge transformations of $\mathscr{a}_{p+1}$ and $\mathscr{b}_{q+1}$ produce a phase that contains both operators and background gauge fields, with a contribution from $Z_\mathcal{T}$ (see eqs.~\eqref{eq:gauge_tr_1})
\begin{equation}
    \exp(-2\pi i \int_{\mathcal{M}_d}
    \left(\alpha_p \cup_\mathbb{Z} \mathscr{b}_{q+1}+\alpha_p \cup_\mathbb{Z} \delta\beta_{q}+(-1)^{p+1}\mathscr{a}_{p+1}\cup_\mathbb{Z}\beta_q\right)
    \cup t_{n+r+1})\ ,
\end{equation}
and a contribution from the coupling to $\mathcal{D}_{n+r}$
\begin{equation}
\begin{aligned}
    \exp &\bigg(2\pi i\int_{\mathcal{M}_d}
    \left(\delta\alpha_p \cup_\mathbb{Z} \mathscr{b}_{q+1}+\mathscr{a}_{p+1}\cup_\mathbb{Z} \delta\beta_q+\delta\alpha_p\cup_\mathbb{Z} \delta\beta_q\right)
    \cup \mathcal{D}_{n+r} \bigg) \\
    &=\exp(2\pi i \int_{\mathcal{M}_d}\delta
    \left(\alpha_p \cup_\mathbb{Z} \mathscr{b}_{q+1}+\alpha_p\cup_\mathbb{Z} \delta\beta_q+(-1)^{p+1}\mathscr{a}_{p+1}\cup_\mathbb{Z} \beta_q\right)\cup \mathcal{D}_{n+r}) \\
    &=\exp((-1)^{p+q}2\pi i \int_{\mathcal{M}_d}(\alpha_p \cup_\mathbb{Z} \mathscr{b}_{q+1}+\alpha_p\cup_\mathbb{Z} \delta\beta_q+(-1)^{p+1}\mathscr{a}_{p+1}\cup_\mathbb{Z} \beta_q)\cup \delta\mathcal{D}_{n+r}) \ .
\end{aligned}
\end{equation}
We have used that $\delta\mathscr{a}_{p+1}=\delta\mathscr{b}_{q+1}=0$ to eliminate terms arising from integrating by parts.
It follows that the background gauge fields must satisfy the equation
\begin{equation}\label{extEq}
    \delta \mathcal{D}_{n+r}=(-1)^{p+q} \ t_{n+r+1}(C_{r+1})\ .
\end{equation}

Moreover, under background gauge transformations of $C_{r+1}$ and $\mathcal{D}_{n+r}$, the variation of the summand in \eqref{Zthat} gives the following contribution from $Z_\mathcal{T}$
\begin{equation}
    \exp((-1)^{p+q+1}2\pi i\int_{\mathcal{M}_d}\mathscr{a}_{p+1}\cup_\mathbb{Z} \mathscr{b}_{q+1}\cup \tau_{n+r})\ ,
\end{equation}
and a contribution from the coupling to $\mathcal{D}_{n+r}$
\begin{equation}
    \exp(2\pi i\int_{\mathcal{M}_d}\mathscr{a}_{p+1}\cup_\mathbb{Z} \mathscr{b}_{q+1}\cup \Lambda_{n+r})\ .
\end{equation}
Hence, invariance under background gauge transformations imposes a modified transformation of the background $\mathcal{D}_{n+r}$, compatible with \eqref{extEq}, namely 
\begin{equation}\label{extGtransf}
    \mathcal{D}_{n+r} \mapsto \mathcal{D}_{n+r} +  \delta \lambda_{n+r-1} + (-1)^{p+q} \ \tau_{n+r}(C_{r+1},\gamma_r)\ .
\end{equation}
In the following, we discuss different values of $n$ and, whenever possible, compare our results with other analyses in the literature.

\subsubsection{\texorpdfstring{$n=0$}{n=0}, non-invertible defects}
For the $n=0$ case, we restrict the group $C^{(r)}$ to be discrete and Abelian, and its characteristic class to be the discrete gauge field itself, $t_{r+1}(C_{r+1})=\mathcal{C}_{r+1}$. Note that this implies that the groups $D$ and $C$ coincide. The anomaly of the theory $\mathcal{T}$ \eqref{eq:anomaly} specified to this case becomes
\begin{equation}\label{ano-n0}
    S_{\mathrm{inflow}}=2\pi i\int_{\mathcal{Y}_{d+1}} \mathcal{A}_{p+1}\cup_\mathbb{Z} \mathcal{B}_{q+1}\cup 
    \mathcal{C}_{r+1} \ .
\end{equation}
Accordingly, in the theory $\widehat{\mathcal{T}}$ the background gauge fields obey the relation \eqref{extEq}
\begin{equation}\label{eq:bckg-non-inv}
    \delta \mathcal{D}_{r}=(-1)^{p+q}\mathcal{C}_{r+1} \ . 
\end{equation}
This implies that the topological defects of the symmetry $C^{(r)}$ can only be inserted at the boundary of open submanifolds, where topological defects of $D^{(r-1)}$ are inserted. In particular, only trivial background gauge fields can be turned on for the symmetry $C^{(r)}$.\footnote{This behavior is special to the $n=0$ case, while it does not hold for characteristic classes with $n\geq1$.} Since the topological defects are not genuine operators, $C^{(r)}$ is no longer a symmetry of $\widehat{\mathcal{T}}$ in the ordinary sense. By contrast, the topological defects $U_d$ defined in eq.~\eqref{aub}, when inserted on closed submanifolds, give rise to an ordinary $D^{(r-1)}$ symmetry in the theory $\widehat{\mathcal{T}}$. 

Note that, even though not manifest, a remnant of the symmetry $C^{(r)}$ still exists in the theory $\widehat{\mathcal{T}}$ . To see this, one should stack on the $C^{(r)}$ defects a $(d-r-1)$-dimensional TQFT that reabsorbs the operator-valued filling, thus defining genuine topological operators. The additional TQFT renders the dressed topological defects non-invertible \cite{Kaidi:2021xfk, Choi:2022jqy}. Therefore, in this $n=0$ scenario the topological defects of $C^{(r)}$ (on closed manifolds) acquire in $\widehat{\mathcal{T}}$ a non-invertible fusion rule. 

This fact was previously discussed in the literature in some particular cases. In $d=2$ (where the only option is $p=q=r=0$) the anomaly\footnote{The normalization here comes from specifying the pairing to the case of $D=\widehat{D}=\mathbb{Z}_2$, see footnote \ref{foot:normalization}.}
\begin{equation}\label{eq:Tachi2d}
     S_{\mathrm{inflow}}=\frac{2\pi i}{2}\int_{\mathcal{Y}_3} \mathcal{A}_1\cup \mathcal{B}_1 \cup \mathcal{C}_1 \ ,
\end{equation}
for $A^{(0)}=\mathbb{Z}_2^{(0)}, \ B^{(0)}= \mathbb{Z}_2^{(0)}$ and $C^{(0)}=\mathbb{Z}_2^{(0)}$, was considered in
\cite{Kapustin:2014zva, Tachikawa:2017gyf}. In particular,  \cite{Tachikawa:2017gyf} shows that upon gauging two of the $\mathbb Z_2$ factors, the symmetry structure becomes non-invertible. The topological defects obey the fusion rules of Rep($D_8$), with the nontrivial element of the third ungauged $\mathbb{Z}_2$ factor being identified with the two-dimensional representation of $D_8$ (dihedral group of order 8).

In \cite{Kaidi:2021xfk}, the authors studied examples in $d=3$ (with $p=1$, $q=r=0$) and $d=4$ (with $p=q=1,$ $r=0$). In $d=3$ they consider the anomaly
\begin{equation}\label{eq:appKOZ}
    S_{\text{inflow}}=\frac{2\pi i}{2}\int_{\mathcal{Y}_4} \mathcal{A}_2 \cup \mathcal{B}_1 \cup \mathcal{C}_1 \ ,
\end{equation}
with $A^{(1)}=\mathbb{Z}_2^{(1)}$, $B^{(0)}=\mathbb{Z}_2^{(0)}$ and $C^{(0)}=\mathbb{Z}_2^{(0)}$, while in $d=4$ they consider
\begin{equation}\label{eq:KOZ}
    S_{\mathrm{inflow}}=\frac{2\pi i}{4}\int_{\mathcal{Y}_5} \mathcal{P}(\mathcal{A}_2) \cup \mathcal{C}_1 \ ,
\end{equation}
with $A^{(1)} \equiv B^{(1)}=\mathbb{Z}_2^{(1)}$ and $C^{(0)}=\mathbb{Z}_2^{(0)}$.\footnote{In this case $A^{(1)}$ and $B^{(1)}$ coincide, i.e.~the symmetry is $\mathbb{Z}_2^{(1)}\times \mathbb{Z}_2^{(0)}$. The cubic anomaly involves a term quadratic in the $\mathbb{Z}_2^{(1)}$ gauge field $\mathcal{A}_2$. Instead of the naive cup product $\mathcal{A}_2\cup \mathcal{A}_2$ we need to use its refinement given by the Pontryagin square $\mathcal{P}(\mathcal{A}_2)$, see e.g.~the appendix of \cite{Kapustin:2013qsa} for a review for physicists.\label{foot:Pontryagin}} In both cases, they specify the TQFTs that need to be stacked on the $C^{(r)}$ defects, and use them to compute the non-invertible fusion of such topological defects.

\paragraph{\texorpdfstring{$r=0$}{r=0} and self-duality}
It was argued in \cite{Kaidi:2021xfk} that the theory $\widehat{\mathcal{T}}$, in the examples \eqref{eq:appKOZ} and \eqref{eq:KOZ} above, has a notion of $\textit{self-duality}$.\footnote{A similar phenomenon is also discussed in \cite{Lu:2024lzf} for more general symmetries, albeit in $d=2$ dimensions.} This is a duality between the theory $\widehat{\mathcal{T}}$ and a theory obtained from $\widehat{\mathcal{T}}$ with certain manipulations involving the dual symmetries. Using the formalism of this section, we can easily see that this duality exists for any theory with an anomaly of the type \eqref{ano-n0} with $r=0$. In particular, we will now show that the invariance of the partition function under the duality is a consequence of equation \eqref{cond-def}. 

The case $r=0$ is an interesting limiting case of our setup, in which $D$ is a $(-1)$-form symmetry \cite{Vandermeulen:2022edk,Aloni:2024jpb}. After gauging the $A^{(p)}\times B^{(q)}$ symmetry, the operator $U_d$ in eq.~\eqref{aub} can be understood as a new term in the action, associated with a discrete parameter $d$. This parameter thus labels different theories $\widehat{\mathcal{T}}_d$, giving rise to the $(-1)$-form symmetry. The defects of $C^{(0)}$ are now codimension-one, and can be interpreted as topological interfaces between theories with different values of $d$. 
\begin{figure}[t]
    \centering
    \tikzset{every picture/.style={line width=0.75pt}} 

\begin{tikzpicture}[x=0.75pt,y=0.75pt,yscale=-1,xscale=1]


\def\HH{200}
\def\LL{325}
\coordinate (AA) at (335,34);
\coordinate (BB) at ($(AA)+(\LL,0)$);
\coordinate (CC) at ($(AA)+(\LL,\HH)$);
\coordinate (DD) at ($(AA)+(0,\HH)$);
\draw  [draw opacity=0][fill={rgb, 255:red, 155; green, 155; blue, 155 }  ,fill opacity=0.47 ] 
(AA) -- (BB) -- (CC) -- (DD) -- cycle ;
\draw [color={rgb, 255:red, 208; green, 2; blue, 27 }  ,draw opacity=1 ][line width=2.5]    
(AA) -- (DD) ;

\draw (11,92) node [anchor=north west][inner sep=0.75pt] [font=\small]   
{\(\displaystyle
\begin{aligned}
Z & {}_{\widehat{\mathcal{T}}_{0}}
\left[\widehat{\mathcal{A}}_{d-p-1} ,\widehat{\mathcal{B}}_{d-q-1}\right] \ \propto \sum _{\mathscr{a}_{p+1} ,\ \mathscr{b}_{q+1}} \! Z_{\mathcal{T}}[ \mathscr{a}_{p+1} ,\mathscr{b}_{q+1}]\\[1em]
& \exp\left( 2\pi i\ \int \mathscr{a}_{p+1} \cup \widehat{\mathcal{A}}_{d-p-1} +\mathscr{b}_{q+1} \cup \widehat{\mathcal{B}}_{d-q-1}\right)
\end{aligned}
\)
};

\draw (343,60) node [anchor=north west][inner sep=0.75pt] [font=\small]
{
\(\displaystyle
\begin{aligned}
Z & {}_{\widehat{\mathcal{T}}_{d}}
\left[\widehat{\mathcal{A}}_{d-p-1} ,\widehat{\mathcal{B}}_{d-q-1}\right] \ \propto \sum _{\mathscr{a}_{p+1} ,\ \mathscr{b}_{q+1}} \! Z_{\mathcal{T}}[ \mathscr{a}_{p+1} ,\mathscr{b}_{q+1}]\\[1em]
&\exp\left( 2\pi i\ \int \mathscr{a}_{p+1} \cup \widehat{\mathcal{A}}_{d-p-1} +\mathscr{b}_{q+1} \cup \widehat{\mathcal{B}}_{d-q-1}\right)\\[1em]
&
\exp\left( 2\pi i\ d\cdot \int \mathscr{a}_{p+1} \cup _{\mathbb{Z}} \mathscr{b}_{q+1}\right)
\end{aligned}
\)
};

\end{tikzpicture}
    \caption{The red line denotes the insertion of a codimension-1 defect of the $C^{(0)}$ symmetry. We show explicitly the expressions of the partition functions on the two sides of the defect, to point out the insertion of the operator $U_d$ on the right half-space.}
    \label{fig:duality}
\end{figure}

Indeed, recall from the discussion above that the defects of $C^{(0)}$ need to be filled with those of $D^{(-1)}$. In the present context this means that one is forced to insert the operators $U_d$ integrated on a half-space, only on one side of the defect of $C^{(0)}$, see figure \ref{fig:duality}. The existence of a topological interface between the two theories can be interpreted as the statement of a duality between the theories with different values of $d$. To show that this is a self-duality, we will now show that the theory with a certain $d$ can be obtained from $d=0$ with a topological operation.

As previously discussed, $U_d$ operators can be expressed as condensation defects\footnote{Note that in this limiting case the gauging operation is not restricted to a submanifold but rather it operates in the same number of dimensions as the spacetime, thus changing the theory rather than defining a defect.} of the $\widehat{A}^{(d-p-2)}\times \widehat{B}^{(d-q-2)}$ symmetry. Using \eqref{cond-def} to rewrite the insertions of these operators, the partition function of $\widehat{\mathcal{T}}_d$ (see the right side of the interface in figure \ref{fig:duality}) becomes
\begin{equation}
    \begin{split}
        \sum_{\substack{\mathscr{a}_{p+1},\ \mathscr{b}_{q+1}\\\widehat{a}_{d-p-1},\ \widehat{b}_{d-q-1}\\\mathscr{a}'_{p+1},\ \mathscr{b}'_{q+1}}}& Z_{\mathcal{T}}[\mathscr{a}_{p+1},\mathscr{b}_{q+1}]\exp\left(2\pi i\int (\mathscr{a}_{p+1}-\mathscr{a'}_{p+1})\cup \widehat{\mathscr{a}}_{d-p-1}+(\mathscr{b}_{q+1}-\mathscr{b}'_{q+1})\cup \widehat{\mathscr{b}}_{d-q-1}\right) \\
        &\exp\left(2\pi i \ d\cdot\int \mathscr{a}'_{p+1} \cup_\mathbb{Z} \mathscr{b}'_{q+1}\right)\exp\left(2\pi i \int\mathscr{a}'_{p+1}\cup \widehat{\mathcal{A}}_{d-p-1}+ \mathscr{b}'_{q+1}\cup \widehat{\mathcal{B}}_{d-q-1}\right) \ . 
    \end{split}
\end{equation}
Integrating out $\widehat{a}_{d-p-1}, \ \widehat{b}_{d-q-1}$ one goes back to the previous expression for the partition function of $\widehat{\mathcal{T}}_d$. On the other hand, if we sum over $\mathscr{a}_{p+1}, \ \mathscr{b}_{q+1}$, we obtain a new expression in which the partition function of $\widehat{\mathcal{T}}_{0}$ (i.e.~the theory with trivial discrete torsion, in the left side of figure \ref{fig:duality}) appears in the summand, namely
\begin{equation}\label{eq::self-duality-final}
    \begin{split}
        \sum_{\substack{\widehat{a}_{d-p-1},\ \widehat{b}_{d-q-1}\\\mathscr{a}'_{p+1},\ \mathscr{b}'_{q+1}}} &Z_{\widehat{\mathcal{T}}_0}\ [\widehat{\mathscr{a}}_{d-p-1},\widehat{\mathscr{b}}_{d-q-1}] 
        \exp \left( 2\pi i \int 
        \mathscr{a}'_{p+1} \cup( \widehat{\mathcal{A}}_{d-p-1}-\widehat{\mathscr{a}}_{q+1})
        \right.\\ 
        &\left.+ \mathscr{b}'_{q+1}\cup( \widehat{\mathcal{B}}_{d-q-1}-\widehat{\mathscr{b}}_{p+1})\right)
        \exp\left( 2\pi i \ d\cdot  \int \mathscr{a}_{p+1}'\cup_\mathbb{Z} \mathscr{b}'_{q+1} \right) \ .
    \end{split}
\end{equation}
This formula can be interpreted as the partition function of a theory obtained from $\widehat{\mathcal{T}}_0$ with a topological manipulation. We thus proved the existence of a topological interface between this manipulation of $\widehat{\mathcal{T}}_0$ and $\widehat{\mathcal{T}}_0$ itself, i.e.~the two theories are dual.

\subsubsection{\texorpdfstring{$n=1$}{n=1}, extension by condensation defect}
\label{sec: n=1_ext}

As a second example, we focus on the $n=1$ case, where the anomaly \eqref{eq:anomaly} becomes
\begin{equation}\label{ano-n1}
    S_{\mathrm{inflow}}=2\pi i \int_{\mathcal{Y}_{d+1}} \mathcal{A}_{p+1}\cup_\mathbb{Z} \mathcal{B}_{q+1}\cup 
    t_{r+2}(C_{r+1}) \ .
\end{equation}
In this case, the relationship \eqref{extEq} between backgrounds in the $\widehat{\mathcal{T}}$ theory is
\begin{equation}\label{ExtEq_n1}
    \delta \mathcal{D}_{r+1}=(-1)^{p+q}t_{r+2}(C_{r+1}) \ .
\end{equation}
The cohomology class $t_{r+2}(C_{r+1})$ is supported on a codimension-$(r+2)$ locus
\begin{equation}\label{eq:tr2supp}
t_{r+2}(C_{r+1}) = \delta^{(r+2)}(\Sigma_{d-r-2})\,\,d~,~~d\in D~.
\end{equation}
Plugging \eqref{eq:tr2supp} into \eqref{ExtEq_n1} we obtain that a defect of the $D^{(r)}$ symmetry, associated with the group element $d$, needs to be inserted on a cycle ending on $\Sigma_{d-r-2}$. 

When $r=0$, the most general class $t_2 \in H^2(BC,D)$ is activated on the submanifold $\Sigma_{d-2}$, singled out by the intersection of two codimension-one topological defects of the $C^{(0)}$ symmetry, labeled by $c_1,\, c_2 \in C$. The element $d$ emanating from such intersection is determined by the group cohomology class evaluated on the two elements of $C^{(0)}$, i.e.~$d=t_2(c_1,c_2)$.
For $r\geq1$ the most general characteristic class $t_{r+2}\in H^{r+2}(B^{r+1}C,D)$ is generically activated by more complicated configurations of topological defects. However, we expect that the simple, non-generic junction that corresponds to adding $r$ transverse dimensions to the $r=0$ case still exists. To understand what class might be activated by such a non-generic configuration we rely on the homomorphism\footnote{To show the existence of this homomorphism, one needs to use the equivalence \cite{Davis:2001woh}
 \begin{equation}
 H^n(X,D) \cong [X,K(D,n)]~.
 \end{equation}
Here, the notation $[X_1,X_2]$ denotes homotopy classes of maps between topological spaces $X_1$ and $X_2$. Next, we need to consider the notion of based loop space $\Omega X$ of a topological space $X$, defined by $\Omega X=\{\gamma:[0,1]\rightarrow X, \ \gamma(0)=\gamma(1)=x_0\}$. The operation $\Omega$ induces a map 
  \begin{equation}
      [X,Y]\rightarrow [\Omega X, \Omega Y] \ .
  \end{equation}
Using that classifying spaces for higher-form symmetries can always be seen as Eilenberg-MacLane spaces, and $\Omega K(-,n)\cong K(-,n-1)$ \cite{Davis:2001woh}, we obtain the desired map
\begin{equation}
\begin{split}
     H^n(B^{k+1}C,D)\cong[K(\widehat{C},k'+1),K(D,n)] \rightarrow &[\Omega K(\widehat{C},k'+1), \Omega K(D,n)]\\
     &\cong[ K(\widehat{C},k'), K(D,n-1)]\cong H^{n-1}(B^kC,D) \ ,
\end{split}
\end{equation}
where $k'=k$ and $\widehat{C}=C$ if $C$ is discrete, while $k'=k+1$ and $\widehat{C}=\mathbb{Z}$ if $C=U(1)$, see footnote \ref{EML}. 
 }
\begin{equation}\label{eq:Phihom}
    \Phi^k:H^n(B^{k+1}C,D)\rightarrow H^{n-k}(BC,D) \ .
\end{equation}
Using this homomorphism, a characteristic class in $H^{r+2}(B^{r+1}C,D)/\text{Ker}(\Phi^{r})$ can be activated by the non-generic junction mentioned above, i.e.~by the intersection $\Sigma_{d-r-2}$ of two codimension-$(r+1)$ topological defects of the $C^{(r)}$ symmetry. The defect of 
the $D^{(r)}$ symmetry emanating from the intersection is determined by the group cohomology class $d=\Phi^r[t_{r+2}](c_1,c_2)$.
\begin{figure}[t]
    \centering
    \tikzset{every picture/.style={line width=0.75pt}} 

\begin{tikzpicture}[x=0.75pt,y=0.75pt,yscale=-1,xscale=1]

\draw [color={rgb, 255:red, 245; green, 166; blue, 35 }  ,draw opacity=1 ]   (240.22,130.38) -- (186.22,206.38) ;
\draw [color={rgb, 255:red, 245; green, 166; blue, 35 }  ,draw opacity=1 ]   (240.22,130.38) -- (289.89,205.05) ;
\draw [color={rgb, 255:red, 245; green, 166; blue, 35 }  ,draw opacity=1 ]   (235.56,54.38) -- (240.22,130.38) ;
\draw [color={rgb, 255:red, 74; green, 144; blue, 226 }  ,draw opacity=1 ]   (288.89,67.05) -- (240.22,130.38) ;
\draw  [draw opacity=0][fill={rgb, 255:red, 74; green, 144; blue, 226 }  ,fill opacity=1 ] (236.73,130.88) .. controls (236.73,128.95) and (238.21,127.38) .. (240.03,127.38) .. controls (241.86,127.38) and (243.33,128.95) .. (243.33,130.88) .. controls (243.33,132.81) and (241.86,134.38) .. (240.03,134.38) .. controls (238.21,134.38) and (236.73,132.81) .. (236.73,130.88) -- cycle ;

\draw (175,172) node [anchor=north west][inner sep=0.75pt]   [align=left] {$\displaystyle \textcolor[rgb]{0.96,0.65,0.14}{c_{1}}$};
\draw (291,177) node [anchor=north west][inner sep=0.75pt]   [align=left] {$\displaystyle \textcolor[rgb]{0.96,0.65,0.14}{c_{2}}$};
\draw (200,58) node [anchor=north west][inner sep=0.75pt]   [align=left] {$\displaystyle \textcolor[rgb]{0.96,0.65,0.14}{c}\textcolor[rgb]{0.96,0.65,0.14}{_{1} c_{2}}$};
\draw (274,97) node [anchor=north west][inner sep=0.75pt]  [color={rgb, 255:red, 74; green, 144; blue, 226 }  ,opacity=1 ] [align=left] {$d$};

\end{tikzpicture}
    \caption{Illustration of a junction between topological defects that reflects the existence of a group extension. The yellow lines are the topological defects associated with $C^{(r)}$. When we fuse two of them, we generate a third $C^{(r)}$ defect and a defect for $D^{(r)}$ (in blue) which is fixed by $d=\Phi^r[t_{r+2}](c_1,c_2)$.
    }
    \label{local-def-ext}
\end{figure}
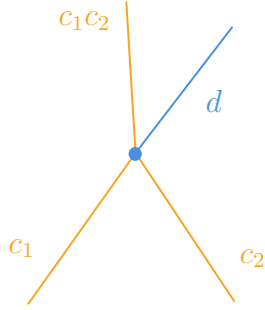

Reading from bottom to top, figure \ref{local-def-ext} illustrates that the fusion of two $C^{(r)}$ defects produces both $C^{(r)}$ and $D^{(r)}$ defects. As a result, the two $r$-form symmetries give rise to a larger $r$-form symmetry $\Gamma^{(r)}$ via the short exact sequence
\begin{equation}
    1 \rightarrow D^{(r)} \rightarrow \Gamma^{(r)} \rightarrow C^{(r)} \rightarrow 1 \ ,
\end{equation}
described by the class $t_{r+2}\in H^{r+2}(B^{r+1}C,D)$.
Hence, in the $n=1$ scenario the topological defects of $C^{(r)}$ (on closed manifolds) are still present in the $\widehat{\mathcal{T}}$ theory as invertible topological defects, but as part of a larger symmetry. That is to say, the $r$-form symmetry $C^{(r)}$ of the $\mathcal{T}$ theory is no longer a symmetry by itself, but rather it is extended by the $D^{(r)}$ symmetry.

In the $n=1$ case, the symmetry structure of the $\widehat{\mathcal{T}}$ theory in $d=3$ (where the only option is $p=q=r=0$) was considered in \cite{Tachikawa:2017gyf}, which however did not discuss the extension by the condensation defects. Here we do not attempt a description of the full set of topological defects and their fusions, but rather we focus on the fate of topological defects of the $C^{(r)}$ symmetry, leaving $d,p,q,r$ arbitrary. An example of the anomaly \eqref{ano-n1} was discussed in \cite{BenettiGenolini:2020doj} for five-dimensional gauge theories and, when tractable, their UV completions. For instance, for pure $SU(N)$ Yang-Mills theory the anomaly inflow reads 
\begin{equation}\label{eq:inflow-YM5d}
    S_{\mathrm{inflow}}=\frac{2\pi i }{N}\int_{\mathcal{Y}_6}\frac{\mathcal{P}(\mathcal{B}_2)}{2}\cup \left[ \frac{dA_1}{2\pi}\right]_N\ ,
\end{equation}
where $\mathcal{B}_2 \in H^2(\mathcal{Y}_{d+1},\mathbb Z_N)$ is the background for the electric 1-form symmetry and $A_1$ is the background for the $U(1)_I$ instantonic 0-form symmetry. The above anomaly is of the form \eqref{ano-n1} in $d=5$, with $A^{(1)}\equiv B^{(1)}=\mathbb{Z}_N^{(1)}$ (see footnote \ref{foot:Pontryagin}), $C^{(0)}=U(1)^{(0)}$ and with $t_2$ being the first Chern class of the $U(1)^{(0)}$ bundle modulo $N$. In \cite{Bertolini:2025wyj} the authors consider the gauging of the electric 1-form symmetry and argue that the instantonic symmetry is indeed extended to $\widetilde{U(1)}^{(0)}_I$\footnote{See also \cite{Sheckler:2025fql} for non-invertible fusion rules of topological defects on open submanifolds.}, that sits in the short exact sequence
\begin{equation}
    1 \rightarrow \mathbb{Z}_N^{(0)} \rightarrow  \widetilde{U(1)}^{(0)} \rightarrow U(1)^{(0)} \rightarrow 1 \ ,
\end{equation}
i.e.~its periodicity is multiplied by a factor of $N$. Such group extension can be seen at the level of backgrounds from equation \eqref{ExtEq_n1}, which in this example becomes
\begin{equation}
    \delta \mathcal{D}_1=\left[\frac{dA_1}{2\pi}\right]_N \ .
\end{equation}

\subsubsection{\texorpdfstring{$n\geq 2$}{n>=2}, (\textit{r}+\textit{n})-group by condensation defect}
The generalization to larger values of $n$ is straightforward. The equation between background gauge fields \eqref{extEq} means that the $C^{(r)}$ symmetry must sit inside a larger symmetry structure together with the $D^{(n+r-1)}$ symmetry. When a theory has several higher-form symmetries, they can combine to form a higher-group structure. If the symmetry of highest degree is a $k$-form symmetry, then the combined symmetry structure can generically form a $(k+1)$-group, see e.g.~\cite{Cordova:2018cvg,Benini:2018reh,Bhardwaj:2023kri}. Therefore, in this case the $D^{(n+r-1)}$ and the $C^{(r)}$ symmetry form a $(r+n)$-group
\begin{equation}
    1 \rightarrow D^{(n+r-1)} \rightarrow \underline{\Gamma} \rightarrow C^{(r)} \rightarrow 1 \ ,
\end{equation}
where the interplay is governed by the class $t_{n+r+1}\in H^{n+r+1}(B^{r+1}C,D)$. The relationship between background gauge fields is given by the pullback of such characteristic class as in equation \eqref{extEq}. 

For example, for $n=2$ the symmetries 
mix to form a $(r+2)$-group, where the cohomology class 
\begin{equation}\label{eq:tr3supp}
t_{r+3}(C_{r+1}) = \delta^{(r+3)}(\Sigma_{d-r-3})\,\, d~,~~d\in D~
\end{equation}
is supported on a codimension-$(r+3)$ submanifold. Unlike the cases of $n=0,1$, to our knowledge this structure with $n\geq2$ was not considered previously in the literature. 

\begin{figure}[t]
    \centering
    \tikzset{every picture/.style={line width=0.75pt}} 

\begin{tikzpicture}[x=0.75pt,y=0.75pt,yscale=-1,xscale=1]

\draw  [draw opacity=0] (415.33,82.23) .. controls (387.77,99.15) and (351.52,109.48) .. (311.78,109.6) .. controls (274.05,109.72) and (239.41,100.62) .. (212.32,85.37) -- (311.44,1.4) -- cycle ; \draw  [color={rgb, 255:red, 245; green, 166; blue, 35 }  ,draw opacity=1 ] (415.33,82.23) .. controls (387.77,99.15) and (351.52,109.48) .. (311.78,109.6) .. controls (274.05,109.72) and (239.41,100.62) .. (212.32,85.37) ;  
\draw  [draw opacity=0][dash pattern={on 4.5pt off 4.5pt}] (415.33,236.51) .. controls (387.77,253.43) and (351.52,263.76) .. (311.78,263.89) .. controls (274.05,264.01) and (239.41,254.91) .. (212.32,239.66) -- (311.44,155.69) -- cycle ; \draw  [color={rgb, 255:red, 245; green, 166; blue, 35 }  ,draw opacity=1 ][dash pattern={on 4.5pt off 4.5pt}] (415.33,236.51) .. controls (387.77,253.43) and (351.52,263.76) .. (311.78,263.89) .. controls (274.05,264.01) and (239.41,254.91) .. (212.32,239.66) ;  
\draw [color={rgb, 255:red, 245; green, 166; blue, 35 }  ,draw opacity=1 ]   (212.32,85.37) -- (212.32,239.66) ;
\draw [color={rgb, 255:red, 245; green, 166; blue, 35 }  ,draw opacity=1 ]   (415.32,82.37) -- (415.32,236.66) ;
\draw [color={rgb, 255:red, 245; green, 166; blue, 35 }  ,draw opacity=1 ] [dash pattern={on 4.5pt off 4.5pt}]  (354.8,104.74) .. controls (366,166.34) and (282,208.74) .. (283.6,262.34) ;
\draw [color={rgb, 255:red, 245; green, 166; blue, 35 }  ,draw opacity=1 ] [dash pattern={on 4.5pt off 4.5pt}]  (318,28.74) .. controls (329.2,90.34) and (245.2,132.74) .. (246.8,186.34) ;
\draw [color={rgb, 255:red, 245; green, 166; blue, 35 }  ,draw opacity=1 ] [dash pattern={on 4.5pt off 4.5pt}]  (246.8,186.34) -- (283.6,262.34) ;
\draw [color={rgb, 255:red, 245; green, 166; blue, 35 }  ,draw opacity=1 ]   (318,28.74) -- (354.8,104.74) ;
\draw [color={rgb, 255:red, 245; green, 166; blue, 35 }  ,draw opacity=1 ]   (320.4,109.34) -- (355.6,181.34) ;
\draw [color={rgb, 255:red, 245; green, 166; blue, 35 }  ,draw opacity=1 ]   (319.6,263.74) -- (354.8,335.74) ;
\draw [color={rgb, 255:red, 245; green, 166; blue, 35 }  ,draw opacity=1 ]   (320.4,109.34) -- (319.6,263.74) ;
\draw [color={rgb, 255:red, 245; green, 166; blue, 35 }  ,draw opacity=1 ]   (355.6,181.34) -- (354.8,335.74) ;
\draw [color={rgb, 255:red, 74; green, 144; blue, 226 }  ,draw opacity=1 ]   (320.4,190.6) .. controls (308,226.67) and (302.8,281) .. (309.2,310.6) ;
\draw  [draw opacity=0][fill={rgb, 255:red, 74; green, 144; blue, 226 }  ,fill opacity=1 ] (316.73,191.3) .. controls (316.73,189.37) and (318.21,187.8) .. (320.03,187.8) .. controls (321.86,187.8) and (323.33,189.37) .. (323.33,191.3) .. controls (323.33,193.24) and (321.86,194.8) .. (320.03,194.8) .. controls (318.21,194.8) and (316.73,193.24) .. (316.73,191.3) -- cycle ;
\draw [color={rgb, 255:red, 245; green, 166; blue, 35 }  ,draw opacity=1 ]   (318,28.74) .. controls (325.2,59.94) and (299.6,94.34) .. (287.8,107.94) ;
\draw  [draw opacity=0] (320.11,263.71) .. controls (317.35,263.82) and (314.57,263.88) .. (311.78,263.89) .. controls (274.05,264.01) and (239.41,254.91) .. (212.32,239.66) -- (311.44,155.69) -- cycle ; \draw  [color={rgb, 255:red, 245; green, 166; blue, 35 }  ,draw opacity=1 ] (320.11,263.71) .. controls (317.35,263.82) and (314.57,263.88) .. (311.78,263.89) .. controls (274.05,264.01) and (239.41,254.91) .. (212.32,239.66) ;  
\draw  [draw opacity=0] (415.32,236.66) .. controls (398.44,247.02) and (378.31,254.91) .. (356.11,259.5) -- (311.43,155.83) -- cycle ; \draw  [color={rgb, 255:red, 245; green, 166; blue, 35 }  ,draw opacity=1 ] (415.32,236.66) .. controls (398.44,247.02) and (378.31,254.91) .. (356.11,259.5) ;  

\draw (187.6,91.74) node [anchor=north west][inner sep=0.75pt]   [align=left] {$\displaystyle \textcolor[rgb]{0.96,0.65,0.14}{c_{1}}$};
\draw (333.2,29.34) node [anchor=north west][inner sep=0.75pt]   [align=left] {$\displaystyle \textcolor[rgb]{0.96,0.65,0.14}{c}\textcolor[rgb]{0.96,0.65,0.14}{_{2}}$};
\draw (427.6,79.74) node [anchor=north west][inner sep=0.75pt]   [align=left] {$\displaystyle \textcolor[rgb]{0.96,0.65,0.14}{c}\textcolor[rgb]{0.96,0.65,0.14}{_{3}}$};
\draw (360.6,310.14) node [anchor=north west][inner sep=0.75pt]   [align=left] {$\displaystyle \textcolor[rgb]{0.96,0.65,0.14}{c}\textcolor[rgb]{0.96,0.65,0.14}{_{1} c_{2} c_{3}}$};
\draw (280,291.74) node [anchor=north west][inner sep=0.75pt]   [align=left] {$\displaystyle \textcolor[rgb]{0.29,0.56,0.89}{d}$};

\end{tikzpicture}
    \caption{Illustration of a junction between topological defects that reflects the existence of a $(r+2)$-group structure. The yellow planes are the topological defects associated with $C^{(r)}$. When we fuse three of them we generate a fourth $C^{(r)}$ defect and a defect for $D^{(r+1)}$ (represented as a blue line) which is fixed by $d=\Phi^r\left[t_{r+3}\right](c_1,c_2,c_3)$.}
    \label{local-def-2grp}
\end{figure}
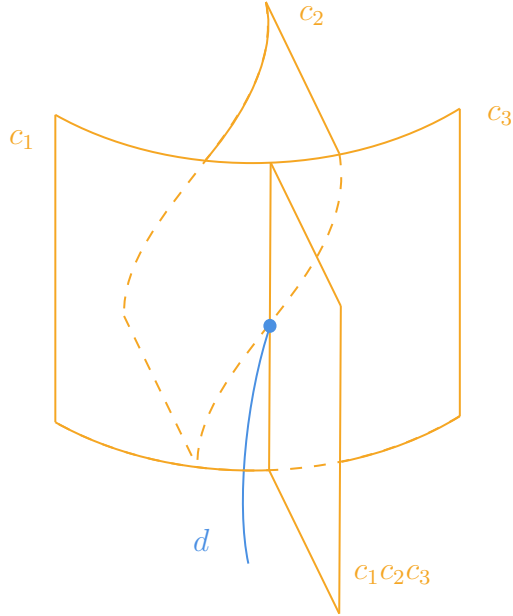

As discussed in section \ref{sec: n=1_ext}, when $r=0$ this submanifold can generically be identified with the intersection of three codimension-1 topological defects of the $C^{(0)}$ symmetry. From such a submanifold a topological defect of $D^{(1)}$ determined by the third group cohomology class emanates. A three-dimensional illustration is provided in figure \ref{local-def-2grp}. Instead, when $r\geq 1$ we rely on the homomorphism \eqref{eq:Phihom} in order to interpret the classes in $H^{r+3}(B^{r+1}C,D)/\text{Ker}(\Phi^r)$ as locally sourced by  the intersection $\Sigma_{d-r-3}$ of three codimension-$(r+1)$ topological defects $c_1, \, c_2, \, c_3$ of $C^{(r)}$. The defect of $D^{(r+1)}$ emanating from such intersection is determined by $d=\Phi^r\left[t_{r+3}\right](c_1,c_2,c_3)$.

\subsection{Charged operators and states}
\label{States}

After discussing the symmetry structure in the theory $\widehat{\mathcal{T}}$ in terms of background gauge fields, we now move on to the description of its action on charged operators, i.e.~its (higher-) representations \cite{Bartsch:2023pzl,Bhardwaj:2023wzd}. Our goal is to generalize the notions of charged operators and states from the standard cases of extensions and higher-groups, to the present case, where these structures are generated by condensation defects. The standard case is reviewed in appendix \ref{App-Reps}. There, we show how to use configurations of (non-topological) charged operators to prepare, in suitable Hilbert spaces, states charged under the given symmetry structure.

We will proceed by first discussing charged operators and states for the symmetry generated by the condensation defect. Then, building on this, we will discuss charged operators and states when the symmetry generated by the condensation defect is embedded, due to the anomaly, in a larger structure. We will do it first in the simpler case of the extension, and then for the higher-group.

\subsubsection{Operators and states charged under the condensation defect}
 The action of the $D^{(n+r-1)}$ symmetry defects $U_d$ is largely determined by the fact that they are condensation defects of $\widehat{A}^{(d-p-2)}\times \widehat{B}^{(d-q-2)}$, as we now discuss. This action is independent of the presence of the anomaly in the parent theory $\mathcal{T}$, whose consequence will be discussed later. 

\begin{figure}[t]
    \centering
    \tikzset{every picture/.style={line width=0.75pt}} 

\begin{tikzpicture}[x=0.75pt,y=0.75pt,yscale=-1,xscale=1]

\draw    (105.13,97.73) -- (179.33,96.89) ;
\draw    (201.13,96.73) -- (275.33,96.89) ;
\draw [color={rgb, 255:red, 189; green, 16; blue, 224 }  ,draw opacity=1 ]   (189,36.22) -- (189.33,145.56) ;
\draw   (146.67,92.89) -- (159.33,96.78) -- (146.67,100.67) ;
\draw  [color={rgb, 255:red, 189; green, 16; blue, 224 }  ,draw opacity=1 ] (185.28,77.21) -- (188.83,64.45) -- (193.05,77.01) ;
\draw    (369.13,97.73) -- (533.33,96.89) ;
\draw   (418.67,93.89) -- (431.33,97.78) -- (418.67,101.67) ;
\draw [color={rgb, 255:red, 189; green, 16; blue, 224 }  ,draw opacity=1 ]   (451,36.22) -- (451.33,89.56) ;
\draw  [color={rgb, 255:red, 189; green, 16; blue, 224 }  ,draw opacity=1 ] (447.28,77.21) -- (450.83,64.45) -- (455.05,77.01) ;
\draw [color={rgb, 255:red, 189; green, 16; blue, 224 }  ,draw opacity=1 ]   (451,104.22) -- (451.33,144.56) ;

\draw (249.67,78) node [anchor=north west][inner sep=0.75pt]   [align=left] {$\displaystyle \alpha $};
\draw (172,41.67) node [anchor=north west][inner sep=0.75pt]  [color={rgb, 255:red, 189; green, 16; blue, 224 }  ,opacity=1 ] [align=left] {$\displaystyle \widehat{\textcolor[rgb]{0.74,0.06,0.88}{a}}$};
\draw (434,41.67) node [anchor=north west][inner sep=0.75pt]   [align=left] {$\displaystyle \textcolor[rgb]{0.74,0.06,0.88}{\widehat{a}}$};
\draw (283.33,86) node [anchor=north west][inner sep=0.75pt]   [align=left] {$\displaystyle =\ e^{2\pi i\, \widehat{a}\cdot \varphi _{\alpha }}$};
\draw (507.67,78) node [anchor=north west][inner sep=0.75pt]   [align=left] {$\displaystyle \alpha $};

\end{tikzpicture}
    \caption{Two-component linking action of $U_{\widehat{a}}[\Sigma_{p+1}]$ on $\mathcal{O}_\alpha[\gamma_{d-p-2}]$.}
    \label{fig:action-hatA}
\end{figure}
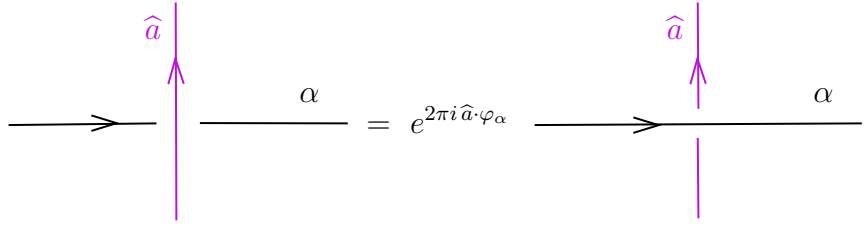

Let us first fix some notation. We label with an index $\alpha$ operators $\mathcal{O}_\alpha(\gamma_{d-p-2})$ charged under the symmetry $\widehat{A}^{(d-p-2)}$, which is generated by topological operators 
\begin{equation}
    U_{\widehat{a}}(\Sigma_{p+1})=\exp(2\pi i\ \widehat{a}\cdot\int_{\Sigma_{p+1}}\mathscr{a}_{p+1})\ .
\end{equation}
The action of $U_{\widehat{a}}$ on the charged operators ${\cal O}_\alpha$ is via two-component linking as illustrated in figure \ref{fig:action-hatA}.
Equivalently, we can say that in the presence of the operator $\mathcal{O}_\alpha$, the flatness condition of the gauge field $\mathscr{a}_{p+1}$ is modified into 
\begin{equation}
    \delta \mathscr{a}_{p+1}=\varphi_\alpha \delta(\gamma_{d-p-2})\ .
\end{equation}
Here the coefficient $\varphi_\alpha \in A$ denotes the charge of the operator $\mathcal{O}_\alpha$. The cocycle $\delta(\gamma_{d-p-2})\in H^{p+2}(\mathcal{M}_d,\mathbb{Z})$ is defined by the property that it associates $1$ to $(p+2)$-cycles that link once with $\gamma_{d-p-2}$, and vanishes elsewhere. Its product with $\varphi_\alpha$ is defined by viewing $A$ as a $\mathbb{Z}$ module. We then have
\begin{equation}\label{eq:UO}
\begin{split}
     U_{\widehat{a}}(\Sigma_{p+1}) \mathcal{O}_\alpha(\gamma_{d-p-2})&=
     \exp(
     2\pi i \ \widehat{a}\cdot \int_{D_{p+2}}\varphi_\alpha \delta(\gamma_{d-p-2}) 
     )
     \mathcal{O}_\alpha(\gamma_{d-p-2})\\
     &=
     e^{-2\pi i\, \widehat{a}\cdot  \varphi_\alpha L_2(\gamma_{d-p-2},\Sigma_{p+1})}
     \mathcal{O}_\alpha(\gamma_{d-p-2})\ ,
\end{split}
\end{equation}
where $\partial D_{p+2}=\Sigma_{p+1}$, and
\begin{equation}
    L_2(\Sigma_{p+1},\gamma_{d-p-2})=\int_{\mathcal{M}_d}\delta(D_{p+2})\wedge 
    d\delta(M_{d-p-1})
\end{equation}
with $\partial M_{d-p-1}=\gamma_{d-p-2}$, is the integer counting the two-component link invariant between the two submanifolds, and we switched to the differential form definition of $\delta$. Equation \eqref{eq:UO} is an operator equation that can be used inside correlation functions. Analogous equations hold for the $\widehat{B}^{(d-q-2)}$ symmetry and its charged operators.

\paragraph{Operators}  As suggested by their formulation \eqref{cond-def} as condensation defects, the topological operators $U_d$ do not act on $(n+r-1)$-dimensional operators by two-component linking as a standard $(n+r-1)$-form symmetry. Instead, we will show that they act nontrivially through triple linking in configurations involving two operators charged under the symmetries participating in the condensation defect. 

Given operators $\mathcal{O}_\alpha[\gamma_{d-p-2}]$ and $\mathcal{O}_\beta[\gamma_{d-q-2}]$, charged respectively under $\widehat{A}$ and $\widehat{B}$, the following operator equation holds
\begin{equation}\label{TLK}  
\begin{split}
&\exp \bigg(2\pi i \ d \cdot\int_{\Sigma_{p+q+2}}\mathscr{a}_{p+1}\cup_\mathbb{Z} \mathscr{b}_{q+1}\bigg)  \mathcal{O}_\alpha[\gamma_{d-p-2}] \, \mathcal{O}_\beta[\gamma_{d-q-2}]\\
& =
\exp\bigg((-1)^{p+q} \,2\pi i
\  d \cdot\int_{\Sigma_{p+q+2}}\!\!\varphi_\alpha\delta^{(p+1)}(M_{d-p-1})\cup_\mathbb{Z} \varphi_\beta \delta^{(q+1)}(M_{d-q-1})\bigg) \\
& \hspace{8.8cm}\times\mathcal{O}_\alpha[\gamma_{d-p-2}] \, \mathcal{O}_\beta[\gamma_{d-q-2}] \\
& =
\exp\bigg(-2\pi i \ d \cdot (\varphi_\alpha\otimes_\mathbb{Z}\varphi_\beta )L_3(\Sigma_{p+q+2},\gamma_{d-p-2},\gamma_{d-q-2})\bigg) \mathcal{O}_\alpha[\gamma_{d-p-2}] \, \mathcal{O}_\beta[\gamma_{d-q-2}] \ ,
\end{split}
\end{equation}
where $\partial M_{d-p-1} =\gamma_{d-p-2}, \,\partial M_{d-q-1} =\gamma_{d-q-2} $, and 
\begin{equation}\label{TLformula}
     L_3(\Sigma_{p+q+2},\gamma_{d-p-2},\gamma_{d-q-2})=\int_{\mathcal{M}_d} \delta(M_{d-p-1}) \wedge \delta(M_{d-q-1}) \wedge d \delta (D_{p+q+3})
\end{equation}
is the three-component link invariant\footnote{This is the three-component link invariant of type 1, see the appendix A of \cite{Kaidi:2023maf}. Symmetry actions with higher link invariants appeared in \cite{Antinucci:2024zjp, Arbalestrier:2025poq}.} between the respective submanifolds. 
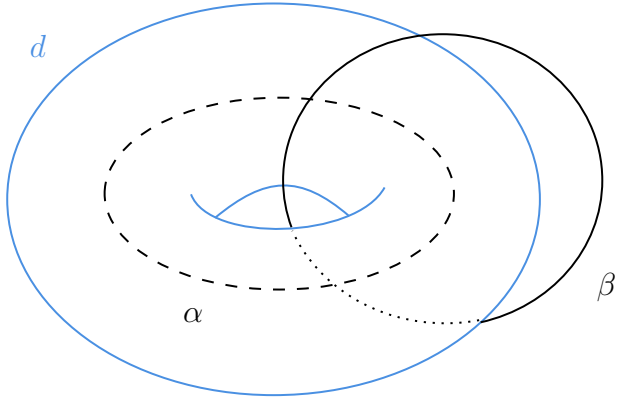
\begin{figure}[t]
    \centering
    \tikzset{every picture/.style={line width=0.75pt}} 

\begin{tikzpicture}[x=0.75pt,y=0.75pt,yscale=-.9,xscale=.9]

\draw  [color={rgb, 255:red, 0; green, 0; blue, 0 }  ,draw opacity=1 ][dash pattern={on 4.5pt off 4.5pt}][line width=0.75]  (241.4,149.32) .. controls (241.4,119.77) and (285.01,95.82) .. (338.8,95.82) .. controls (392.59,95.82) and (436.2,119.77) .. (436.2,149.32) .. controls (436.2,178.87) and (392.59,202.82) .. (338.8,202.82) .. controls (285.01,202.82) and (241.4,178.87) .. (241.4,149.32) -- cycle ;
\draw  [color={rgb, 255:red, 74; green, 144; blue, 226 }  ,draw opacity=1 ] (187.09,152.5) .. controls (187.09,92.19) and (253.58,43.3) .. (335.61,43.3) .. controls (417.64,43.3) and (484.13,92.19) .. (484.13,152.5) .. controls (484.13,212.81) and (417.64,261.7) .. (335.61,261.7) .. controls (253.58,261.7) and (187.09,212.81) .. (187.09,152.5) -- cycle ;
\draw [color={rgb, 255:red, 74; green, 144; blue, 226 }  ,draw opacity=1 ]   (289.6,149.7) .. controls (299.69,176.95) and (381.39,175.07) .. (397.53,145.88) ;
\draw [color={rgb, 255:red, 74; green, 144; blue, 226 }  ,draw opacity=1 ]   (303.02,162.8) .. controls (331.27,139.44) and (350.4,138.5) .. (377.6,161.7) ;
\draw  [draw opacity=0] (345.61,168.06) .. controls (342.58,159.86) and (340.93,151.06) .. (340.93,141.91) .. controls (340.93,96.91) and (380.78,60.43) .. (429.93,60.43) .. controls (479.09,60.43) and (518.93,96.91) .. (518.93,141.91) .. controls (518.93,180.25) and (490.02,212.4) .. (451.07,221.09) -- (429.93,141.91) -- cycle ; \draw   (345.61,168.06) .. controls (342.58,159.86) and (340.93,151.06) .. (340.93,141.91) .. controls (340.93,96.91) and (380.78,60.43) .. (429.93,60.43) .. controls (479.09,60.43) and (518.93,96.91) .. (518.93,141.91) .. controls (518.93,180.25) and (490.02,212.4) .. (451.07,221.09) ;  
\draw  [draw opacity=0][dash pattern={on 0.84pt off 2.51pt}] (450.26,219.88) .. controls (444.16,221.03) and (437.83,221.63) .. (431.35,221.63) .. controls (392.75,221.63) and (359.71,200.29) .. (346.06,170.04) -- (431.35,139.08) -- cycle ; \draw  [dash pattern={on 0.84pt off 2.51pt}] (450.26,219.88) .. controls (444.16,221.03) and (437.83,221.63) .. (431.35,221.63) .. controls (392.75,221.63) and (359.71,200.29) .. (346.06,170.04) ;  

\draw (283.6,211.6) node [anchor=north west][inner sep=0.75pt]   [align=left] {$\displaystyle \alpha $};
\draw (514.2,191.8) node [anchor=north west][inner sep=0.75pt]   [align=left] {$\displaystyle \beta $};
\draw (198,59) node [anchor=north west][inner sep=0.75pt]   [align=left] {$\displaystyle \textcolor[rgb]{0.29,0.56,0.89}{d}$};

\end{tikzpicture}
    \caption{The configuration of charged operators $\mathcal{O}_{\alpha}[\gamma_{d-p-2}]$, $\mathcal{O}_{\beta}[\gamma_{d-q-2}]$ and topological defect $U_d[\Sigma_{p+q+2}]$ that activates the triple linking \eqref{TLformula}. This implies that $U_d[\Sigma_{p+q+2}]$ measures a charge, as in equation \eqref{TLK}.}
    \label{fig:TLK}
\end{figure} 
The above equation shows that the operators $U_d$ act by triple linking on the already existing charged operators of the $\widehat{A}^{(d-p-2)}\times \widehat{B}^{(d-q-2)}$ symmetry.

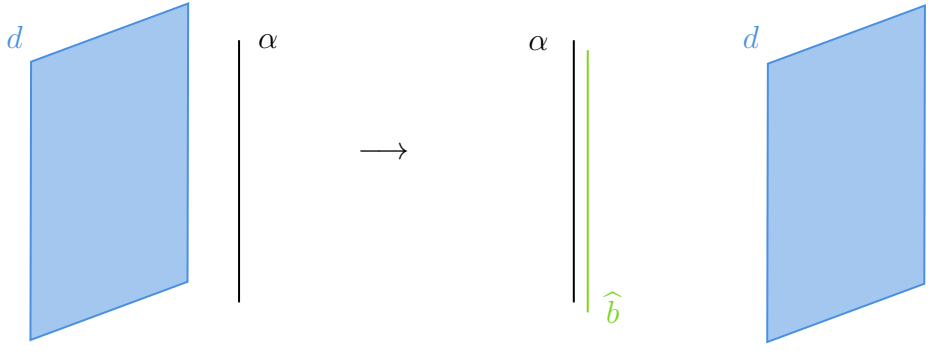
\begin{figure}[t]
     \centering
     \tikzset{every picture/.style={line width=0.75pt}} 

\begin{tikzpicture}[x=0.75pt,y=0.75pt,yscale=-1,xscale=1]

\draw  [color={rgb, 255:red, 74; green, 144; blue, 226 }  ,draw opacity=1 ][fill={rgb, 255:red, 74; green, 144; blue, 226 }  ,fill opacity=0.5 ] (184.38,54.1) -- (184.07,193.77) -- (105.24,222.99) -- (105.55,83.32) -- cycle ;
\draw  [color={rgb, 255:red, 74; green, 144; blue, 226 }  ,draw opacity=1 ][fill={rgb, 255:red, 74; green, 144; blue, 226 }  ,fill opacity=0.5 ] (554.15,55.1) -- (553.83,194.77) -- (475,223.99) -- (475.32,84.32) -- cycle ;
\draw    (209.93,72.54) -- (209.93,204.14) ;
\draw    (377.93,72.54) -- (377.93,204.14) ;
\draw [color={rgb, 255:red, 126; green, 211; blue, 33 }  ,draw opacity=1 ]   (384.93,77.54) -- (384.93,209.14) ;

\draw (91.7,63.87) node [anchor=north west][inner sep=0.75pt]  [color={rgb, 255:red, 74; green, 144; blue, 226 }  ,opacity=1 ] [align=left] {$\displaystyle d$};
\draw (461.32,63.87) node [anchor=north west][inner sep=0.75pt]  [color={rgb, 255:red, 74; green, 144; blue, 226 }  ,opacity=1 ] [align=left] {$\displaystyle d$};
\draw (218,68.06) node [anchor=north west][inner sep=0.75pt]   [align=left] {$\displaystyle \alpha $};
\draw (392.68,197.36) node [anchor=north west][inner sep=0.75pt]  [color={rgb, 255:red, 245; green, 166; blue, 35 }  ,opacity=1 ] [align=left] {$\displaystyle \textcolor[rgb]{0.49,0.83,0.13}{\widehat{b}}$};
\draw (267.98,124.61) node [anchor=north west][inner sep=0.75pt]   [align=left] {$\displaystyle \longrightarrow $};
\draw (353.69,69.06) node [anchor=north west][inner sep=0.75pt]   [align=left] {$\displaystyle \alpha $};

\end{tikzpicture}
     \caption{As we slide the condensation defect $U_d[\Sigma_{p+q+2}] $ through the operator $\mathcal{O}_{\alpha}[\gamma_{d-p-2}]$, a charge $U_{\widehat{b}}[\widetilde{\Sigma}_{p+q+3} \cap \gamma_{d-p-2}]$ is deposited, where $\widehat{b}=(-1)^{p(d+1)}d\cdot(\varphi_\alpha \otimes_\mathbb{Z} -)$.}
     \label{fig:charge-deposit}
\end{figure}

Another way to understand the action of the symmetry $D^{(n+r-1)}$ is to analyze what happens when we move its charge operator across an operator charged under either $\widehat{A}^{(d-p-2)}$ or $\widehat{B}^{(d-q-2)}$, say for definiteness $\mathcal{O}_\alpha[\gamma_{d-p-2}]$. This process is described by the equation
\begin{equation}\label{charge_dep}
 \begin{split}
     \exp & \bigg(2\pi i \ d \cdot\int_{\Sigma_{p+q+2}}\mathscr{a}_{p+1}\cup_\mathbb{Z} \mathscr{b}_{q+1}\bigg)  \mathcal{O}_\alpha[\gamma_{d-p-2}]\\
    & = \exp \bigg(2\pi i \ d \cdot\int_{\Sigma_{p+q+2}'}\mathscr{a}_{p+1}\cup_\mathbb{Z} \mathscr{b}_{q+1}\bigg) \\
    &  \hspace{0.4cm}\times \exp \bigg(2\pi i \ d \cdot\int_{\widetilde{\Sigma}_{p+q+3}}\varphi_\alpha\delta^{(p+2)}(\gamma_{d-p-2})\cup_\mathbb{Z} \mathscr{b}_{q+1}\bigg)  \mathcal{O}_\alpha[\gamma_{d-p-2}] \ ,
\end{split}
\end{equation}
where $\partial \widetilde{\Sigma}_{p+q+3}=\Sigma_{p+q+2}-\Sigma_{p+q+2}'$, see figure \ref{fig:charge-deposit} for a cartoon. We do not cross operators charged under $\widehat{B}^{(d-q-2)}$ when deforming $\Sigma_{p+q+2}$, hence we used $\delta\mathscr{b}_{q+1} = 0$ in the manipulations above.
We interpret the additional factor on the right-hand side as a charge operator of the $\widehat{B}^{(d-q-2)}$ symmetry.
The associated element of $\widehat{B}$ is specified by the function $(-1)^{p(d+1)}d\cdot(\varphi_\alpha \otimes_\mathbb{Z} -)$ from $B$ to $U(1)$. Equivalently, as we move $U_d$ across $\mathcal{O}_\alpha$ we deposit a charge of $\widehat{B}^{(d-q-2)}$ on top of it.
When we consider a triple-linking configuration as in \eqref{TLK}, the charge deposited on $\mathcal{O}_\alpha$ then acts by (standard) two-component linking on the operator $\mathcal{O}_\beta$. This type of action --whereby the topological defect associated with a $k$-form symmetry acts on an extended operator of dimension greater than $k$, modifying the nature of the operator as it slides through it-- was termed a \textit{generalized charge} in \cite{Bhardwaj:2023wzd}. The phenomenon described here then shows that operators charged under $\widehat{A}^{(d-p-2)}$ and $\widehat{B}^{(d-q-2)}$ carry generalized charges of the symmetry generated by the condensation defect, with a nontrivial interplay of topological and non-topological operators. This interplay is such that, in this setup, the action by a higher-link invariant can be reinterpreted as a generalized charge. 

\paragraph{States} We now want to construct states in Hilbert spaces that transform under the action of this symmetry. In order to do so, we consider the spatial slice $S^{p+1}\times S^{q+1}\times S^{d-p-q-3}$.\footnote{When $d-p-q-3=0$, the $S^0$ factor can be dropped.} We define two families of states in the following ways:
\begin{itemize}
\item{States $\ket{\alpha,\widehat{b}}$: we fill \(S^{p+1}\) and we insert, at a point on the ball $B^{p+2}$, both an operator $\mathcal{O}_\alpha[S^{q+1}\times S^{d-p-q-3}]$, charged under the \(\widehat{A}^{(d-p-2)}\) symmetry, and a charge operator $U_{\widehat{b}}[S^{q+1}]$ for the $\widehat{B}^{(d-q-2)}$ symmetry.}
\item{States $\ket{\beta,\widehat{a}}$: we fill $S^{q+1}$ and we insert, at a point on the ball $B^{q+2}$, both an operator $\mathcal{O}_\beta[S^{p+1}\times S^{d-p-q-3}]$, charged under the $\widehat{B}^{(d-q-2)}$ symmetry, and a charge operator $U_{\widehat{a}}[S^{p+1}]$ for the $\widehat{A}^{(d-p-2)}$ symmetry.}
\end{itemize}
By equation \eqref{charge_dep}, the operators $U_d$ act independently on the two families of states, by permuting the label of the charge operator. Explicitly
\begin{equation}
\begin{split}
    U_d\ket{\alpha,\widehat{b}}&=\ket{\alpha, \widehat{b}+d \cdot (\varphi_\alpha \otimes -)}~,\\
    U_d\ket{\beta,\widehat{a}}&=\ket{\beta, \widehat{a}+d \cdot (- \otimes \varphi_\beta)}~.
\end{split}
\end{equation}
This means that the $D^{(n+r-1)}$ symmetry does not act diagonally on this set of states. Its eigenvalues are pure phases and the respective eigenvectors are complex linear combinations of states prepared by the operator insertions described above. Note that, in general, the eigenvectors cannot themselves be prepared via the insertion of some operator, given that this would require taking linear combinations with coefficients that are not positive integers.

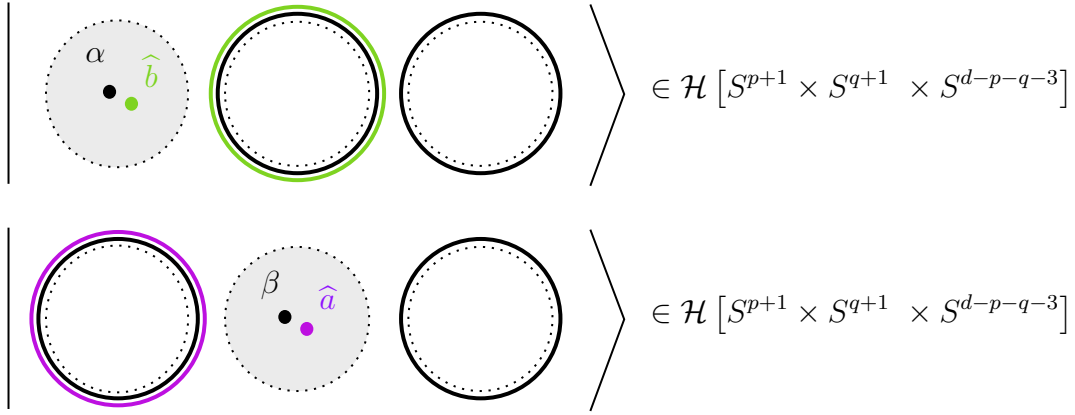
\begin{figure}[t]
        \centering
        \tikzset{every picture/.style={line width=0.75pt}} 

\begin{tikzpicture}[x=0.75pt,y=0.75pt,yscale=-1,xscale=1]

\draw  [color={rgb, 255:red, 0; green, 0; blue, 0 }  ,draw opacity=1 ][dash pattern={on 0.84pt off 2.51pt}] (184,56.83) .. controls (184,36.86) and (200.19,20.67) .. (220.17,20.67) .. controls (240.14,20.67) and (256.33,36.86) .. (256.33,56.83) .. controls (256.33,76.81) and (240.14,93) .. (220.17,93) .. controls (200.19,93) and (184,76.81) .. (184,56.83) -- cycle ;
\draw  [fill={rgb, 255:red, 155; green, 155; blue, 155 }  ,fill opacity=0.2 ][dash pattern={on 0.84pt off 2.51pt}] (94.13,56.83) .. controls (94.13,36.49) and (110.09,20) .. (129.79,20) .. controls (149.49,20) and (165.46,36.49) .. (165.46,56.83) .. controls (165.46,77.18) and (149.49,93.67) .. (129.79,93.67) .. controls (110.09,93.67) and (94.13,77.18) .. (94.13,56.83) -- cycle ;
\draw  [color={rgb, 255:red, 0; green, 0; blue, 0 }  ,draw opacity=1 ][dash pattern={on 0.84pt off 2.51pt}] (276,56.83) .. controls (276,36.86) and (292.19,20.67) .. (312.17,20.67) .. controls (332.14,20.67) and (348.33,36.86) .. (348.33,56.83) .. controls (348.33,76.81) and (332.14,93) .. (312.17,93) .. controls (292.19,93) and (276,76.81) .. (276,56.83) -- cycle ;
\draw    (75.33,11) -- (75.33,102) ;
\draw    (367.33,12) -- (384.33,57) -- (367.33,103) ;
\draw  [draw opacity=0][fill={rgb, 255:red, 0; green, 0; blue, 0 }  ,fill opacity=1 ] (122.73,55.83) .. controls (122.73,53.9) and (124.21,52.33) .. (126.03,52.33) .. controls (127.86,52.33) and (129.33,53.9) .. (129.33,55.83) .. controls (129.33,57.77) and (127.86,59.33) .. (126.03,59.33) .. controls (124.21,59.33) and (122.73,57.77) .. (122.73,55.83) -- cycle ;
\draw  [draw opacity=0][fill={rgb, 255:red, 126; green, 211; blue, 33 }  ,fill opacity=1 ] (133.73,61.83) .. controls (133.73,59.9) and (135.21,58.33) .. (137.03,58.33) .. controls (138.86,58.33) and (140.33,59.9) .. (140.33,61.83) .. controls (140.33,63.77) and (138.86,65.33) .. (137.03,65.33) .. controls (135.21,65.33) and (133.73,63.77) .. (133.73,61.83) -- cycle ;
\draw  [color={rgb, 255:red, 0; green, 0; blue, 0 }  ,draw opacity=1 ][line width=1.5]  (180.33,56.67) .. controls (180.33,34.58) and (198.24,16.67) .. (220.33,16.67) .. controls (242.42,16.67) and (260.33,34.58) .. (260.33,56.67) .. controls (260.33,78.76) and (242.42,96.67) .. (220.33,96.67) .. controls (198.24,96.67) and (180.33,78.76) .. (180.33,56.67) -- cycle ;
\draw  [color={rgb, 255:red, 0; green, 0; blue, 0 }  ,draw opacity=1 ][line width=1.5]  (272.33,56.67) .. controls (272.33,34.58) and (290.24,16.67) .. (312.33,16.67) .. controls (334.42,16.67) and (352.33,34.58) .. (352.33,56.67) .. controls (352.33,78.76) and (334.42,96.67) .. (312.33,96.67) .. controls (290.24,96.67) and (272.33,78.76) .. (272.33,56.67) -- cycle ;
\draw  [color={rgb, 255:red, 126; green, 211; blue, 33 }  ,draw opacity=1 ][line width=1.5]  (176.83,56.67) .. controls (176.83,32.64) and (196.31,13.17) .. (220.33,13.17) .. controls (244.36,13.17) and (263.83,32.64) .. (263.83,56.67) .. controls (263.83,80.69) and (244.36,100.17) .. (220.33,100.17) .. controls (196.31,100.17) and (176.83,80.69) .. (176.83,56.67) -- cycle ;
\draw  [color={rgb, 255:red, 0; green, 0; blue, 0 }  ,draw opacity=1 ][fill={rgb, 255:red, 155; green, 155; blue, 155 }  ,fill opacity=0.2 ][dash pattern={on 0.84pt off 2.51pt}] (184,169.83) .. controls (184,149.86) and (200.19,133.67) .. (220.17,133.67) .. controls (240.14,133.67) and (256.33,149.86) .. (256.33,169.83) .. controls (256.33,189.81) and (240.14,206) .. (220.17,206) .. controls (200.19,206) and (184,189.81) .. (184,169.83) -- cycle ;
\draw  [dash pattern={on 0.84pt off 2.51pt}] (94.13,169.83) .. controls (94.13,149.49) and (110.09,133) .. (129.79,133) .. controls (149.49,133) and (165.46,149.49) .. (165.46,169.83) .. controls (165.46,190.18) and (149.49,206.67) .. (129.79,206.67) .. controls (110.09,206.67) and (94.13,190.18) .. (94.13,169.83) -- cycle ;
\draw  [color={rgb, 255:red, 0; green, 0; blue, 0 }  ,draw opacity=1 ][dash pattern={on 0.84pt off 2.51pt}] (276,169.83) .. controls (276,149.86) and (292.19,133.67) .. (312.17,133.67) .. controls (332.14,133.67) and (348.33,149.86) .. (348.33,169.83) .. controls (348.33,189.81) and (332.14,206) .. (312.17,206) .. controls (292.19,206) and (276,189.81) .. (276,169.83) -- cycle ;
\draw    (75.33,124) -- (75.33,215) ;
\draw    (367.33,125) -- (384.33,170) -- (367.33,216) ;
\draw  [draw opacity=0][fill={rgb, 255:red, 0; green, 0; blue, 0 }  ,fill opacity=1 ] (210.73,168.83) .. controls (210.73,166.9) and (212.21,165.33) .. (214.03,165.33) .. controls (215.86,165.33) and (217.33,166.9) .. (217.33,168.83) .. controls (217.33,170.77) and (215.86,172.33) .. (214.03,172.33) .. controls (212.21,172.33) and (210.73,170.77) .. (210.73,168.83) -- cycle ;
\draw  [draw opacity=0][fill={rgb, 255:red, 189; green, 16; blue, 224 }  ,fill opacity=1 ] (221.73,174.83) .. controls (221.73,172.9) and (223.21,171.33) .. (225.03,171.33) .. controls (226.86,171.33) and (228.33,172.9) .. (228.33,174.83) .. controls (228.33,176.77) and (226.86,178.33) .. (225.03,178.33) .. controls (223.21,178.33) and (221.73,176.77) .. (221.73,174.83) -- cycle ;
\draw  [color={rgb, 255:red, 0; green, 0; blue, 0 }  ,draw opacity=1 ][line width=1.5]  (90.33,169.67) .. controls (90.33,147.58) and (108.24,129.67) .. (130.33,129.67) .. controls (152.42,129.67) and (170.33,147.58) .. (170.33,169.67) .. controls (170.33,191.76) and (152.42,209.67) .. (130.33,209.67) .. controls (108.24,209.67) and (90.33,191.76) .. (90.33,169.67) -- cycle ;
\draw  [color={rgb, 255:red, 0; green, 0; blue, 0 }  ,draw opacity=1 ][line width=1.5]  (272.33,169.67) .. controls (272.33,147.58) and (290.24,129.67) .. (312.33,129.67) .. controls (334.42,129.67) and (352.33,147.58) .. (352.33,169.67) .. controls (352.33,191.76) and (334.42,209.67) .. (312.33,209.67) .. controls (290.24,209.67) and (272.33,191.76) .. (272.33,169.67) -- cycle ;
\draw  [color={rgb, 255:red, 189; green, 16; blue, 224 }  ,draw opacity=1 ][line width=1.5]  (86.83,169.67) .. controls (86.83,145.64) and (106.31,126.17) .. (130.33,126.17) .. controls (154.36,126.17) and (173.83,145.64) .. (173.83,169.67) .. controls (173.83,193.69) and (154.36,213.17) .. (130.33,213.17) .. controls (106.31,213.17) and (86.83,193.69) .. (86.83,169.67) -- cycle ;

\draw (396,43) node [anchor=north west][inner sep=0.75pt]   [align=left] {$\displaystyle \in \mathcal{H}\left[ S^{p+1} \times S^{q+1} \ \times S^{d-p-q-3}\right]$};
\draw (396,156) node [anchor=north west][inner sep=0.75pt]   [align=left] {$\displaystyle \in \mathcal{H}\left[ S^{p+1} \times S^{q+1} \ \times S^{d-p-q-3}\right]$};
\draw (112.4,32.4) node [anchor=north west][inner sep=0.75pt]   [align=left] {$\displaystyle \alpha $};
\draw (142,36) node [anchor=north west][inner sep=0.75pt]   [align=left] {$\displaystyle \textcolor[rgb]{0.49,0.83,0.13}{\widehat{b}}$};
\draw (200.4,144.4) node [anchor=north west][inner sep=0.75pt]   [align=left] {$\displaystyle \beta $};
\draw (230,152) node [anchor=north west][inner sep=0.75pt]   [align=left] {$\displaystyle \textcolor[rgb]{0.56,0.07,1}{\widehat{a}}$};

\end{tikzpicture}
        \caption{Cartoon of the states charged under the condensation defect, prepared via operator insertions. Each dotted circle represents one of the spherical factors of the spatial slice. The shaded disks indicate which sphere is filled in order to prepare the given state. The black dots and circles represent the insertions of the operators charged under $\widehat{A}^{(d-p-2)}$ or $\widehat{B}^{(d-q-2)}$. Finally, the colored dots and circles represent the insertions of topological defects of $\widehat{B}^{(d-q-2)}$ (green) or $\widehat{A}^{(d-p-2)}$ (purple).}
        \label{States-cond-def}
\end{figure}

\subsubsection{Extension}
\label{sec:states-extension}
\begin{figure}[t]
    \centering
    \tikzset{every picture/.style={line width=0.75pt}} 

\begin{tikzpicture}[x=0.75pt,y=0.75pt,yscale=-1,xscale=1]

\draw  [color={rgb, 255:red, 0; green, 0; blue, 0 }  ,draw opacity=1 ][dash pattern={on 4.5pt off 4.5pt}][line width=0.75]  (251.4,149.09) .. controls (251.4,119.55) and (295.01,95.6) .. (348.8,95.6) .. controls (402.59,95.6) and (446.2,119.55) .. (446.2,149.09) .. controls (446.2,178.64) and (402.59,202.59) .. (348.8,202.59) .. controls (295.01,202.59) and (251.4,178.64) .. (251.4,149.09) -- cycle ;
\draw  [color={rgb, 255:red, 74; green, 144; blue, 226 }  ,draw opacity=1 ] (197.09,152.27) .. controls (197.09,91.96) and (263.58,43.07) .. (345.61,43.07) .. controls (427.64,43.07) and (494.13,91.96) .. (494.13,152.27) .. controls (494.13,212.58) and (427.64,261.47) .. (345.61,261.47) .. controls (263.58,261.47) and (197.09,212.58) .. (197.09,152.27) -- cycle ;
\draw [color={rgb, 255:red, 74; green, 144; blue, 226 }  ,draw opacity=1 ]   (299.6,149.48) .. controls (309.69,176.73) and (391.39,174.85) .. (407.53,145.65) ;
\draw [color={rgb, 255:red, 74; green, 144; blue, 226 }  ,draw opacity=1 ]   (313.02,162.57) .. controls (341.27,139.21) and (360.4,138.28) .. (387.6,161.48) ;
\draw  [color={rgb, 255:red, 245; green, 166; blue, 35 }  ,draw opacity=1 ] (243.8,149.09) .. controls (243.8,114.02) and (290.81,85.6) .. (348.8,85.6) .. controls (406.79,85.6) and (453.8,114.02) .. (453.8,149.09) .. controls (453.8,184.16) and (406.79,212.59) .. (348.8,212.59) .. controls (290.81,212.59) and (243.8,184.16) .. (243.8,149.09) -- cycle ;
\draw  [color={rgb, 255:red, 245; green, 166; blue, 35 }  ,draw opacity=1 ] (211.33,147.9) .. controls (211.33,100.28) and (271.51,61.68) .. (345.73,61.68) .. controls (419.96,61.68) and (480.13,100.28) .. (480.13,147.9) .. controls (480.13,195.53) and (419.96,234.13) .. (345.73,234.13) .. controls (271.51,234.13) and (211.33,195.53) .. (211.33,147.9) -- cycle ;
\draw  [draw opacity=0][dash pattern={on 4.5pt off 4.5pt}] (311.26,163.17) .. controls (304.67,182.46) and (282.04,196.68) .. (255.14,196.68) .. controls (223.17,196.68) and (197.23,176.59) .. (197.08,151.77) -- (255.14,151.56) -- cycle ; \draw  [color={rgb, 255:red, 74; green, 144; blue, 226 }  ,draw opacity=1 ][dash pattern={on 4.5pt off 4.5pt}] (311.26,163.17) .. controls (304.67,182.46) and (282.04,196.68) .. (255.14,196.68) .. controls (223.17,196.68) and (197.23,176.59) .. (197.08,151.77) ;  
\draw  [draw opacity=0] (197.09,149.27) .. controls (197.08,149.04) and (197.08,148.81) .. (197.08,148.58) .. controls (197.08,130.42) and (209.02,114.77) .. (226.25,107.61) -- (247.4,148.58) -- cycle ; \draw  [color={rgb, 255:red, 74; green, 144; blue, 226 }  ,draw opacity=1 ] (197.09,149.27) .. controls (197.08,149.04) and (197.08,148.81) .. (197.08,148.58) .. controls (197.08,130.42) and (209.02,114.77) .. (226.25,107.61) ;  
\draw  [draw opacity=0] (265.9,110.66) .. controls (292.41,114.62) and (312.4,132.27) .. (312.4,153.45) .. controls (312.4,155.91) and (312.13,158.33) .. (311.61,160.68) -- (254.63,153.45) -- cycle ; \draw  [color={rgb, 255:red, 74; green, 144; blue, 226 }  ,draw opacity=1 ] (265.9,110.66) .. controls (292.41,114.62) and (312.4,132.27) .. (312.4,153.45) .. controls (312.4,155.91) and (312.13,158.33) .. (311.61,160.68) ;  
\draw  [draw opacity=0] (227.42,108.28) .. controls (229.79,101.56) and (236.99,96.68) .. (245.5,96.68) .. controls (255.02,96.68) and (262.89,102.78) .. (264.21,110.73) -- (245.5,113.08) -- cycle ; \draw  [color={rgb, 255:red, 245; green, 166; blue, 35 }  ,draw opacity=1 ] (227.42,108.28) .. controls (229.79,101.56) and (236.99,96.68) .. (245.5,96.68) .. controls (255.02,96.68) and (262.89,102.78) .. (264.21,110.73) ;  
\draw  [draw opacity=0][dash pattern={on 4.5pt off 4.5pt}] (260.78,114.09) .. controls (257.89,118.57) and (252.24,121.61) .. (245.73,121.61) .. controls (237.16,121.61) and (230.06,116.32) .. (228.82,109.42) -- (245.73,107.27) -- cycle ; \draw  [color={rgb, 255:red, 245; green, 166; blue, 35 }  ,draw opacity=1 ][dash pattern={on 4.5pt off 4.5pt}] (260.78,114.09) .. controls (257.89,118.57) and (252.24,121.61) .. (245.73,121.61) .. controls (237.16,121.61) and (230.06,116.32) .. (228.82,109.42) ;  
\draw [color={rgb, 255:red, 245; green, 166; blue, 35 }  ,draw opacity=1 ] [dash pattern={on 4.5pt off 4.5pt}]  (227.66,107.71) -- (261.84,110.8) ;

\end{tikzpicture}
    \caption{The junction of topological defects in fig.~\ref{local-def-ext}, associated with the extension, is used to wrap an operator charged under $\widehat{A}^{(d-p-2)}$ (or $\widehat{B}^{(d-q-2)}$), depicted by a black dashed line. Moving the defect of $D^{(r)}$ (blue) through the charged operator, we get a nontrivial action because we deposit a charge of $\widehat{B}^{(d-q-2)}$ (or $\widehat{A}^{(d-p-2)}$) on it. This implies that also moving the three topological defects of $C^{(r)}$ (orange) we must get a nontrivial action. The transverse section of this configuration corresponds to fig.~\ref{defects-ext} in Appendix \ref{App-Reps}.}
    \label{cond-def-ext}
\end{figure}
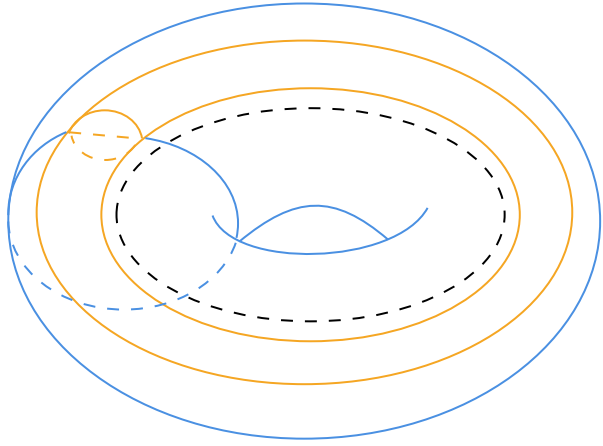

We have already discussed how the $D^{(n+r-1)}$ symmetry acts on extended operators by triple linking, and how to construct representations of such symmetry in suitable Hilbert spaces. We now discuss its interplay with the $C^{(r)}$ symmetry and the operators charged under it. We begin by considering the $n=1$ case, where $D^{(r)}$ sits in a group extension with $C^{(r)}$, captured by the short exact sequence
\begin{equation}\label{eq:ext-std}
    1 \rightarrow D^{(r)} \rightarrow \Gamma^{(r)} \rightarrow C^{(r)} \rightarrow 1 \ ,
\end{equation}
described by the class $t_{r+2}\in H^{r+2}(B^{r+1}C,D)$.
The condensation defects $U_d$ act on extended operators charged under $\widehat{A}^{(d-p-2)}$ (or $\widehat{B}^{(d-q-2)}$) by depositing a charge operator, as in equation \eqref{charge_dep}. Moreover, they can end on a configuration of topological defects of the $C^{(r)}$ symmetry, as shown in figure \ref{local-def-ext}. This implies that also the defects of $C^{(r)}$ must act non-trivially on operators charged under $\widehat{A}^{(d-p-2)}$ (or $\widehat{B}^{(d-q-2)}$). We show in figure \ref{cond-def-ext} a junction of topological defects that can be used to derive the action of the whole $C^{(r)}$ group. This is inspired by the analogous configuration in the case of a standard group extension discussed in appendix \ref{App-Reps}, see figure \ref{defects-ext}. 
As for standard group extensions, the states constructed as in figure~\ref{States-cond-def} organize into projective representations of $C^{(r)}$, i.e.~representations of $\Gamma^{(r)}$. 

It is worth stressing that representations of $\Gamma^{(r)}$ are not uniquely identified by the charges under $\widehat{A}^{(d-p-2)}$ and $\widehat{B}^{(d-q-2)}$. As a consequence, the charged operators will have to carry additional labels that specify the charge under the extended symmetry. 
For instance, there can be a nontrivial operator $\mathcal{O}_{\alpha_0}$ which is neutral under $\widehat{A}^{(d-p-2)}$ (i.e.~$\varphi_{\alpha_0}=0$) but charged under $C^{(r)}$. The corresponding states $\ket{\alpha_0,\widehat{b}}$ (see fig.~\ref{States-cond-def}) can have nontrivial overlap only with states with the same $C^{(r)}$ charge. In particular, there can be an overlap with conventional $C^{(r)}$-charged states (i.e.~states created by the $r$-dimensional extended operators charged under $C^{(r)}$). This is an instance of a selection rule encompassing both standard and generalized charges. Notice that, as explained above, these projective representations only arise in specific Hilbert spaces, e.g.~they do not appear in Hilbert spaces that capture standard charges of the $r$-form symmetry $C^{(r)}$. We will see a concrete manifestation of this phenomenon in section \ref{sec:index-E1}.

\subsubsection{Higher-group}
\label{sec:states-Higher-group}

Finally, we discuss the more intricate scenario of $n=2$, where $D^{(r+1)}$ and $C^{(r)}$ form an $(r+2)$-group $\underline{\Gamma}$, which can be formally associated with the sequence
\begin{equation}
    1 \rightarrow D^{(r+1)} \rightarrow \underline{\Gamma} \rightarrow C^{(r)} \rightarrow 1 \ ,
\end{equation}
governed by the class $t_{r+3}\in H^{r+3}(B^{r+1}C,D)$. In the case of a standard extension, see \eqref{eq:ext-std}, a representation of the structure $\Gamma^{(r)}$ was associated with projective representations of $C^{(r)}$, carried by configurations of extended operators charged under $D^{(r)}$. Similarly, in the present case, a representation of $\underline{\Gamma}$ will be associated with objects that carry projective representations of $C^{(r)}$. The latter should have some interplay with the charged objects of $D^{(r+1)}$.

The case of a standard $(r+2)$-group charge, reviewed in appendix \ref{App-Reps}, suggests that the objects charged projectively under $C^{(r)}$ should be interfaces between defects $\mathcal{O}_\alpha$ charged under $\widehat{A}^{(d-p-2)}$ (or $\mathcal{O}_\beta$ charged under $\widehat{B}^{(d-q-2)}$), namely operators of codimension one inside $\mathcal{O}_\alpha$ (or $\mathcal{O}_\beta$). More precisely, these interfaces will be allowed to carry a nontrivial projective representation whenever the underlying configuration of $\mathcal{O}_\alpha$ and $\mathcal{O}_\beta$ supports a $D^{(r+1)}$ charge, i.e.~if $\mathcal{O}_\alpha$ and $\mathcal{O}_\beta$ can triple-link with a charge defect $U_d$. The charge defect of $C^{(r)}$ can then triple-link with $\mathcal{O}_\beta$ and with an interface on $\mathcal{O}_\alpha$ (or vice versa). In this configuration, the charge defects of $C^{(r)}$ will necessarily also intersect $\mathcal{O}_\alpha$ (or $\mathcal{O}_\beta$), and this allows the presence of additional projective phases in their fusion. 

We will now briefly discuss how to see that the interfaces can indeed be projectively charged under $C^{(r)}$, precisely when inserted within the triple-linking configuration as outlined above.
One needs to use that the condensation defect can end on larger configurations of $C^{(r)}$ defects, as shown in figure \ref{local-def-2grp}. In the same spirit of the discussion for $n=1$ above, around figure \ref{cond-def-ext}, one needs to consider a junction of $C^{(r)}$ and $D^{(r+1)}$ defects wrapping $\mathcal{O}_\alpha$ (or $\mathcal{O}_\beta$). This leads to a straightforward adaptation of the argument for a standard $(r+2)$-group action to the present case of the condensation defect. 
  
\begin{figure}[t]
    \centering
    \input{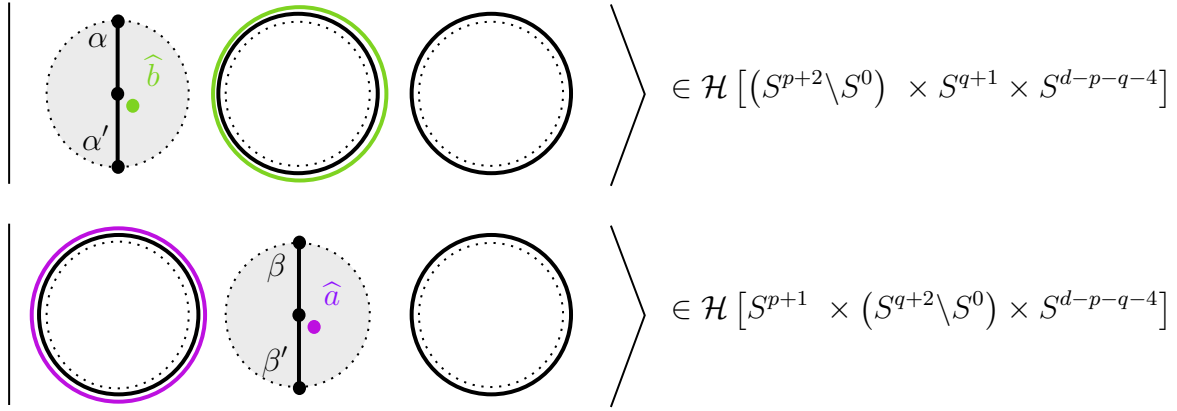}
    \caption{Cartoon of the states charged under the $(r+2)$-group by condensation defect, prepared via operator insertions. Each dotted circle represents one of the spherical factors of the spatial slice. The shaded disks indicate which sphere is filled in order to prepare the given state. The black lines and circles represent the insertions of the operators charged under $\widehat{A}^{(d-p-2)}$ or $\widehat{B}^{(d-q-2)}$. The black dots in the middle of the lines represent the interfaces between the operators charged under $\widehat{A}^{(d-p-2)}$ or $\widehat{B}^{(d-q-2)}$.
    The black dots at the end of the lines represent the removed $S^0$. Finally, the colored dots and circles represent the insertions of topological defects of $\widehat{B}^{(d-q-2)}$ (green) or $\widehat{A}^{(d-p-2)}$ (purple).
    }
    \label{cond-def-2grp}
\end{figure}

Let us now discuss the defect Hilbert space that, by analogy with more standard higher-groups, will carry a projective representation of $C^{(r)}$. 
This is the Hilbert space on a spatial slice pierced by the defect. As an example, we pick the spatial slice $\Sigma_{d-1}=(S^{p+2}\backslash S^0)\times S^{q+1}\times S^{d-p-q-4}$, where the two removed subspaces $S^0\times S^{q+1}\times S^{d-p-q-4}$ are the intersection with the defect.\footnote{When $d-p-q-4=0$, the $S^0$ factor can be dropped.} As illustrated in figure \ref{cond-def-2grp}, the states are prepared by filling the pierced $S^{p+2}$ and inserting (i) a charge operator $U_{\widehat{b}}[S^{q+1}]$ for the $\widehat{B}^{(d-q-2)}$ symmetry; and (ii) operators charged under $\widehat{A}^{(d-p-2)}$\footnote{The same argument can be repeated for $\widehat{B}^{(d-q-2)}$, as done in fig.~\ref{States-cond-def}.} as follows: $(d-p-3)$ dimensions wrap $S^{q+1}\times S^{d-p-q-4}$, and the remaining dimension stretches along an interval connecting the two removed points of $S^{p+2}$. A $(d-p-3)$-dimensional interface on the charged operator is then a point on the interval. Different interfaces label different states on this slice.

\section{Examples and Applications}
\label{SEC:examples}

In this section, we provide several examples in which, upon gauging certain internal symmetries, some of the structures discussed in the previous section arise. We will focus, in particular, on scalar $\text{QED}_3$, on ${\cal N}=4$ $\text{SQED}_3$ and on  five-dimensional SYM with $\mathfrak{su}(2)$ gauge algebra and its SCFT UV completions. 

\subsection{Gauging a discrete subgroup in scalar \texorpdfstring{$\text{QED}_3$}{QED3} }\label{subsec:gadsiq3}

The action of scalar QED$_{3}$ is
\begin{equation}
    S = \int_{\mathcal{M}_3} \left( 
    \overline{(\nabla^\mu z_I)} \nabla_\mu z^I + \frac{1}{4e^2} f^{\mu\nu} f_{\mu\nu} + V(z,\overline{z})
    \right)~,
\end{equation}
where $z^I$ are $N$ complex scalar fields, $I=1,\dots, N$, $a_\mu$ is an Abelian gauge field with field strength $f_{\mu\nu}$ and $V(z,\overline{z})$ a gauge-invariant potential.

The gauge coupling $e^2$ has the dimension of energy, and therefore it sets an energy scale in the theory. As a result, the theory is asymptotically free and flows to strong coupling in the IR. The gauge transformations are
\begin{equation}
    z^I \to e^{i\alpha} z^I
    \quad , \quad
    \bar{z}_I \to e^{-i\alpha} \bar{z}_I
    \quad , \quad
    a_\mu \to a_\mu + \partial_\mu\alpha \ ,
\end{equation}
the covariant derivative is
\begin{equation}\label{cov-der-1}
    \nabla_\mu z^I = \partial_\mu z^I - i a_\mu z^I \ .
\end{equation}
If the potential is a function only of $\Tr \bar{z}z \equiv \overline{z}_I z^I$ this theory has a 0-form symmetry group $PSU(N)=SU(N)/\mathbb{Z}_{N}$. The scalars $z^I$ transform in the fundamental representation of $SU(N)$, but the global symmetry is reduced to $PSU(N)$ because the center $\mathbb{Z}_{N}$ of $SU(N)$ is contained in the gauge group. This theory also has a magnetic $U(1)^{(0)}_M$ symmetry with current $J^{(1)}_M = \star \frac{f}{2\pi}$.

There is a mixed 't Hooft anomaly between $PSU(N)^{(0)}$ and $U(1)^{(0)}_M$  given by the following anomaly inflow \cite{Komargodski:2017dmc}
\begin{equation}\label{anomaly-QED3}
    S_{\mathrm{inflow}}=  \frac{2\pi i}{N} \int_{\mathcal{Y}_4} w_2(PSU(N)) \cup \left[\frac{dA_M}{2\pi}\right]_N \ ,
\end{equation}
where $A_M$ is a background gauge field for $U(1)^{(0)}_M$, $[\cdot]_N$ denotes mod $N$ reduction, and $w_2(PSU(N))$ is the $\mathbb{Z}_N$-valued characteristic class of $PSU(N)$ bundles that quantifies the obstruction to lift them to $SU(N)$ bundles. The anomaly comes from the fact that, in the presence of a background for $PSU(N)$, the fluxes of $\frac{f}{2\pi}$ are no longer integer, because they need to satisfy the constraint 
\begin{equation}\label{fract-fluxes}
    \int_{\Sigma_2} \frac{f}{2\pi}
    =
    -\frac{1}{N} \int_{\Sigma_2}\! w_2
    \ \ \text{ mod } 1 \ .
\end{equation}
To obtain an anomaly of the form studied in section \ref{sec:CUBIC} we restrict to a $\mathbb{Z}_N\times\mathbb{Z}_N$ subgroup of $PSU(N)$. Whenever we need to distinguish the two $\mathbb{Z}_N$ factors, we use the subscripts $A$ and $B$. The action of the $\mathbb{Z}_N$ generators on the scalars is 
\begin{equation}\label{ZNxZN}
    z^I \to (\rho_A)^I_{~J} z^J \quad , \quad
    z^I \to (\rho_B)^I_{~J} z^J
\end{equation}
where
\begin{equation}\label{Gamma-N}
    \rho_A = 
    \begin{cases}
        Q & N \text{ odd} \\
        \zeta Q & N \text{ even} \\
    \end{cases}
    \ , \quad 
    \rho_B = 
    \begin{cases}
    P & N \text{ odd} \\
    \zeta P & N \text{ even} \\
    \end{cases}
    \ ,
\end{equation}
with $\eta=\exp(\frac{2\pi i}{N})$, $\zeta=\exp(\frac{2\pi i}{2N})$ and
\begin{equation}\label{PQ-def}
    Q = \text{diag}(1,\eta,\eta^2,\dots,\eta^{N-1})
    \quad , \quad
    P = \begin{pmatrix}
        0 & 1 \\
        \mathds{1}_{N-1} & 0
    \end{pmatrix}
    \ ,
\end{equation}
where $\mathds{1}_{k}$ denotes the $k\times k$ identity matrix. Note that, while for $N$ odd $\rho_A$ and $\rho_B$ are indeed generators of $\mathbb{Z}_N$, meaning that $\rho_A^N = \rho_B^N =\mathds{1}_{N}$, for $N$ even we have $\rho_A^N = \rho_B^N =-\mathds{1}_{N}$ due to the additional phase factor $\zeta$. This phase factor is required to ensure that $\rho_{A,B}$ are elements of $SU(N)$, i.e.~the group acting on the scalar fields $z,\bar{z}$. In both cases, since the phase factor $\zeta$ can be reabsorbed by acting with a transformation in the center of $SU(N)$, these transformations generate a $\mathbb{Z}_N\times \mathbb{Z}_N$ subgroup of $PSU(N)$. The reason for the choice of this particular subgroup is that this $\mathbb{Z}_N\times \mathbb{Z}_N$ undergoes a nontrivial extension when lifted from $PSU(N)$ to $SU(N)$. This means that the obstruction $w_2(PSU(N))$ remains a nontrivial class if we take a $PSU(N)$ bundle that restricts to a $\mathbb{Z}_N\times\mathbb{Z}_N$ bundle. In particular, denoting with $\mathcal{A}_1$ and $\mathcal{B}_1$ the background gauge fields of $(\mathbb{Z}_N)_A$ and $(\mathbb{Z}_N)_B$, we show in Appendix \ref{appendix-obstruction} that the restriction gives
\begin{equation}\label{eq:w2-N-reduction}
    w_2(PSU(N))
    \vert_{\mathbb{Z}_N\times \mathbb{Z}_N} = 
    \begin{cases}
    \mathcal{A}_1 \cup \mathcal{B}_1 & \text{for } N \text{ odd}~,   \\ 
    \tfrac{N}{2}\beta_2(\mathcal{A}_1) + 
    \tfrac{N}{2}\beta_2(\mathcal{B}_1) + \mathcal{A}_1 \cup \mathcal{B}_1 & \text{for } N \text{ even}~.  \\ 
    \end{cases}
\end{equation}
Here $\beta_2$ is the Bockstein map associated with the non-split exact sequence
\begin{equation}\label{ses-2N}
    1 \to \mathbb{Z}_2 \to \mathbb{Z}_{2N}
    \to \mathbb{Z}_N \to 1 \ .
\end{equation}
We can write it explicitly as 
\begin{equation}
\beta_2(\mathcal{A}_1) = \left[\frac{1}{N}\delta\widetilde{\mathcal{A}_1}\right]_2~,
\end{equation}
where $\widetilde{\mathcal{A}_1}$ denotes a lift of the $\mathbb{Z}_N$ valued cochain to $\mathbb{Z}_{2N}$, whose $\delta$ must be a multiple of $N$, and similarly for $\mathcal{B}_1$. Note that $\beta_2(\mathcal{A}_1)$ and  $\beta_2(\mathcal{B}_1)$ are $\mathbb{Z}_2$ valued, and multiplying them by $N/2$ we obtain $\mathbb{Z}_N$ valued cocycles. In any case, the extra $\beta_2$ terms for $N$ even do not contribute to the anomaly as they can be canceled by a counterterm. To see this, note that 
\begin{align}
\begin{split}
    \frac{2\pi i}{N} \int_{\mathcal{Y}_4} \left(\frac{N}{2}\beta_2(\mathcal{A}_1) \right)\cup \left[\frac{dA_M}{2\pi}\right]_N 
    & = 
    \frac{2\pi i}{2N} \int_{\mathcal{Y}_4}
    \delta \widetilde{\mathcal{A}_1}
    \cup \left[\frac{dA_M}{2\pi}\right]_{2N} \\
    &  =
    \frac{2\pi i}{2N} \int_{\mathcal{Y}_4}
    \delta \left(
    \widetilde{\mathcal{A}_1}
    \cup \left[\frac{dA_M}{2\pi}\right]_{2N}
    \right) \\
    & 
    = \frac{2\pi i}{2N} \int_{\mathcal{M}_3}
    \widetilde{\mathcal{A}_1}
    \cup \left[\frac{dA_M}{2\pi}\right]_{2N}~,
\end{split}
\end{align}
and similarly for $\mathcal{B}_1$. In order to collect $\delta$ when going from the first to the second line, we used that the $\mathbb{Z}_N$ valued cocycle $\left[\frac{dA_M}{2\pi}\right]_N$ remains closed when lifted to $\mathbb{Z}_{2N}$. In what follows we assume that such counterterms are added, so that the mixed anomaly between $U(1)^{(0)}_M$ and $\mathbb{Z}^{(0)}_N\times\mathbb{Z}^{(0)}_N$ is
\begin{equation}\label{eq:anoZNZNU1}
S_{\mathrm{inflow}}= \frac{2\pi i}{N} \int_{\mathcal{Y}_4} (\mathcal{A}_1\cup\mathcal{B}_1)  \cup \left[\frac{dA_M}{2\pi}\right]_N~, 
\end{equation}
for any $N$. This is precisely of the form of \eqref{eq:anomaly},  with $n=1$, $p=q=r=0$, $A=B=\mathbb{Z}_N$, $C=U(1)$ and $t_2=[c_1]_N$.

Denoting the topological defect of the magnetic symmetry as
\begin{equation}
    U^M_\alpha[\Sigma_2] = \exp \left(
    \frac{i\alpha}{2\pi} \int_{\Sigma_2} f
    \right)
\end{equation}
and using \eqref{fract-fluxes}, we see that the effect 
of turning on the $\mathbb{Z}_N^{(0)}\times\mathbb{Z}_N^{(0)}$ background gauge fields is that 
\begin{equation}\label{ext-bkg}
    U^M_{\alpha+2\pi}[\Sigma_2,\mathcal{A}_1,\mathcal{B}_1]
    =
    U^M_{\alpha}[\Sigma_2,\mathcal{A}_1,\mathcal{B}_1]
    \exp \left(-\frac{2\pi i}{N}
    \int_{\Sigma_2} \mathcal{A}_1\cup\mathcal{B}_1
    \right) \ .
\end{equation}
The periodicity of the parameter $\alpha$ is then extended to $\alpha\sim \alpha + 2\pi N$. Clearly, in QED$_3$ this extension is only visible when $\mathcal{A}_1$ and $\mathcal{B}_1$ are turned on, while the correct identification in any correlation function without background sources  is still $\alpha\sim\alpha+2\pi$.

\subsubsection{Gauging \texorpdfstring{$\mathbb{Z}_N^{(0)}\times\mathbb{Z}_N^{(0)}\subset PSU(N)^{(0)}$}{ZNxZN in PSU(N)}}\label{sec:gaugZnZn}

In order to obtain an example of extension by condensation defect, we gauge the subgroup $\mathbb{Z}_N^{(0)}\times\mathbb{Z}_N^{(0)}$ of the $PSU(N)^{(0)}$ global symmetry, which enters the cubic anomaly \eqref{eq:anoZNZNU1}. 

It will be convenient to embed the discrete background gauge fields $\mathcal{A}_1,\mathcal{B}_1$ into continuous ones $A_1,B_1$, with the addition of Lagrange multipliers. To this end, it is useful to introduce a background gauge field $C_\mu^\alpha$ for $PSU(N)^{(0)}$, where the index $\alpha$ runs over the Lie algebra generators ${(T^\alpha)^I}_J$.  The action coupled to the background is 
\begin{align}
\begin{split}\label{cov-der-2}
    S_{\text{QED}}[A_1,B_1] &= \int_{\mathcal{M}_3}  \left( \Tr \left(
    \overline{(D^\mu z)} D_\mu z\right) + \frac{1}{4e^2} f^{\mu\nu} f_{\mu\nu} + V(\Tr \bar{z}z)
    \right)~,\\
    (D_\mu z)^I & \equiv \left({\delta^I}_J(\partial_\mu - i a_\mu) - i C^\alpha_\mu {(T^\alpha)^I}_J  \right) z^J~,
\end{split}
\end{align}
with
\begin{equation}
    C^\alpha T^\alpha = 
    A_1 T_A + B_1 T_B~,
\end{equation}
so that the background field $C_\mu^\alpha$ has nonvanishing components only along the Lie algebra elements $T_A$ and $T_B$. The latter are defined by 
\begin{equation}
    \rho_A \equiv \exp\left(\frac{2\pi i}{N} T_A\right) \ , \quad
    \rho_B \equiv \exp\left(\frac{2\pi i}{N} T_B\right) \ ,
\end{equation}
in terms of the generators $\rho_A$ and $\rho_B$ of $\mathbb{Z}_N\times \mathbb{Z}_N$. For example, for $N=2$ we have 
\begin{equation}
    T_A = 
    \frac{1}{2}
    \begin{pmatrix}
    1 & 0 \\
    0  & -1 
    \end{pmatrix}
    = \frac{1}{2} \sigma_3
    \ , \quad \quad
    T_B = 
    \frac{1}{2}
    \begin{pmatrix}
    0 & 1 \\
    1 & 0 
    \end{pmatrix}
    = \frac{1}{2} \sigma_1 \ ,
\end{equation}
and for $N=3$ 
\begin{equation}
    T_A = 
    \begin{pmatrix}
    0 & 0 & 0\\
    0 & 1 & 0\\
    0 & 0 & -1\\
    \end{pmatrix}
    \ , \quad \quad
    T_B = 
    \frac{i}{\sqrt{3}}
    \begin{pmatrix}
    0 & 1 & -1 \\
    -1 & 0 & 1 \\
    1 & -1 & 0
    \end{pmatrix}
    \ .
\end{equation}
In order to reduce $A_1,B_1$ to $\mathbb{Z}_N$ gauge fields, we further introduce additional dynamical fields $\omega_{A,B}$ and $\phi_{A,B}$, respectively $U(1)$ gauge fields and compact scalars. They enter the following additional terms in the action
\begin{equation}
\begin{split}
\frac{i}{2\pi}\int_{\mathcal{M}_3} 
    \omega_A \wedge (NA_1-d\phi_A)+ \frac{i}{2\pi}\int_{\mathcal{M}_3} 
    \omega_B \wedge (NB_1-d\phi_B) \ ~.
\end{split}
\end{equation}
The associated gauge transformations are
\begin{equation}\label{eq:gtAB}
\begin{split}
    A_1 & \to A_1 + d\Lambda_{A}
    ~ ,~~
    \phi_{A} \to \phi_{A} + N \Lambda_{A}~,~~z  \to e^{i \Lambda_A T_A} z \\
    B_1 & \to B_1 + d\Lambda_{B}~, ~~
    \phi_{B} \to \phi_{B} + N \Lambda_{B}~,~~z  \to e^{i \Lambda_B T_B} z \ .
\end{split}
\end{equation}
After using the equations of motion of $\omega_{A,B}$, the fields $A_1$ and $B_1$ can be mapped to the discrete gauge fields $ \mathcal{A}_1, \mathcal{B}_1$ via
\begin{equation}
\begin{split}
    \mathcal{A}_1 = \frac{N}{2\pi} A_1~,~~\mathcal{B}_1 = \frac{N}{2\pi} B_1 \ .
\end{split}
\end{equation}
Next, in order to gauge the $\mathbb{Z}_N^{(0)}\times\mathbb{Z}_N^{(0)}$ symmetry, we promote the background gauge fields $A_1$ and $B_1$ to dynamical gauge fields $a_1$ and $b_1$.  The resulting path integral is
\begin{equation}\label{eq:gaugedtheory}
\begin{split}
    Z_{\text{QED}/{\mathbb{Z}_N \times \mathbb{Z}_N}}
    =
    \int &\mathcal{D}[z,\bar{z},a, a_1,b_1,\phi_A,\phi_B,\omega_A,\omega_B]\
    \exp\left(-S_{\text{QED}}[a_1,b_1]\right)\\
    \ & 
    \times\exp
   \left( \frac{i}{2\pi}
    \int_{\mathcal{M}_3}
    \omega_A \wedge (Na_1 -d\phi_A)
    +
    \omega_B \wedge (Nb_1 -d\phi_B)
     \right)   \ .
\end{split}
\end{equation}
After gauging $\mathbb{Z}_N^{(0)}\times\mathbb{Z}_N^{(0)}$, only an $SL(2,\mathbb{Z}_N)^{(0)}$ subgroup of $PSU(N)^{(0)}$ remains a symmetry, as explained in Appendix \ref{app-normalizer}. Moreover, as a consequence of the above discussion, the magnetic $U(1)^{(0)}_M$ is extended, as we discuss in more detail below. 

\subsubsection{Symmetry extension and charged objects}
\begin{figure}[t]
    \centering
    \tikzset{every picture/.style={line width=0.75pt}} 

\begin{tikzpicture}[x=0.75pt,y=0.75pt,yscale=-1,xscale=1]

\draw [color={rgb, 255:red, 245; green, 166; blue, 35 }  ,draw opacity=1 ]   (323.42,96.78) -- (269.42,172.78) ;
\draw [color={rgb, 255:red, 245; green, 166; blue, 35 }  ,draw opacity=1 ]   (323.42,96.78) -- (373.09,171.45) ;
\draw [color={rgb, 255:red, 245; green, 166; blue, 35 }  ,draw opacity=1 ]   (318.76,20.78) -- (323.42,96.78) ;
\draw [color={rgb, 255:red, 74; green, 144; blue, 226 }  ,draw opacity=1 ]   (372.09,33.45) -- (323.42,96.78) ;
\draw  [draw opacity=0][fill={rgb, 255:red, 74; green, 144; blue, 226 }  ,fill opacity=1 ] (319.93,97.28) .. controls (319.93,95.35) and (321.41,93.78) .. (323.23,93.78) .. controls (325.06,93.78) and (326.53,95.35) .. (326.53,97.28) .. controls (326.53,99.21) and (325.06,100.78) .. (323.23,100.78) .. controls (321.41,100.78) and (319.93,99.21) .. (319.93,97.28) -- cycle ;

\draw (258.2,138.4) node [anchor=north west][inner sep=0.75pt]   [align=left] {$\displaystyle \textcolor[rgb]{0.96,0.65,0.14}{\alpha }$};
\draw (374.2,143.4) node [anchor=north west][inner sep=0.75pt]   [align=left] {$\displaystyle \textcolor[rgb]{0.96,0.65,0.14}{\beta }$};
\draw (250.2,24.4) node [anchor=north west][inner sep=0.75pt]   [align=left] {$\displaystyle \textcolor[rgb]{0.96,0.65,0.14}{[\alpha +\beta]_{2\pi} }$};
\draw (370.2,47.4) node [anchor=north west][inner sep=0.75pt]  [color={rgb, 255:red, 74; green, 144; blue, 226 }  ,opacity=1 ] [align=left] {$\displaystyle \left\lfloor \frac{\alpha +\beta }{2\pi }\right\rfloor $};

\end{tikzpicture}
    \caption{Fusion between magnetic defects in the gauged theory, giving rise to the extension. The orange lines denote the defects of the original $U(1)^{(0)}_M$ symmetry. The parameters $\alpha$ and $\beta$ are angles $\in(0,2\pi)$, and so is their sum mod $2\pi$ indicated with $[\alpha + \beta]_{2\pi}$. The blue line denotes the condensation defect giving a $\mathbb{Z}_N^{(0)}$ symmetry in the gauged theory.  The junction is then saying that, whenever $\alpha+\beta> 2\pi$, the fusion of $\alpha$ and $\beta$ defects is not just the $[\alpha + \beta]_{2\pi}$ defect, but rather its dressing with the generator of $\mathbb{Z}_N^{(0)}$. This is the fusion of the $\mathbb{Z}_N$ extension of $U(1)_M$.}
    \label{fig::ext-QED}
\end{figure}
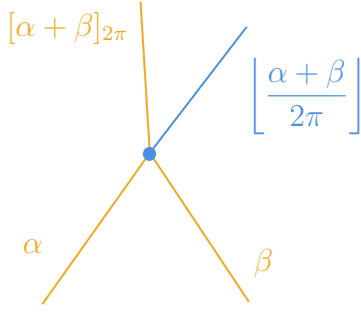
In the gauged theory \eqref{eq:gaugedtheory}, the relation \eqref{ext-bkg} becomes an operator equation 
\begin{equation}\label{eq:2piN-ext}
    U^M_{\alpha+2\pi k}[\Sigma_2]
    =
    U^M_{\alpha}[\Sigma_2]\,
    U_{k}[\Sigma_2] \ ,
\end{equation}
where 
\begin{align}
\begin{split}\label{eq:phiAphiB}
    U_{k}[\Sigma_2]
   =
    \exp \left(-\frac{2\pi i k}{N}
    \int_{\Sigma_2} 
    \frac{d\phi_A}{2\pi}
    \wedge
    \frac{d\phi_B}{2\pi}
    \right) \ .
\end{split}    
\end{align}
Equation \eqref{eq:2piN-ext} implies that for generic $\alpha, \beta \in (0,2\pi)$ we have the following fusion (the related junction of topological defects is shown in figure \ref{fig::ext-QED})
\begin{equation}
U^M_{\alpha}[\Sigma_2]U^M_{\beta}[\Sigma_2] = U^M_{[\alpha+\beta]_{2\pi}}[\Sigma_2] U_{\left \lfloor \frac{\alpha + \beta}{2\pi} \right\rfloor}[\Sigma_2]~.
\end{equation}
When the sum of the two angles is larger than $2\pi$, the floor function $\left\lfloor \cdot \right\rfloor$ gives $1$, meaning that we need to insert the generator of $\mathbb{Z}_N^{(0)}$, signaling the group extension of $U(1)$ by $\mathbb{Z}_N$.  This is a specific instance of the general discussion in section \ref{sec:CUBIC}. As shown in \eqref{cond-def}, the $\mathbb{Z}_N^{(0)}$ defects $U_k$ are condensation defects  of the
``dual'' $\mathbb{Z}_N^{(1)}\times \mathbb{Z}_N^{(1)}$ one-form symmetry.   

Let us now explain the configurations of extended operators that carry charge under the extension. Applying the general discussion in section \ref{States}, the objects charged under the $U_k$'s are links of two line operators. Both of them are twist defects, one for the $\mathbb{Z}_N$ symmetry gauged by $a_1$, and one for the $\mathbb{Z}_N$ symmetry gauged by $b_1$. We denote them as $t^{n_A}_A$, $t^{n_B}_B$, respectively, where $n_A,n_B$ are integer labels. They can be defined by assigning a prescribed monodromy for the fields $z$ around the curve on which they are supported, namely
\begin{equation}\label{eq:gaugedtheory-2}
\begin{split}
   &\langle t^{n_{A,B}}_{A,B}[\gamma_1] \dots \rangle \\
    &=
   \frac{1}{ Z_{\text{QED}/{\mathbb{Z}_N \times \mathbb{Z}_N}}} \left.\int \mathcal{D}[z,\bar{z},a, a_1,b_1,\phi_A,\phi_B,\omega_A,\omega_B]\right\vert_{\overset{\displaystyle z^{\circlearrowleft_{\gamma_1}
   } = \exp(i \tfrac{2\pi n_{A,B}}{N} T_{A,B})z~,}{\displaystyle\bar{z}^{\circlearrowleft_{\gamma_1}
   } = \exp(-i \tfrac{2\pi n_{A,B}}{N} T_{A,B})\bar{z}}}\\
    & \times (\dots) \exp\left(-S_{\text{QED}}[a_1,b_1]\right)\\
    \ & 
    \times\exp
   \left( \frac{i}{2\pi}
    \int_{\mathcal{M}_3}
    \omega_A \wedge (Na_1 -d\phi_A)
    +
    \omega_B \wedge (Nb_1 -d\phi_B)
    \right)   \ .
\end{split}
\end{equation}
Here the dots denote arbitrary operator insertions away from the curve $\gamma_1$. The notation $z^{\circlearrowleft_{\gamma_1}}, \bar{z}^{\circlearrowleft_{\gamma_1}}$ denotes the monodromy around the codimension-two locus specified by the curve $\gamma_1$. More precisely, denoting with $\theta_{A,B}$ the phases of the fields $z,\bar{z}$ associated with the two $U(1)$ transformations \eqref{eq:gtAB}, the integers $ n_{A,B}$ are the winding numbers of $\theta_{A,B}$ around $\gamma_1$. Only their reductions modulo $N$, i.e.~$[n_{A,B}]_N$, are protected by the topology. Correspondingly, these line operators are charged under the respective ``dual'' $\mathbb{Z}_N^{(1)}$ symmetry, with charges $[n_A]_N,[n_B]_N\in \mathbb{Z}_N$. Another presentation of the same operators can be given by noticing that the nontrivial periodicity can be reabsorbed by a singular (not properly quantized) gauge transformation \eqref{eq:gtAB}, with parameter $\Lambda_{A,B}$ satisfying
\begin{equation}
d \Lambda_{A,B} = \frac{2 \pi n_{A,B}}{N}\,\delta^{(1)}(D_2)~,~~\partial D_2 = \gamma_1~.
\end{equation}
The location of the surface $D_2$ that fills $\gamma_1$ is not physical because it can be changed by a nonsingular (properly quantized) gauge transformation. After this transformation, the defect can be described as a disorder operator for the compact scalars $\phi_{A,B}$, namely in the path integral we prescribe a singular behavior for this field on the curve $\gamma_1$ such that
\begin{equation}\label{eq:twistdisorder}
d \phi_{A,B} = 2 \pi n_{A,B}\,\delta^{(1)}(D_2)~,~~\partial D_2 = \gamma_1~.
\end{equation}

Having defined the twist defects, we can now see explicitly that the link between $t^{n_A}_A$ and $t^{n_B}_B$ carries charge under the $\mathbb{Z}_N^{(0)}$ defects $U_k$. This is most easily seen in the description of $t^{n_A}_A$ and $t^{n_B}_B$ as disorder operators for $\phi_A$ and $\phi_B$, respectively. Indeed, using \eqref{eq:phiAphiB}, together with \eqref{eq:twistdisorder}, we obtain
\begin{equation}
U_k[\Sigma_2]
    t^{n_A}_A[\gamma_1]
    t^{n_B}_B[\gamma'_1] = \exp\left(-\frac{2\pi i k\, n_A n_B}{N}L_3(\Sigma_2,\gamma_1,\gamma_1')\right) t^{n_A}_A[\gamma_1]
    t^{n_B}_B[\gamma'_1]~,  
\end{equation}
where $L_3(\Sigma_2,\gamma_1,\gamma_1')$ denotes the triple linking number \eqref{TLformula}, see fig.~\ref{fig:TLK} for an illustration of the defect measuring the charge of the configuration. Note that the phase depends on the integers $n_A,n_B$ only through the $\mathbb{Z}_N$ projections $[n_A]_N,[n_B]_N$. As a result, the $\mathbb{Z}_N^{(0)}$ charge, defined as the coefficient of the triple-linking number, is given by the product $[n_A]_N[n_B]_N$ of the charges under the respective $\mathbb{Z}_N^{(1)}$ symmetries.

Since the link of twist defects carries charge under the extension (in the sense of triple-linking, as we just explained), it must also be charged under the whole extended $U(1)^{(0)}_M$, or equivalently it must carry fractional charge (i.e.~quantized in units of $1/N$) under the magnetic symmetry. To see this, and to determine the charge, consider using the operator equation \eqref{eq:2piN-ext} in a correlation function containing also the twist defects. We obtain
\begin{equation}
U_{\alpha+2\pi k}^M[\Sigma_2]
    t^{n_A}_A[\gamma_1]
    t^{n_B}_B[\gamma'_1] = \exp\left(-\frac{2\pi i k\, n_A n_B}{N}L_3(\Sigma_2,\gamma_1,\gamma_1')\right) U_{\alpha}^M[\Sigma_2]
    t^{n_A}_A[\gamma_1]
    t^{n_B}_B[\gamma'_1]~.
\end{equation}
The most general solution to this equation is
\begin{equation}
U_{\alpha}^M[\Sigma_2]
    t^{n_A}_A[\gamma_1]
    t^{n_B}_B[\gamma'_1] = \exp\left(-\frac{i \alpha \,\widetilde{n}_A \widetilde{n}_B}{N}L_3(\Sigma_2,\gamma_1,\gamma_1')\right) 
    t^{n_A}_A[\gamma_1]
    t^{n_B}_B[\gamma'_1]~,
\end{equation}
where $\widetilde{n}_A,\widetilde{n}_B$ are integers such that $[\widetilde{n}_{A,B}]_N = [n_{A,B}]_N$. For a generic angle $\alpha$ we see that the phase factor is no longer just a function of the one-form symmetry charges $[n_A]_N$ and $[n_B]_N$, but rather it depends on an integer lift $\widetilde{n}_{A,B}$. The constraint coming from the symmetry structure simply tells us that such integer lifts must appear but, by itself, it does not specify the values of $\widetilde{n}_A$ and $\widetilde{n}_B$. The particular lifts will depend on the properties of the line operators involved, which clearly are not uniquely specified by their charges under the $\mathbb{Z}^{(1)}_N$ symmetry. In the case at hand, since the operators are precisely defined by an integer lift $n_{A,B}$ of the $\mathbb{Z}^{(1)}_N$ charge, it is natural to conjecture that $\widetilde{n}_{A,B} = n_{A,B}$. 

By exploiting the detailed structure of the $U(1)^{(0)}_M$ symmetry in the theory under consideration, and in particular on the origin of the extension, eq.~\eqref{fract-fluxes}, we can in fact do better and prove the conjecture, with some additional minimal assumption. To this end, let us first go back to the theory before gauging $\mathbb{Z}_N^{(0)}\times \mathbb{Z}_N^{(0)}$. In this theory we can turn on background $PSU(N)$ bundles, in which case the characteristic class $w_2\in H^2(\mathcal{M}_3,\mathbb{Z}_N)$ is a nonzero $c$-number. It is convenient to embed $w_2$ in a background $U(1)$ two-form gauge field $B_2$. This can be achieved by enlarging the field content with the addition of $B_2$, together with the standard Lagrange multiplier fields to ensure that on-shell it becomes a $\mathbb{Z}_N$ gauge field, and then imposing the background gauge invariance
\begin{equation}\label{gauge-symm-L-1}
\begin{split}
    B_2 & \rightarrow B_2 + d\Lambda_1 \\
    f & \rightarrow f + d\Lambda_1 \ , \\
\end{split}
\end{equation}
upon which one can identify $w_2=\frac{N}{2\pi} B_2$. This forces us to define the gauge-invariant quantity $\tilde{f}=f-B_2$ as the current for the magnetic symmetry. As a result, the topological charge operator for $U(1)_M^{(0)}$ is
\begin{equation}
    U^M_{\alpha}[\Sigma_2,B_2] = 
    \exp
    \left(
    \frac{i\alpha}{2\pi}\int_{\Sigma_2}\tilde{f}
    \right)
    = 
    \exp
    \left(
    \frac{i\alpha}{2\pi}\int_{\Sigma_2}(f-B_2)
    \right) \ .
\end{equation}
If we compare the defect at $\alpha+2\pi k$ with the one at $\alpha$ we get the expected result
\begin{equation}
    U^M_{\alpha+2\pi k}[\Sigma_2,B_2]
    =
    U^M_{\alpha}[\Sigma_2,B_2]
    \exp
    \left(
    -ik\int_{\Sigma_2}B_2
    \right)
    =
    U^M_{\alpha}[\Sigma_2,B_2]
    \exp
    \left(
    -\frac{2\pi i k}{N}\int_{\Sigma_2}w_2
    \right)
    \ .
\end{equation}
The restriction from background $PSU(N)$ bundles to $\mathbb{Z}_N\times\mathbb{Z}_N$ ones is achieved by 
\begin{equation}
    \frac{NB_2}{2\pi} = \frac{NA_1}{2\pi} \wedge \frac{NB_1}{2\pi}
    =
    \frac{d\phi_{A}}{2\pi} \wedge \frac{d\phi_{B}}{2\pi}~,
\end{equation}
where $A_1$ and $B_1$ are the background fields introduced in section \ref{sec:gaugZnZn}. This is the only possible identification upon the restriction, if we assume that $B_2$ is entirely determined by $\phi_A$ and $\phi_B$.\footnote{A more general possibility would be $\frac{NB_2}{2\pi} =
    (1+ m N)\frac{d\phi_{A}}{2\pi} \wedge \frac{d\phi_{B}}{2\pi}$ for an arbitrary integer $m$. This would lead to either $\widetilde{n}_A$ or $\widetilde{n}_B$ being a multiple of $1+m N$. If we further assume that $\widetilde{n}_A$,$\widetilde{n}_B$ should take all possible integer values, the only possibility is $m=0$.} The parameter of the gauge symmetry \eqref{gauge-symm-L-1} is then identified with those of the symmetries acting on $\phi_{A,B}$ as follows
\begin{equation}
    d\Lambda_1 = 
    d\Lambda_A 
    \wedge d\phi_B
    +
    d\phi_A
    \wedge d\Lambda_B
    + N
    d\Lambda_A 
    \wedge d\Lambda_B \ .
\end{equation}
The gauge-invariant magnetic defect which satisfies \eqref{eq:2piN-ext} is
\begin{equation}
    U^M_{\alpha}[\Sigma_2]
    =
    \exp
    \left(
    \frac{i\alpha}{2\pi}\int_{\Sigma_2}\tilde{f}
    \right)
    = 
    \exp
    \left(
    \frac{i\alpha}{2\pi}\int_{\Sigma_2}f
    \right)
    \exp
    \left(
    -\frac{i\alpha}{N}\int_{\Sigma_2}
    \frac{d\phi_A}{2\pi}
    \wedge
    \frac{d\phi_B}{2\pi}
    \right) \ .
\end{equation}
Notice that for $\alpha\notin 2\pi \mathbb{Z}$ the two terms are not separately gauge invariant, while their product is. This last expression shows that, when the operator triple-links with the twist defects, the coefficient of the triple-linking number is controlled precisely by the integers $n_{A,B}$ that define the twist defects. Once again this is manifest in the presentation \eqref{eq:twistdisorder} of the twist defects. As a result, we showed that the value of the fractional $U(1)_M^{(0)}$ charge of a link of twist defects  $t^{n_A}_A[\gamma_1]
    t^{n_B}_B[\gamma'_1]$ is precisely $\frac{n_A n_B}{N}$.

\subsection{Extension of non-Abelian symmetry in \texorpdfstring{$3d$}{3d} supersymmetric theories} \label{sec:susy-non-Abelian}
Consider $3d$ $\mathcal{N}=4$ supersymmetric QED (SQED) with $N$ charged hypermultiplets $\mathcal{Q}$. In terms of $\mathcal{N}=2$ supersymmetry the hypermultiplets decompose into pairs of chiral superfields $Q,\widetilde{Q}$ of charge $1,-1$.
This theory has a $PSU(N)^{(0)}$ flavor symmetry acting on $Q_i$ (and $\widetilde{Q}^i$) and a $U(1)^{(0)}_M$ magnetic symmetry with current $\star \frac{f}{2\pi}$. Like in the non-supersymmetric case, these two symmetries have the mixed 't Hooft anomaly  \eqref{anomaly-QED3}. The theory is strongly coupled in the IR. For $N=2$ it flows to a SCFT, where  $U(1)^{(0)}_M$ is enhanced to $SO(3)^{(0)}_M$ and the anomaly inflow is matched by \cite{Genolini:2022mpi}
\begin{equation}
     \frac{2\pi i}{2} \int_{\mathcal{Y}_4} w_2(SO(3)_f) \cup w_2(SO(3)_M) \ .
\end{equation}
For $N>2$ instead, there is no non-Abelian enhancement of the magnetic symmetry. As a result, the behavior in this case is analogous to the non supersymmetric theory studied in the previous section, and here we focus on the case of $N=2$.

If we now consider the same $\mathbb{Z}_2^{(0)} \times \mathbb{Z}_2^{(0)}$ subgroup of $PSU(2)^{(0)}=SO(3)_f^{(0)}$ as the one discussed in scalar QED in section \ref{subsec:gadsiq3}, the anomaly reduces to
\begin{equation}
     \frac{2\pi i}{2} \int_{\mathcal{Y}_4} \bigg( \beta(\mathcal{A}_1) +
    \beta(\mathcal{B}_1) + \mathcal{A}_1 \cup \mathcal{B}_1\bigg) \cup w_2(SO(3)_M) \ .
\end{equation}
A difference with the case of scalar QED is that, due to the non-Abelian enhancement, in this case we cannot discard the Bockstein terms because $\beta(w_2(SO(3)_M))=w_3(SO(3)_M)\neq 0$. Nevertheless, we can still rewrite this inflow, up to a counterterm, as
\begin{equation}\label{eq:ano-susy-qed-ZNxZN}
     \frac{2\pi i}{2} \int_{\mathcal{Y}_4} 
    \bigg( 
    \mathcal{A}_1 \cup \mathcal{B}_1 \cup  w_2(SO(3)_M) 
    +
    \mathcal{A}_1 \cup w_3(SO(3)_M)
    +
    \mathcal{B}_1 \cup w_3(SO(3)_M) 
    \bigg) \ .
\end{equation}
This rewriting is useful to discuss the consequences of this anomaly if we gauge $\mathbb{Z}_2^{(0)} \times \mathbb{Z}_2^{(0)}$. Let us examine the consequences of the various terms in \eqref{eq:ano-susy-qed-ZNxZN}:
\begin{itemize}
    \item Suppose we had only the first term in \eqref{eq:ano-susy-qed-ZNxZN}. In this case the $SO(3)^{(0)}_M$ symmetry would be extended to $SU(2)_M^{(0)}$ by the condensation defect of the dual $\widehat{\mathbb{Z}}_2^{(1)}\times \widehat{\mathbb{Z}}_2^{(1)}$ symmetry.
    \item If we had only the other two terms $\big(\mathcal{A}_1 + \mathcal{B}_1 \big) \cup w_3(SO(3)_M)$, instead, we could change the basis of $\mathbb{Z}_2^{(0)} \times \mathbb{Z}_2^{(0)}$ to
    \begin{equation}
        \begin{cases}
        \mathcal{A}_1' = \mathcal{A}_1 + \mathcal{B}_1 \\
        \mathcal{B}_1' = \mathcal{B}_1 \\
    \end{cases}
    \end{equation}
    so that the anomaly becomes 
    $\mathcal{A}'_1 \cup w_3(SO(3)_M)$. After gauging $\mathbb{Z}_2^{(0)}\times\mathbb{Z}_2^{(0)}$, the second $\mathbb{Z}_2^{(0)}$ (corresponding to $\mathcal{B}_1'$) gives rise to an independent $\widehat{\mathbb{Z}}_2^{(1)}$, while the other $\widehat{\mathbb{Z}}_2^{(1)}$ (corresponding to $\mathcal{A}_1'$) is combined with $SO(3)^{(0)}_M$ in a $2$-group with
    \begin{equation}\label{2-group-ext-susy}
        \delta \widehat{\mathcal{A}}'_2
        = - w_3(SO(3)_M) \ .
    \end{equation}
    As usual, one can see the 2-group structure by analyzing the gauge transformations of the partition function before gauging, as follows: the anomaly term $i\pi \int_{\mathcal{Y}_4} \mathcal{A}_1' \cup w_3$ signals that under $\mathcal{A}_1' \to \mathcal{A}_1' + \delta \Lambda_0$ the partition function picks a phase
    \begin{equation}
        Z[\mathcal{A}'_1+\delta\Lambda_0,\dots] = Z[\mathcal{A}'_1,\dots] \exp \left( -i\pi \int_{\mathcal{M}_3} \Lambda_0 w_3  \right) \ .
    \end{equation}
    Naively, this would suggest that if we gauge $\mathbb{Z}_2^{(0)}$, $SO(3)^{(0)}_M$ is broken, meaning that we cannot consistently couple it. However, as a result of the gauging procedure, we get a dual $\mathbb{Z}_2^{(1)}$ that we can couple through a term $i\pi \int_{\mathcal{M}_3} \mathscr{a}_1'
    \cup \widehat{\mathcal{A}}_2'$ so that the new partition function is
    \begin{equation}
    \sim \sum_{\mathscr{a}_1'} Z[\mathscr{a}_1',\dots] \exp\left(i\pi\int_{\mathcal{M}_3} \mathscr{a}_1' \cup \widehat{\mathcal{A}}_2' \right) \ .  
    \end{equation}
    Now, under a gauge transformation $\mathscr{a}_1' \to \mathscr{a}_1' + \delta \lambda_0$, we generate a phase
    \begin{equation}
        \exp \left( -i\pi \int_{\mathcal{M}_3} \lambda_0 w_3 \right)
        \exp \left( i\pi \int_{\mathcal{M}_3} \delta \lambda_0 \cup \widehat{\mathcal{A}}_2' \right)
        =
        \exp \left( -i\pi \int_{\mathcal{M}_3} \lambda_0 (w_3 
        + \delta \widehat{\mathcal{A}}_2') \right) \ ,
    \end{equation}
    which vanishes if we impose \eqref{2-group-ext-susy}. Thus, the naive conclusion that $SO(3)^{(0)}_M$ was broken is reinterpreted as the fact that $SO(3)^{(0)}_M$ is not a good subgroup of the full $2$-group symmetry \cite{Cordova:2018cvg}. 
    \item 
    If we now consider the whole anomaly \eqref{eq:ano-susy-qed-ZNxZN}, the anomalous variation of the partition function is 
    \begin{equation}
    \begin{split}
    Z&[\mathcal{A}_1+\delta\Lambda^A_0, \mathcal{B}_1+\delta\Lambda^B_0\dots] \\
    = & Z[\mathcal{A}_1,\mathcal{B}_1,\dots] \exp \left(- i\pi \int_{\mathcal{M}_3} 
        (
        \Lambda^A_0\mathcal{B}_1
        -\mathcal{A}_1\Lambda^B_0
    - \delta \Lambda^A_0 \Lambda^B_0
        ) \cup w_2+ \Lambda^A_0 w_3 + \Lambda^B_0 w_3
        \right) \ .
    \end{split}
    \end{equation}
    Once we gauge $\mathbb{Z}_2^{(0)} \times \mathbb{Z}_2^{(0)}$
    we can cancel this phase using the backgrounds of the dual symmetry $\widehat{\mathbb{Z}}_2^{(1)}\times \widehat{\mathbb{Z}}_2^{(1)}$ and the one for the composite symmetry $\mathbb{Z}_2^{(0)}$. Indeed, if we couple them, the new partition function is
    \begin{equation}
    \sim \sum_{\mathscr{a}_1,\mathscr{b}_1} Z[\mathscr{a}_1,\mathscr{b}_1,\dots] \exp 
    \left(
    i\pi\int_{\mathcal{M}_3}
    \mathscr{a}_1 \cup \widehat{\mathcal{A}}_2 +
    \mathscr{b}_1 \cup \widehat{\mathcal{B}}_2 +
    \mathscr{a}_1 \cup \mathscr{b}_1 \cup \mathcal{D}_1
    \right) \ .  
    \end{equation}
    Now, under a gauge transformation 
    \begin{equation}
    \mathscr{a}_1 \to \mathscr{a}_1 + \delta \lambda^A_0 \ , \qquad \mathscr{b}_1 \to \mathscr{b}_1 + \delta \lambda^B_0
    ,    
    \end{equation}
     a phase
    \begin{equation}
    \begin{split}
        & \exp \left( -i\pi \int_{\mathcal{M}_3} 
        (
        \lambda^A_0\mathscr{b}_1
        -\mathscr{a}_1\lambda^B_0
        -\delta \lambda^A_0 \lambda^B_0
        )\cup w_2+ \lambda^A_0 w_3 + \lambda^B_0 w_3  
        \right) \\
        \ & \times 
        \exp \left( i\pi \int_{\mathcal{M}_3}  
        (\delta \lambda_0^A \cup \mathscr{b}_1
        +\mathscr{a}_1 \cup \delta \lambda_0^B
        +\delta \lambda_0^A \cup \delta \lambda_0^B) \cup \mathcal{D}_1
        +
        \delta \lambda_0^A \cup \widehat{\mathcal{A}}_2 +
        \delta \lambda_0^B \cup \widehat{\mathcal{B}}_2 
        \right) 
    \end{split}
    \end{equation}
    is generated. Integrating by parts the second line, the phase vanishes if we impose 
    \begin{equation}
    \begin{split}
        \delta \mathcal{D}_1 & = w_2 \\
        \delta \widehat{\mathcal{A}}_2 & = -w_3 \\
        \delta \widehat{\mathcal{B}}_2 & = -w_3 \ . \\
    \end{split}
    \end{equation}
    Given that the first equation imposes that $w_2$ is exact, $w_3=\beta_2(w_2)$ is also exact, which means that in this case there is no $2$-group since the putative Postnikov class $w_3$ vanishes in cohomology. The conclusion is that, as a consequence of \eqref{eq:ano-susy-qed-ZNxZN}, the $SO(3)_M^{(0)}$ symmetry is extended to $SU(2)_M^{(0)}$ by the condensation defect of the dual $\widehat{\mathbb{Z}}_2^{(1)}\times \widehat{\mathbb{Z}}_2^{(1)}$.
\end{itemize}

\subsubsection{Generalization to \texorpdfstring{$T(SU(N))$}{TSU(N)} with \texorpdfstring{$N>2$}{N>2}}
The SCFT in the previous section can be seen as the Gaiotto-Witten $T(SU(N))$ theory \cite{Gaiotto:2008ak} for $N=2$. The generalization to $N>2$  is given by the IR fixed point of the gauge theory described by the following quiver diagram 
\begin{equation}\label{TSUNquiver}
\begin{tikzpicture}[baseline,scale=0.85]
\tikzstyle{every node}=[font=\scriptsize]
\node[draw, circle] (p0) at (0,0) {\fontsize{14pt}{14pt}\selectfont $1$};
\node[draw, circle] (p1) at (3,0) {\fontsize{14pt}{14pt}\selectfont $2$};
\node (p2) at (5,0) {\fontsize{12pt}{12pt}\selectfont $\cdots$};
\node[draw, circle] (p3) at (7,0) {\fontsize{8pt}{8pt}\selectfont $N-1$};
\node[draw, rectangle] (p4) at (10,0) {\fontsize{14pt}{14pt}\selectfont $N$};

\draw (p0) to (p1);
\draw (p1) to (p2);
\draw (p2) to (p3);
\draw (p3) to (p4);
\end{tikzpicture}~~~.
\end{equation}
The $T(SU(N))$ theory has symmetry group  $PSU(N)_f^{(0)} \times PSU(N)_M^{(0)}$ with anomaly \cite{Mekareeya:2022spm}
\begin{equation}
     \frac{2\pi i}{N} \int_{\mathcal{Y}_4} w_2(PSU(N)_f) \cup w_2(PSU(N)_M) \ .
\end{equation}
If we pick again the aforementioned  $\mathbb{Z}_N^{(0)} \times \mathbb{Z}_N^{(0)}$ subgroup of $PSU(N)_f^{(0)}$, $w_2(PSU(N)_f)$ reduces as in \eqref{eq:w2-N-reduction}.  Thus, if we gauge it, the $PSU(N)^{(0)}_M$ symmetry is extended to $SU(N)_M^{(0)}$. Given that in the $N$ odd case there are no $\beta_2$ terms in the reduction of $w_2(PSU(N)_f)$, the analysis does not have the difficulties it had in the $N=2$ case. In the $N$ even case, those are instead present and their resolution follows the same treatment as in the $N=2$ case. 

\subsection{Extension in \texorpdfstring{$5d$}{5d} from the supersymmetric partition function}
\label{sec:index-E1}

We now focus on $\mathcal{N}=1$ five-dimensional SYM with $\mathfrak{su}(2)$ gauge algebra. As already noticed in \cite{Bertolini:2025wyj}, due to the mixed anomaly \eqref{eq:BGTanointro} of the theory with gauge group $SU(2)$, in the theory with gauge group $SO(3)$ the $U(1)^{(0)}_I$ instantonic symmetry is extended. As explained in section \ref{States}, this has the implication of introducing projective representations of $U(1)^{(0)}_I$ in suitable Hilbert spaces $\mathcal{H}[\Sigma_4]$. 

This statement can be explicitly checked by computing partition functions on $S^1_R \times \Sigma_4$, where the subscript refers to the radius $R$ of the circle, with a fugacity 
\begin{equation}
    \theta = \int_{S^1_R}A_I \ ,
\end{equation}
for the instantonic symmetry. We follow the computation carried out in \cite{Kim:2025fpz} for $\mathcal{N}=1$ SYM with gauge group $SU(2)$. The partition function generically depends on the metric of $\Sigma_4$. A (compact oriented) four-manifold is equipped with an intersection form, i.e.~a quadratic form on the second de Rham cohomology
\begin{equation}
    B:H^2(\Sigma_4,\mathbb R)\times H^2(\Sigma_4,\mathbb R) \rightarrow \mathbb{R} \ .
\end{equation}
The intersection form is positive definite on self-dual two-forms, and negative definite on anti-self-dual two-forms. We denote its signature with $(b_2^+,b_2^-)$. If $b_2^+=1$, the geometric dependence of the partition function is expressed through the so-called period point, i.e.~a normalized self-dual two-form
\begin{equation}
    J \in H^2(\Sigma_4,\mathbb{R})\ , \ \ \star J=J\ , \ \ B(J,J)=1 \ . 
\end{equation}
The dependence on the gauge coupling $g^2$ and the fugacity for the instantonic symmetry $\theta$ can be packed in the complex parameter $\mathcal{R}$, where
\begin{equation}
    \mathcal{R}^4=\exp(-\frac{8\pi^2 R}{g^2}+i\theta) \ .
\end{equation}

We are interested in turning on background gauge fields for the electric $\mathbb{Z}_2^{(1)}$ 1-form symmetry. Following the conventions of \cite{Kim:2025fpz}, these can be parametrized by an element of $H^2(\Sigma_4,\mathbb{Q})$, denoted as $\boldsymbol{\mu}$. In order to restrict to $\mathbb{Z}_2^{(1)}$, one takes $2\boldsymbol{\mu}$ to be an integral lift of the second Stiefel Whitney class $w_2$ of the gauge bundle. Similarly, the flux for the instantonic $U(1)^{(0)}_I$ 0-form symmetry is denoted as $\boldsymbol{n}\in H^2(\Sigma_4,\mathbb{Z})$. The partition function can be expanded in a power series for small $\mathcal{R}$
\begin{equation}
    ^{SU(2)}Z^{J}_{\boldsymbol{\mu},\boldsymbol{n}}(\mathcal{R})=\sum_l d_{\boldsymbol{\mu},\boldsymbol{n}}(l)\mathcal{R}^l\ ,
\end{equation}
where we added a left superscript specifying the global form of the gauge group. The summation index runs over (see equations (4.58) and (4.60) in \cite{Kim:2025fpz})
\begin{equation}\label{Z:Rpower}
    l= -4B(\boldsymbol{\mu},\boldsymbol{\mu})-3\chi_h \text{   mod } 4 \ .
\end{equation}
The norm of the electric background $B(\boldsymbol{\mu},\boldsymbol{\mu})$ gives a multiple of $1/4$, and a multiple of $1/2$ on spin manifolds.  
Notice that by turning off $\boldsymbol{\mu}$, the powers of $\mathcal{R}$ appearing in the expansion are all multiples of 4, modulo an overall offset $-3\chi_h$, that depends on the Euler characteristic $\chi$ and the signature $\sigma$ of $\Sigma_4$ as $\chi_h=\frac{\chi+\sigma}{4}$. This means that the result is periodic in $\theta$ with period $2\pi$, i.e.~the correct  identification is $\theta\sim \theta + 2\pi$. Turning on electric backgrounds generically shifts the powers of $\mathcal{R}$ appearing in the expansion by an arbitrary integer. We will restrict to spin manifolds $\Sigma_4$, where the power shift is constrained to be an even integer. In this case, summing over all possible background electric fluxes will generically produce all even integer powers of $\mathcal{R}$, therefore replacing the $2\pi$ identification with a larger $4\pi$ identification for $\theta$. This sum is precisely reconstructing the partition function of the $SO(3)$ gauge theory as it is implementing the gauging of the electric symmetry
\begin{equation}
    ^{SO(3)}Z^{J}_{\boldsymbol{n}}(\mathcal{R})=\sum_{\boldsymbol{\mu}} \ ^{SU(2)}Z^{J}_{\boldsymbol{\mu},\boldsymbol{n}}(\mathcal{R})\ .
\end{equation}
From this fact we can deduce that generically the fugacity of the $U(1)_I^{(0)}$ instantonic symmetry is extended to have a $4\pi$ periodicity (in the spin case).

Below, we verify this expectation by explicitly extracting the coefficients of the small $\mathcal{R}$ expansion, using the formulas derived in \cite{Kim:2025fpz}. As shown there, the partition function can be decomposed as a sum of a $U$-plane integral contribution $\Phi^J_{\boldsymbol{\mu},\boldsymbol{n}}(\mathcal{R})$ and a sum over Seiberg-Witten contributions $Z_{SW,\boldsymbol{\mu},\boldsymbol{n},j}^J$
\begin{equation}
    ^{SU(2)}Z^{J}_{\boldsymbol{\mu},\boldsymbol{n}}(\mathcal{R})=\Phi^J_{\boldsymbol{\mu},\boldsymbol{n}}(\mathcal{R})+\sum_j Z_{SW,\boldsymbol{\mu},\boldsymbol{n},j}^J\ .
\end{equation}
The dependence on the period point $J$ of the $U$-plane integral is discontinuous, and the discontinuity
\begin{equation}\label{eq:wallcross}
    \Delta\Phi^{JJ'}_{\boldsymbol{\mu},\boldsymbol{n}}(\mathcal{R})=\Phi^{J}_{\boldsymbol{\mu},\boldsymbol{n}}(\mathcal{R})-\Phi^{J'}_{\boldsymbol{\mu},\boldsymbol{n}}(\mathcal{R}) \ ,
\end{equation}
can be expanded around $\mathcal{R}=0$ using a wall-crossing formula. The contribution to the wall-crossing formula from the weak-coupling cusp in the $U$ plane, denoted with an additional subscript $\infty$, is given by formula (5.74) in
\cite{Kim:2025fpz}.

\begin{table}[ht]
\begin{center}
    \begin{tabular}{|>{$}c<{$} | >{$}c<{$} | >{$}c<{$}|>{$}c<{$}|>{$}c<{$}|}
    \hline
    $\diagbox[width=17mm,
    innerrightsep=11pt,innerleftsep=11pt]{$\boldsymbol{n}$}{$\boldsymbol{\mu}$}$ & \left(0,0\right) 
    & \left(\tfrac{1}{2},0\right) & \left(0,\tfrac{1}{2}\right) 
    & \left(\tfrac{1}{2},\tfrac{1}{2}\right) \\
    \hline
    (0,0) & 0 & 0 & 0 & 0 \\ 
    (0,2) & 0 & -\mathcal{R} & 0 & -4\mathcal{R}^3 \\  
    (0,4) 
    & -6\mathcal{R}^5 + 10\mathcal{R}^9
    & -2\mathcal{R} + 6\mathcal{R}^5
    & 0 
    & -35\mathcal{R}^3 + 28\mathcal{R}^7 \\
    (1,-5) 
    & 56\mathcal{R}^5 - 400\mathcal{R}^9
    & 0 
    & 0
    & 84\mathcal{R}^3 + 64\mathcal{R}^7 \\
    (1,-3)
    & 6\mathcal{R}^5 - 4\mathcal{R}^9
    & 0
    & 0
    & 20\mathcal{R}^3 \\
    (1,-1)
    & 0
    & 0
    & 0
    & \mathcal{R}^3 \\
    (1,1)
    & 0
    & 0
    & 0
    & 0 \\
    (2,-4)
    & 56\mathcal{R}^5 - 184\mathcal{R}^9
    & 4\mathcal{R} - 36\mathcal{R}^5 - 56\mathcal{R}^9
    & 55\mathcal{R}^9
    & 56\mathcal{R}^3 + 120\mathcal{R}^7 \\
    (2,-2)
    & 6\mathcal{R}^5 - 2\mathcal{R}^9
    & 3\mathcal{R} - 3\mathcal{R}^5
    & \mathcal{R}^9
    & 10\mathcal{R}^3 \\
    (2,2)
    & 0
    & \mathcal{R} + \mathcal{R}^5 + \mathcal{R}^9
    & 0
    & -\mathcal{R}^3 - \mathcal{R}^7 \\
    (2,4)
    & 0
    & 0
    & 0
    & -20\mathcal{R}^3 \\
    (3,-5)
    & 252\mathcal{R}^5 - 2200\mathcal{R}^9
    & 0
    & 0
    & 120\mathcal{R}^3 + 1100\mathcal{R}^7 \\
    (3,-3)
    & 56\mathcal{R}^5 - 112\mathcal{R}^9
    & 0
    & 0
    & 35\mathcal{R}^3 + 56\mathcal{R}^7 \\
    (3,3)
    & 0
    & 0
    & 0
    & -4\mathcal{R}^3 - 10\mathcal{R}^7 \\
    (4,-4)
    & 252\mathcal{R}^5 - 1540\mathcal{R}^9
    & 6\mathcal{R} - 90\mathcal{R}^5 - 1316\mathcal{R}^9
    & 2002\mathcal{R}^9
    & 84\mathcal{R}^3 + 603\mathcal{R}^7 \\
    (4,4)
    & 0
    & 2\mathcal{R} + 18\mathcal{R}^5 + 90\mathcal{R}^9
    & 0
    & -10\mathcal{R}^3 - 54\mathcal{R}^7 \\
    \hline
    \end{tabular}
    \end{center}
    \caption{Some values of $\Delta\Phi^{JJ'}_{\boldsymbol{\mu},\boldsymbol{n}}(\mathcal{R})$ expanded up to $\mathcal{O}(\mathcal{R}^{9})$.
    Here $J=(1+\delta,1)$, $J'=(\epsilon,1)$ and $\delta=10^{-8}$, $\epsilon=10^{-1}$.
    Notice that for $S^2\times S^2$ we have $\chi=4$ and $\sigma=0$, so there is an overall prefactor $\mathcal{R}^{-3}$ in the expansion. Taking this into account, the table represents an expansion in {\it even} powers of ${\cal R}$, as detailed in the main text.}
    \label{tab:DeltaPhi}
\end{table}

We pick the spin manifold $\Sigma_4=S^2 \times S^2$, as it is a candidate to detect fractional instantonic charges according to the discussion in section \ref{sec:states-extension} (setting $p=q=1 $ and $d=5$). The discussion in that section also makes clear that the superconformal index, with $\Sigma_4=S^4$, is not sensitive to fractional instantonic charges. This is equivalent to the statement that point-like operators do not feel the extension. From the point of view of the gauge theory path integral, the useful property of the background $\Sigma_4=S^2 \times S^2$ is that, thanks to the nontrivial two-cycles, it supports bundles of $SO(3)$ which are not bundles of $SU(2)$, while $\Sigma_4=S^4$ does not. With $\Sigma_4=S^2 \times S^2$ we have several simplifications \cite{Kim:2025fpz}
\begin{itemize}
    \item the Seiberg-Witten contributions $Z_{SW,\boldsymbol{\mu},\boldsymbol{n},j}^J$ vanish,
    \item the wall-crossing formula is entirely determined by the weak-coupling cusp in the $U$ plane integral.
\end{itemize}
Hence, it is sufficient to evaluate formula (5.74) in \cite{Kim:2025fpz}.
We display the evaluation of the wall-crossing formula for this background in Table \ref{tab:DeltaPhi}, for fixed values of the period points $J,J'$ and for several values of electric and instantonic background fluxes. From these evaluations we can see that there are several choices of background instantonic fluxes that contain powers of $\mathcal{R}$ that enforce an extended $4\pi$ periodicity of the fugacity.

\subsection{Gauging a discrete subgroup in the \texorpdfstring{$E_1$}{E1} theory}
\label{subsec::3-group-E1}

In \cite{BenettiGenolini:2020doj}, the authors analyzed the anomaly between the $SO(3)_I^{(0)}$ and $\mathbb{Z}_2^{(1)}$  symmetries of Seiberg's $E_1$ theory, a strongly coupled SCFT that can be viewed as a UV completion of five-dimensional  $\mathcal{N}=1$ SYM with gauge group $SU(2)$ \cite{Seiberg:1996bd}.\footnote{See \cite{DeWolfe:1999hj,Kim:2012gu, Bashkirov:2012re,Rodriguez-Gomez:2013dpa,Tachikawa:2015mha,Cremonesi:2015lsa,Closset:2018bjz,Apruzzi:2021vcu,BenettiGenolini:2019zth,Bertolini:2021cew,Bertolini:2022osy,Akhond:2023vlb,DelZotto:2022joo,Cvetic:2022imb,Closset:2023pmc,Acharya:2024bnt,Akhond:2024nyr} for a partial list of works where the symmetries of the $E_1$ theory and some of its siblings have been discussed.}
The anomaly, determined by matching with the IR expression \eqref{eq:BGTanointro}, reads
\begin{equation}\label{eq:anoinfUV} 
   \exp \left( \frac{2\pi i}{2} \int_{\mathcal{Y}_6} 
    w_2(SO(3)_I)
    \cup
    \frac{\mathcal{P}(\mathcal{B}_2)}{2} \right) 
    \ .
\end{equation} 
The result of gauging the $\mathbb{Z}_2^{(1)}$ is already discussed in \cite{Bertolini:2025wyj}: the $SO(3)_I^{(0)}$ symmetry is extended to $SU(2)_I^{(0)}$ by the condensation defect associated with the dual $\mathbb{Z}_2^{(2)}$ symmetry, thereby providing another example of the $n=1$ case. This is the UV SCFT counterpart of what happens in $\mathcal{N}=1$ $SU(2)$ SYM, as we reviewed in the previous section. 

Here we want to discuss instead what happens upon gauging a subgroup of the non-Abelian instantonic symmetry $SO(3)_I^{(0)}$. We will show that this provides also an example of the general phenomenon discussed in section \ref{sec:CUBIC}, though in this case with $n=2$ (i.e.~we will obtain a 3-group involving condensation defects upon gauging). 

We choose the subgroup to be the $\mathbb{Z}_2^{(0)}\times\mathbb{Z}_2^{(0)}$ subgroup of $SO(3)_I^{(0)}$ that we already used multiple times in this section. With this choice, the anomaly \eqref{eq:anoinfUV} reduces to
\begin{equation}\label{eq:anoinfUV-Z2xZ2} 
   \exp \left( \frac{2\pi i}{2} \int_{\mathcal{Y}_6} 
    \bigg( \beta_2(\mathcal{A}_1) +
    \beta_2(\mathcal{B}_1) + \mathcal{A}_1 \cup \mathcal{B}_1\bigg)
    \cup
    \frac{\mathcal{P}(\mathcal{B}_2)}{2} \right) 
    \ .
\end{equation}
If we gauge $\mathbb{Z}_2^{(0)}\times\mathbb{Z}_2^{(0)}$, we get a dual 3-form symmetry $\widehat{\mathbb{Z}}_2^{(3)}\times\widehat{\mathbb{Z}}_2^{(3)}$ with associated background gauge fields $\widehat{\mathcal{A}}_4$ and $\widehat{\mathcal{B}}_4$, and the constraints 
\begin{equation}
\begin{split}\label{eq:3-group}
    \delta \widehat{\mathcal{A}}_4 & = -\beta_2
    \left(
    \frac{\mathcal{P}(\mathcal{B}_2)}{2} 
    \right) \\
    \delta \widehat{\mathcal{B}}_4 & = -\beta_2
    \left(
    \frac{\mathcal{P}(\mathcal{B}_2)}{2} 
    \right) \\
    \delta \mathcal{D}_3 & = 
    \frac{\mathcal{P}(\mathcal{B}_2)}{2} \ ,
\end{split}
\end{equation}
where $\mathcal{D}_3$ is the background gauge field for the $\mathbb{Z}_2^{(2)}$ symmetry generated by the condensation defect of $\widehat{\mathbb{Z}}_2^{(3)}\times\widehat{\mathbb{Z}}_2^{(3)}$. Using the third equation, we see that $\frac{\mathcal{P}(\mathcal{B}_2)}{2}$ is exact, which implies that $\beta_2\left(\frac{\mathcal{P}(\mathcal{B}_2)}{2} \right) = 0$. Thus, we are left with a $3$-group between $\mathbb{Z}_2^{(2)}$ and $\mathbb{Z}_2^{(1)}$, given by the last equation in \eqref{eq:3-group}. 

Following the discussion in section \ref{sec:states-Higher-group}, this symmetry structure allows the existence of two-dimensional interfaces carrying projective charges under $\mathbb{Z}_2^{(1)}$, and that are supported on three-dimensional twist defects charged under $\widehat{\mathbb{Z}}_2^{(3)}\times\widehat{\mathbb{Z}}_2^{(3)}$. Being charged under a 1-form symmetry, these surfaces carry a generalized charge.

\section*{Acknowledgments}
We thank Pierluigi Niro for collaboration in the early stages of the project, and for enlightening discussions thereafter. We also wish to thank Thomas Bartsch, Philip Boyle Smith, Christian Copetti, Matteo Dell'Acqua, Andrea Grigoletto, Pavel Putrov, Elias Riedel G\aa rding and Yunqin Zheng for comments and discussions. We acknowledge support by INFN Iniziativa Specifica ST\&FI.

\appendix

\section{Review of states charged under higher-form symmetries}\label{App-Reps}
In this appendix we provide a warm-up for section \ref{States}, reviewing the action on states of topological operators associated with higher-form symmetries, extensions and higher-groups. We follow arguments along the lines of \cite{Bartsch:2023pzl,Bhardwaj:2023wzd}.

\begin{figure}[t]
    \centering
    \tikzset{every picture/.style={line width=0.75pt}} 

\begin{tikzpicture}[x=0.75pt,y=0.75pt,yscale=-1,xscale=1]

\draw  [color={rgb, 255:red, 0; green, 0; blue, 0 }  ,draw opacity=1 ][dash pattern={on 0.84pt off 2.51pt}] (251,85.88) .. controls (251,65.91) and (267.19,49.72) .. (287.17,49.72) .. controls (307.14,49.72) and (323.33,65.91) .. (323.33,85.88) .. controls (323.33,105.86) and (307.14,122.05) .. (287.17,122.05) .. controls (267.19,122.05) and (251,105.86) .. (251,85.88) -- cycle ;
\draw  [fill={rgb, 255:red, 155; green, 155; blue, 155 }  ,fill opacity=0.2 ][dash pattern={on 0.84pt off 2.51pt}] (161.13,85.88) .. controls (161.13,65.54) and (177.09,49.05) .. (196.79,49.05) .. controls (216.49,49.05) and (232.46,65.54) .. (232.46,85.88) .. controls (232.46,106.22) and (216.49,122.72) .. (196.79,122.72) .. controls (177.09,122.72) and (161.13,106.22) .. (161.13,85.88) -- cycle ;
\draw    (142.33,40.05) -- (142.33,131.05) ;
\draw    (338.33,41.05) -- (355.33,86.05) -- (338.33,132.05) ;
\draw  [draw opacity=0][fill={rgb, 255:red, 0; green, 0; blue, 0 }  ,fill opacity=1 ] (192.73,86.88) .. controls (192.73,84.95) and (194.21,83.38) .. (196.03,83.38) .. controls (197.86,83.38) and (199.33,84.95) .. (199.33,86.88) .. controls (199.33,88.81) and (197.86,90.38) .. (196.03,90.38) .. controls (194.21,90.38) and (192.73,88.81) .. (192.73,86.88) -- cycle ;
\draw  [color={rgb, 255:red, 0; green, 0; blue, 0 }  ,draw opacity=1 ][line width=1.5]  (247.33,85.72) .. controls (247.33,63.62) and (265.24,45.72) .. (287.33,45.72) .. controls (309.42,45.72) and (327.33,63.62) .. (327.33,85.72) .. controls (327.33,107.81) and (309.42,125.72) .. (287.33,125.72) .. controls (265.24,125.72) and (247.33,107.81) .. (247.33,85.72) -- cycle ;

\draw (367,72.05) node [anchor=north west][inner sep=0.75pt]   [align=left] {$\displaystyle \in \mathcal{H}\left[ S^{d-p-1} \ \times S^{p}\right]$};

\end{tikzpicture}
    \caption{State charged under a $p$-form symmetry.}
    \label{states_pform}
\end{figure}
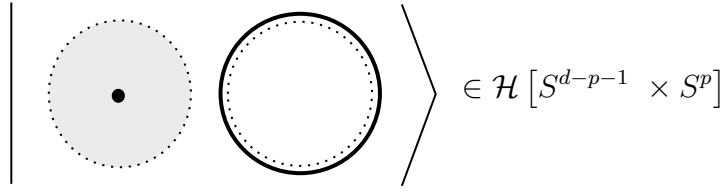

A $p$-form symmetry is a set of codimension-$(p+1)$ topological operators \cite{Gaiotto:2014kfa}. When these topological defects act on $p$-dimensional operators via two-component linking, a candidate state that transforms under this symmetry can be found on the Hilbert space supported on the codimension-1 submanifold $\Sigma_{d-1}=S^{d-p-1}\times S^{p}$. Such a state can be prepared by filling $S^{d-p-1}$, and inserting a charged operator of the $p$-form symmetry wrapping $S^{p}$  as shown in figure \ref{states_pform}.\footnote{In the case $p=0$ we can collapse the $S^0$ factor to a single point. See \cite{Arbalestrier:2026lna} for explicit construction in free theories.}

Two $p$-form symmetries, $A^{(p)}$ and $G^{(p)}$, can form a group extension $\Gamma^{(p)}$ via the exact sequence
\begin{equation}\label{ses-p}
     1\rightarrow A^{(p)}\rightarrow \Gamma^{(p)}\rightarrow G^{(p)} \rightarrow 1 \ ,
\end{equation}
determined by a characteristic class $ t_{p+2}\in H^{p+2}(B^{p+1}G,A)$, where $B^{p+1}G$ is the classifying space of $G^{(p)}$ bundles. Instead, a $p$-form symmetry and a $q$-form symmetry, with $q>p$, can form a higher-group $\underline{\Gamma}$ via the exact sequence
\begin{equation}\label{ses-higher}
    1\rightarrow A^{(q)}\rightarrow \underline{\Gamma}\rightarrow G^{(p)} \rightarrow 1 \ , 
\end{equation}
determined now by a characteristic class $t_{q+2}\in H^{q+2}(B^{p+1}G,A)$. We will restrict to characteristic classes in the image of the homomorphism
\begin{equation}
    \Phi^p:H^{q+2}(B^{p+1}G,A)\rightarrow H^{q-p+2}(BG,A) \ ,
\end{equation}
introduced in section \ref{sec:CUBIC}. This is equivalent to considering classes that are activated by non-generic configurations of topological defects, see the discussion around \eqref{eq:Phihom}.

\subsection{Group extension}
Let us consider two $p$-form symmetries $A^{(p)}$ and $G^{(p)}$ that sit in a nontrivial group extension \eqref{ses-p}.
As we are going to show, states and operators charged under the $A^{(p)}$ symmetry are also {\it projectively} charged under the $G^{(p)}$ symmetry, with the projective phase being determined by the group cohomology class $\Phi^p[t_{p+2}]=\psi \in H^2(BG,A)$. Here we are assuming that $A^{(p)}$ is Abelian for any $p \geq 0$, whereas $G^{(p)}$ may be non-Abelian for $p=0$. 

\begin{figure}[t]
    \centering
    \tikzset{every picture/.style={line width=0.75pt}} 

\begin{tikzpicture}[x=0.75pt,y=0.75pt,yscale=-1,xscale=1]

\draw  [draw opacity=0] (263.65,29.15) .. controls (278.87,16.25) and (298.83,8.38) .. (320.73,8.28) .. controls (368.68,8.05) and (407.72,45.14) .. (407.94,91.1) .. controls (408.16,137.07) and (369.46,174.52) .. (321.52,174.74) .. controls (299.11,174.85) and (278.65,166.81) .. (263.19,153.53) -- (321.12,91.51) -- cycle ; \draw  [color={rgb, 255:red, 74; green, 144; blue, 226 }  ,draw opacity=1 ] (263.65,29.15) .. controls (278.87,16.25) and (298.83,8.38) .. (320.73,8.28) .. controls (368.68,8.05) and (407.72,45.14) .. (407.94,91.1) .. controls (408.16,137.07) and (369.46,174.52) .. (321.52,174.74) .. controls (299.11,174.85) and (278.65,166.81) .. (263.19,153.53) ;  
\draw  [color={rgb, 255:red, 245; green, 166; blue, 35 }  ,draw opacity=1 ] (205.04,91.34) .. controls (205.04,56.76) and (232.64,28.72) .. (266.69,28.72) .. controls (300.74,28.72) and (328.34,56.76) .. (328.34,91.34) .. controls (328.34,125.93) and (300.74,153.97) .. (266.69,153.97) .. controls (232.64,153.97) and (205.04,125.93) .. (205.04,91.34) -- cycle ;
\draw [color={rgb, 255:red, 245; green, 166; blue, 35 }  ,draw opacity=1 ]   (264.16,28.72) -- (264.02,153.97) ;
\draw  [color={rgb, 255:red, 245; green, 166; blue, 35 }  ,draw opacity=1 ] (319.8,100.85) -- (328.69,86.52) -- (335.07,102.14) ;
\draw  [color={rgb, 255:red, 245; green, 166; blue, 35 }  ,draw opacity=1 ] (256.35,94.47) -- (264.12,79.5) -- (271.68,94.58) ;
\draw  [color={rgb, 255:red, 245; green, 166; blue, 35 }  ,draw opacity=1 ] (214.84,82.83) -- (204.85,96.42) -- (199.71,80.36) ;
\draw  [color={rgb, 255:red, 74; green, 144; blue, 226 }  ,draw opacity=1 ] (413.84,102.83) -- (403.85,116.42) -- (398.71,100.36) ;

\draw (171,112.16) node [anchor=north west][inner sep=0.75pt]   [align=left] {$\displaystyle \textcolor[rgb]{0.96,0.65,0.14}{g_{1} g_{2}}$};
\draw (244,120.16) node [anchor=north west][inner sep=0.75pt]   [align=left] {$\displaystyle \textcolor[rgb]{0.96,0.65,0.14}{g_{1}}$};
\draw (329,115.16) node [anchor=north west][inner sep=0.75pt]   [align=left] {$\displaystyle \textcolor[rgb]{0.96,0.65,0.14}{g_{2}}$};
\draw (404,34.16) node [anchor=north west][inner sep=0.75pt]   [align=left] {$\displaystyle \textcolor[rgb]{0.29,0.56,0.89}{\psi }\textcolor[rgb]{0.29,0.56,0.89}{( g}\textcolor[rgb]{0.29,0.56,0.89}{_{1}}\textcolor[rgb]{0.29,0.56,0.89}{,g}\textcolor[rgb]{0.29,0.56,0.89}{_{2}}\textcolor[rgb]{0.29,0.56,0.89}{)}$};

\end{tikzpicture}
    \caption{The junction of topological defects in fig.~\ref{local-def-ext}, associated with the extension, arranged in a configuration that can be used to wrap operators charged under $A^{(p)}$ and $G^{(p)}$.}
    \label{defects-ext}
\end{figure}
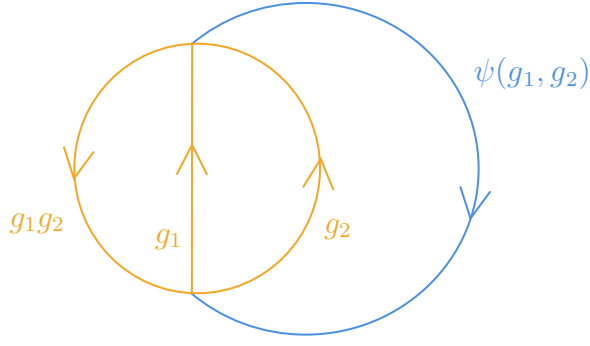

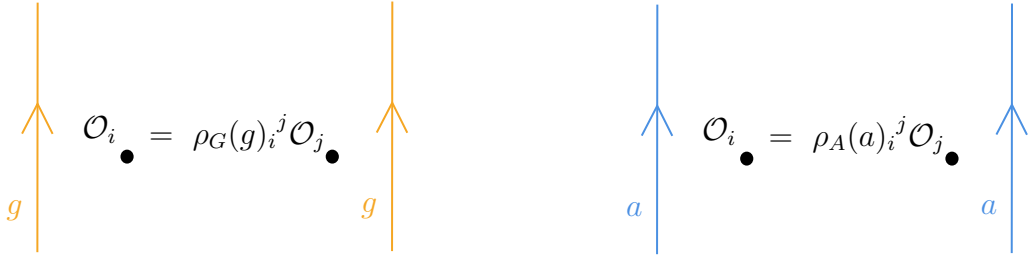
\begin{figure}[t]
    \centering
    \tikzset{every picture/.style={line width=0.75pt}} 

\begin{tikzpicture}[x=0.75pt,y=0.75pt,yscale=-1,xscale=1]

\draw [color={rgb, 255:red, 245; green, 166; blue, 35 }  ,draw opacity=1 ]   (79.5,16.08) -- (79.36,141.33) ;
\draw  [color={rgb, 255:red, 245; green, 166; blue, 35 }  ,draw opacity=1 ] (71.68,81.83) -- (79.45,66.86) -- (87.01,81.94) ;
\draw  [draw opacity=0][fill={rgb, 255:red, 0; green, 0; blue, 0 }  ,fill opacity=1 ] (121.07,93.36) .. controls (121.07,91.43) and (122.54,89.86) .. (124.37,89.86) .. controls (126.19,89.86) and (127.67,91.43) .. (127.67,93.36) .. controls (127.67,95.3) and (126.19,96.86) .. (124.37,96.86) .. controls (122.54,96.86) and (121.07,95.3) .. (121.07,93.36) -- cycle ;
\draw  [draw opacity=0][fill={rgb, 255:red, 0; green, 0; blue, 0 }  ,fill opacity=1 ] (224.07,93.36) .. controls (224.07,91.43) and (225.54,89.86) .. (227.37,89.86) .. controls (229.19,89.86) and (230.67,91.43) .. (230.67,93.36) .. controls (230.67,95.3) and (229.19,96.86) .. (227.37,96.86) .. controls (225.54,96.86) and (224.07,95.3) .. (224.07,93.36) -- cycle ;
\draw [color={rgb, 255:red, 245; green, 166; blue, 35 }  ,draw opacity=1 ]   (257.5,15.53) -- (257.36,140.78) ;
\draw  [color={rgb, 255:red, 245; green, 166; blue, 35 }  ,draw opacity=1 ] (249.68,81.28) -- (257.45,66.31) -- (265.01,81.39) ;
\draw [color={rgb, 255:red, 74; green, 144; blue, 226 }  ,draw opacity=1 ]   (390.5,17.08) -- (390.36,142.33) ;
\draw  [color={rgb, 255:red, 74; green, 144; blue, 226 }  ,draw opacity=1 ] (382.68,82.83) -- (390.45,67.86) -- (398.01,82.94) ;
\draw  [draw opacity=0][fill={rgb, 255:red, 0; green, 0; blue, 0 }  ,fill opacity=1 ] (432.07,94.36) .. controls (432.07,92.43) and (433.54,90.86) .. (435.37,90.86) .. controls (437.19,90.86) and (438.67,92.43) .. (438.67,94.36) .. controls (438.67,96.3) and (437.19,97.86) .. (435.37,97.86) .. controls (433.54,97.86) and (432.07,96.3) .. (432.07,94.36) -- cycle ;
\draw  [draw opacity=0][fill={rgb, 255:red, 0; green, 0; blue, 0 }  ,fill opacity=1 ] (535.07,94.36) .. controls (535.07,92.43) and (536.54,90.86) .. (538.37,90.86) .. controls (540.19,90.86) and (541.67,92.43) .. (541.67,94.36) .. controls (541.67,96.3) and (540.19,97.86) .. (538.37,97.86) .. controls (536.54,97.86) and (535.07,96.3) .. (535.07,94.36) -- cycle ;
\draw [color={rgb, 255:red, 74; green, 144; blue, 226 }  ,draw opacity=1 ]   (568.5,16.53) -- (568.36,141.78) ;
\draw  [color={rgb, 255:red, 74; green, 144; blue, 226 }  ,draw opacity=1 ] (560.68,82.28) -- (568.45,67.31) -- (576.01,82.39) ;

\draw (62.33,113.53) node [anchor=north west][inner sep=0.75pt]   [align=left] {$\displaystyle \textcolor[rgb]{0.96,0.65,0.14}{g}$};
\draw (134,70.53) node [anchor=north west][inner sep=0.75pt]   [align=left] {$\displaystyle =\ \rho _{G}{( g)_i}^{j}\mathcal{O}_{j}$};
\draw (240.33,112.98) node [anchor=north west][inner sep=0.75pt]   [align=left] {$\displaystyle \textcolor[rgb]{0.96,0.65,0.14}{g}$};
\draw (95.33,71.53) node [anchor=north west][inner sep=0.75pt]   [align=left] {$\displaystyle \ \mathcal{O}_{i}$};
\draw (373.33,114.53) node [anchor=north west][inner sep=0.75pt]   [align=left] {$\displaystyle \textcolor[rgb]{0.29,0.56,0.89}{a}$};
\draw (445,71.53) node [anchor=north west][inner sep=0.75pt]   [align=left] {$\displaystyle =\ {\rho _{A}( a)_i}^{j}\mathcal{O}_{j}$};
\draw (551.33,113.98) node [anchor=north west][inner sep=0.75pt]   [align=left] {$\displaystyle \textcolor[rgb]{0.29,0.56,0.89}{a}$};
\draw (406.33,72.53) node [anchor=north west][inner sep=0.75pt]   [align=left] {$\displaystyle \ \mathcal{O}_{i}$};

\end{tikzpicture}
    \caption{Action of $G^{(p)}$ and $A^{(p)}$ on the operators $\mathcal{O}_i$, respectively via matrices $\rho_G$ and $\rho_A$.}
    \label{fig:ExtAction}
\end{figure}

This relation between charged operators can be derived from the corresponding  relation between the topological operators of $G^{(p)}$ and $A^{(p)}$. The topological defects of the $A^{(p)}$ symmetry can terminate on a codimension-$(p+2)$  locus, singled out by the intersection of three topological defects of the $G^{(p)}$ symmetry, as shown in figure \ref{defects-ext}.
Now consider a set of operators $\mathcal{O}_i(\gamma_p)$ with an action via two-component linking of the two symmetries described by unitary matrices $\rho_{G}$ and $\rho_{A}$, as shown in figure \ref{fig:ExtAction}.
While $\rho_A$ furnishes a unitary representation of the Abelian group $A^{(p)}$, the existence of the configuration of topological defects in figure \ref{defects-ext} imposes the constraint
\begin{equation}\label{eq:rGrGrArG}
    \rho_G(g_1)\rho_G(g_2) =\rho_A(\psi(g_1,g_2))\rho_G(g_1g_2) 
\end{equation}
\begin{figure}
    \centering
    \begin{subfigure}{0.79\textwidth}
    \centering
    \tikzset{every picture/.style={line width=0.75pt}} 

\begin{tikzpicture}[x=0.75pt,y=0.75pt,yscale=-1,xscale=1]

\draw  [draw opacity=0] (276.4,50.9) .. controls (290.09,39.36) and (308.03,32.33) .. (327.71,32.24) .. controls (370.88,32.03) and (406.03,65.32) .. (406.23,106.59) .. controls (406.42,147.85) and (371.58,181.47) .. (328.41,181.67) .. controls (308.28,181.77) and (289.89,174.58) .. (275.98,162.7) -- (328.06,106.96) -- cycle ; \draw  [color={rgb, 255:red, 74; green, 144; blue, 226 }  ,draw opacity=1 ] (276.4,50.9) .. controls (290.09,39.36) and (308.03,32.33) .. (327.71,32.24) .. controls (370.88,32.03) and (406.03,65.32) .. (406.23,106.59) .. controls (406.42,147.85) and (371.58,181.47) .. (328.41,181.67) .. controls (308.28,181.77) and (289.89,174.58) .. (275.98,162.7) ;  
\draw  [color={rgb, 255:red, 245; green, 166; blue, 35 }  ,draw opacity=1 ] (223.54,106.8) .. controls (223.54,75.75) and (248.39,50.58) .. (279.05,50.58) .. controls (309.7,50.58) and (334.55,75.75) .. (334.55,106.8) .. controls (334.55,137.85) and (309.7,163.02) .. (279.05,163.02) .. controls (248.39,163.02) and (223.54,137.85) .. (223.54,106.8) -- cycle ;
\draw [color={rgb, 255:red, 245; green, 166; blue, 35 }  ,draw opacity=1 ]   (276.77,50.58) -- (276.65,163.02) ;
\draw  [color={rgb, 255:red, 245; green, 166; blue, 35 }  ,draw opacity=1 ] (327.02,118.16) -- (335.03,105.29) -- (340.78,119.32) ;
\draw  [color={rgb, 255:red, 245; green, 166; blue, 35 }  ,draw opacity=1 ] (269.74,109.61) -- (276.73,96.17) -- (283.54,109.71) ;
\draw  [color={rgb, 255:red, 245; green, 166; blue, 35 }  ,draw opacity=1 ] (232.31,92.73) -- (223.31,104.93) -- (218.69,90.51) ;
\draw  [color={rgb, 255:red, 74; green, 144; blue, 226 }  ,draw opacity=1 ] (411.54,117.12) -- (402.54,129.32) -- (397.92,114.9) ;
\draw  [draw opacity=0][fill={rgb, 255:red, 0; green, 0; blue, 0 }  ,fill opacity=1 ] (376.63,111.31) .. controls (376.63,109.58) and (377.96,108.17) .. (379.6,108.17) .. controls (381.24,108.17) and (382.57,109.58) .. (382.57,111.31) .. controls (382.57,113.05) and (381.24,114.45) .. (379.6,114.45) .. controls (377.96,114.45) and (376.63,113.05) .. (376.63,111.31) -- cycle ;
\draw [color={rgb, 255:red, 155; green, 155; blue, 155 }  ,draw opacity=1 ]   (330,198.84) -- (330,256.04) ;
\draw [shift={(330,258.04)}, rotate = 270] [color={rgb, 255:red, 155; green, 155; blue, 155 }  ,draw opacity=1 ][line width=0.75]    (10.93,-3.29) .. controls (6.95,-1.4) and (3.31,-0.3) .. (0,0) .. controls (3.31,0.3) and (6.95,1.4) .. (10.93,3.29)   ;
\draw  [draw opacity=0] (210.4,272.9) .. controls (224.09,261.36) and (242.03,254.33) .. (261.71,254.24) .. controls (304.88,254.03) and (340.03,287.32) .. (340.23,328.59) .. controls (340.42,369.85) and (305.58,403.47) .. (262.41,403.67) .. controls (242.28,403.77) and (223.89,396.58) .. (209.98,384.7) -- (262.06,328.96) -- cycle ; \draw  [color={rgb, 255:red, 74; green, 144; blue, 226 }  ,draw opacity=1 ] (210.4,272.9) .. controls (224.09,261.36) and (242.03,254.33) .. (261.71,254.24) .. controls (304.88,254.03) and (340.03,287.32) .. (340.23,328.59) .. controls (340.42,369.85) and (305.58,403.47) .. (262.41,403.67) .. controls (242.28,403.77) and (223.89,396.58) .. (209.98,384.7) ;  
\draw  [color={rgb, 255:red, 245; green, 166; blue, 35 }  ,draw opacity=1 ] (157.54,328.8) .. controls (157.54,297.75) and (182.39,272.58) .. (213.05,272.58) .. controls (243.7,272.58) and (268.55,297.75) .. (268.55,328.8) .. controls (268.55,359.85) and (243.7,385.02) .. (213.05,385.02) .. controls (182.39,385.02) and (157.54,359.85) .. (157.54,328.8) -- cycle ;
\draw [color={rgb, 255:red, 245; green, 166; blue, 35 }  ,draw opacity=1 ]   (210.77,272.58) -- (210.65,385.02) ;
\draw  [color={rgb, 255:red, 245; green, 166; blue, 35 }  ,draw opacity=1 ] (261.02,340.16) -- (269.03,327.29) -- (274.78,341.32) ;
\draw  [color={rgb, 255:red, 245; green, 166; blue, 35 }  ,draw opacity=1 ] (203.74,331.61) -- (210.73,318.17) -- (217.54,331.71) ;
\draw  [color={rgb, 255:red, 245; green, 166; blue, 35 }  ,draw opacity=1 ] (167.31,312.73) -- (158.31,324.93) -- (153.69,310.51) ;
\draw  [color={rgb, 255:red, 74; green, 144; blue, 226 }  ,draw opacity=1 ] (345.54,339.12) -- (336.54,351.32) -- (331.92,336.9) ;
\draw  [draw opacity=0][fill={rgb, 255:red, 0; green, 0; blue, 0 }  ,fill opacity=1 ] (376.63,333.31) .. controls (376.63,331.58) and (377.96,330.17) .. (379.6,330.17) .. controls (381.24,330.17) and (382.57,331.58) .. (382.57,333.31) .. controls (382.57,335.05) and (381.24,336.45) .. (379.6,336.45) .. controls (377.96,336.45) and (376.63,335.05) .. (376.63,333.31) -- cycle ;

\draw (193.05,124.58) node [anchor=north west][inner sep=0.75pt]   [align=left] {$\displaystyle \textcolor[rgb]{0.96,0.65,0.14}{g}\textcolor[rgb]{0.96,0.65,0.14}{_{1} g_{2}}$};
\draw (257.77,131.76) node [anchor=north west][inner sep=0.75pt]   [align=left] {$\displaystyle \textcolor[rgb]{0.96,0.65,0.14}{g}\textcolor[rgb]{0.96,0.65,0.14}{_{1}}$};
\draw (324.9,138.94) node [anchor=north west][inner sep=0.75pt]   [align=left] {$\displaystyle \textcolor[rgb]{0.96,0.65,0.14}{g}\textcolor[rgb]{0.96,0.65,0.14}{_{2}}$};
\draw (399.49,139.55) node [anchor=north west][inner sep=0.75pt]   [align=left] {$\displaystyle \textcolor[rgb]{0.29,0.56,0.89}{\psi }\textcolor[rgb]{0.29,0.56,0.89}{(}\textcolor[rgb]{0.29,0.56,0.89}{g}\textcolor[rgb]{0.29,0.56,0.89}{_{1}}\textcolor[rgb]{0.29,0.56,0.89}{,g}\textcolor[rgb]{0.29,0.56,0.89}{_{2}}\textcolor[rgb]{0.29,0.56,0.89}{)}$};
\draw (352.26,90.74) node [anchor=north west][inner sep=0.75pt]   [align=left] {$\displaystyle \ \mathcal{O}^{i}$};
\draw (131.05,351.58) node [anchor=north west][inner sep=0.75pt]   [align=left] {$\displaystyle \textcolor[rgb]{0.96,0.65,0.14}{g}\textcolor[rgb]{0.96,0.65,0.14}{_{1} g_{2}}$};
\draw (191.77,353.76) node [anchor=north west][inner sep=0.75pt]   [align=left] {$\displaystyle \textcolor[rgb]{0.96,0.65,0.14}{g}\textcolor[rgb]{0.96,0.65,0.14}{_{1}}$};
\draw (256.9,360.94) node [anchor=north west][inner sep=0.75pt]   [align=left] {$\displaystyle \textcolor[rgb]{0.96,0.65,0.14}{g}\textcolor[rgb]{0.96,0.65,0.14}{_{2}}$};
\draw (333.49,358.55) node [anchor=north west][inner sep=0.75pt]   [align=left] {$\displaystyle \textcolor[rgb]{0.29,0.56,0.89}{\psi }\textcolor[rgb]{0.29,0.56,0.89}{(}\textcolor[rgb]{0.29,0.56,0.89}{g}\textcolor[rgb]{0.29,0.56,0.89}{_{1}}\textcolor[rgb]{0.29,0.56,0.89}{,g}\textcolor[rgb]{0.29,0.56,0.89}{_{2}}\textcolor[rgb]{0.29,0.56,0.89}{)}$};
\draw (352.26,312.74) node [anchor=north west][inner sep=0.75pt]   [align=left] {$\displaystyle \ \mathcal{O}^{i}$};
\draw (342.8,216.31) node [anchor=north west][inner sep=0.75pt]   [align=left] {$\displaystyle \rho _{A}( \psi ( g_{1} ,g_{2}))$};

\end{tikzpicture}
    \caption{Unlinking the topological defect of $A^{(p)}$ through the defect $\mathcal{O}_i$ gives the action by $\rho_A$, denoted via the grey arrow.}
    \end{subfigure}
    \begin{subfigure}{0.79\textwidth}
    \centering
    \input{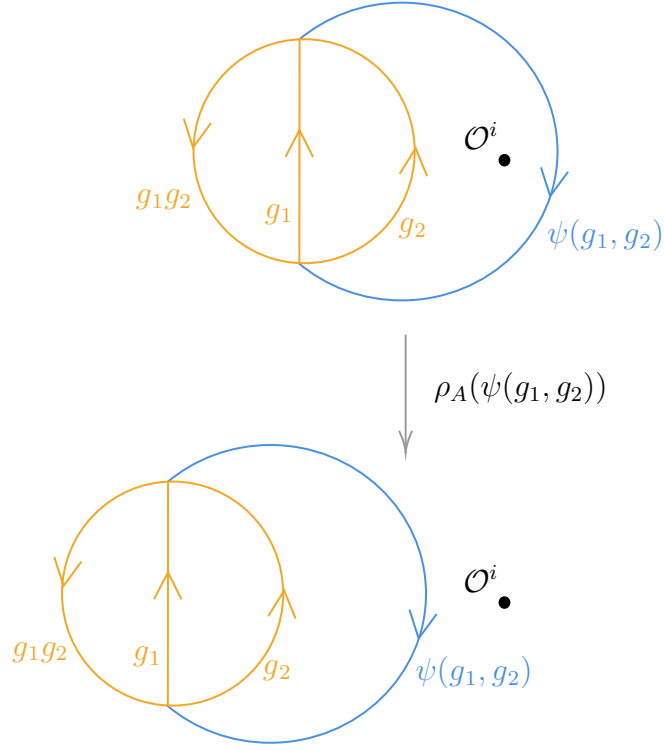}
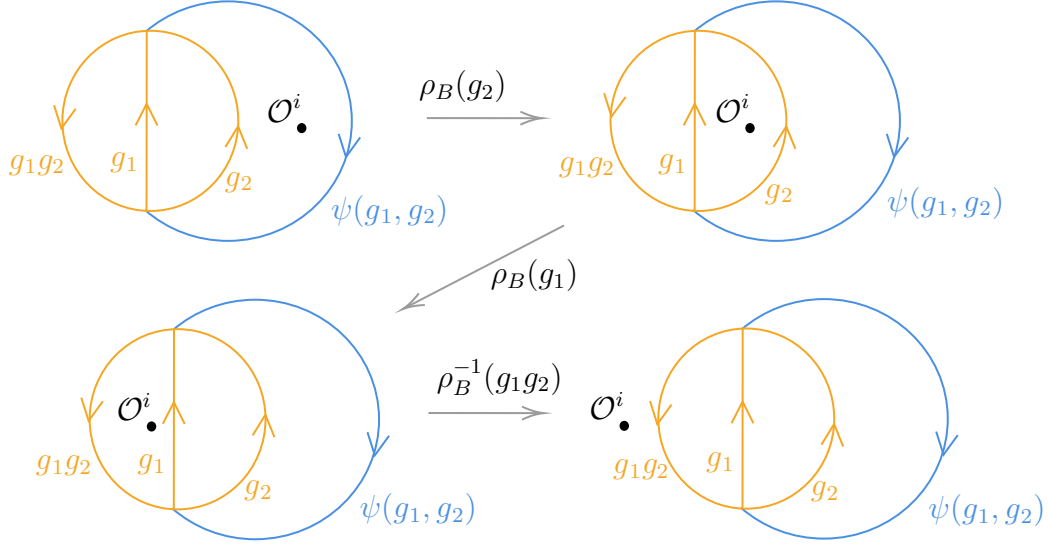
    \caption{Unlinking the topological defects of $G^{(p)}$ through the defect $\mathcal{O}_i$ gives the action by $\rho_G$, denoted via the grey arrow.}
    \end{subfigure}
    \caption{Configuration of topological defects in fig.~\ref{defects-ext} linking with $\mathcal{O}_i$ that leads to the constraint \eqref{eq:rGrGrArG}.}
    \label{fig:Extsweep}
\end{figure}
\hspace{-.15cm}on the action of the two symmetries, as follows from sweeping the operator through either one type of topological defect or the other, see figure \ref{fig:Extsweep}. Taking the extension \eqref{ses-p} to be central (meaning that $A$ sits inside the center of $\Gamma$),\footnote{This is always the case when $p>0$.} the matrices $\rho_A(a)$ commute with the matrices $\rho_G(g)$. 

Since $A$ is Abelian we can restrict to operators $\mathcal{O}'_i$ such that the action of $A$ is simply multiplication by a phase $\varphi:A\rightarrow U(1)$, namely  
\begin{equation}
    \rho'_G(g_1)\rho'_G(g_2)=\varphi(\psi(g_1,g_2))\rho'_G(g_1g_2) \ .
\end{equation}
It then follows that the matrices $\rho'_G$ define a projective representation of $G^{(p)}$ with projective phase determined by $[\varphi \circ \psi] \in H^2(BG,U(1))$. This implies that the corresponding states, prepared as in figure \ref{states_pform}, span a vector space supporting projective representations of $G^{(p)}$. The projective nature of this representation captures the nontriviality of the extension \eqref{ses-p}.

\subsection{Higher-group}
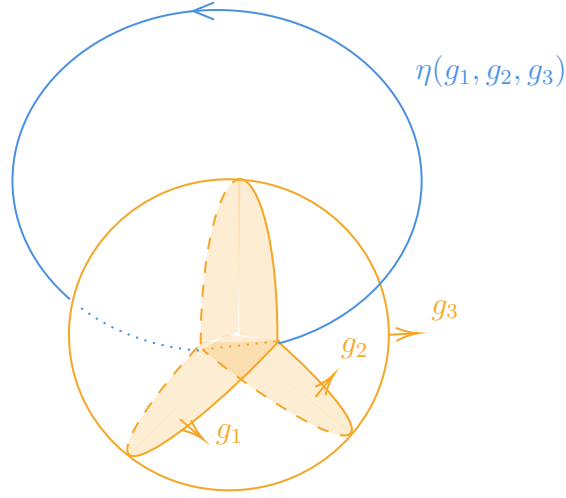
\begin{figure}[t]
    \centering
    \tikzset{every picture/.style={line width=0.75pt}} 

\begin{tikzpicture}[x=0.75pt,y=0.75pt,yscale=-1,xscale=1]

\draw  [color={rgb, 255:red, 245; green, 166; blue, 35 }  ,draw opacity=1 ] (104.11,195.04) .. controls (104.11,151.94) and (140.05,116.99) .. (184.38,116.99) .. controls (228.72,116.99) and (264.66,151.94) .. (264.66,195.04) .. controls (264.66,238.15) and (228.72,273.09) .. (184.38,273.09) .. controls (140.05,273.09) and (104.11,238.15) .. (104.11,195.04) -- cycle ;
\draw  [draw opacity=0][fill={rgb, 255:red, 245; green, 166; blue, 35 }  ,fill opacity=0.2 ][dash pattern={on 4.5pt off 4.5pt}] (170.22,199.91) .. controls (170.21,197.79) and (170.23,195.64) .. (170.26,193.48) .. controls (170.99,151.25) and (179.87,117.11) .. (190.14,116.92) -- (188.99,193.78) -- cycle ; \draw  [color={rgb, 255:red, 245; green, 166; blue, 35 }  ,draw opacity=1 ][dash pattern={on 4.5pt off 4.5pt}] (170.22,199.91) .. controls (170.21,197.79) and (170.23,195.64) .. (170.26,193.48) .. controls (170.99,151.25) and (179.87,117.11) .. (190.14,116.92) ;  
\draw  [draw opacity=0][fill={rgb, 255:red, 245; green, 166; blue, 35 }  ,fill opacity=0.2 ] (188.86,116.97) .. controls (189.12,116.92) and (189.38,116.9) .. (189.64,116.9) .. controls (200.18,116.9) and (208.73,151.87) .. (208.73,195) .. controls (208.73,196.29) and (208.72,197.58) .. (208.71,198.87) -- (189.64,195) -- cycle ; \draw  [color={rgb, 255:red, 245; green, 166; blue, 35 }  ,draw opacity=1 ] (188.86,116.97) .. controls (189.12,116.92) and (189.38,116.9) .. (189.64,116.9) .. controls (200.18,116.9) and (208.73,151.87) .. (208.73,195) .. controls (208.73,196.29) and (208.72,197.58) .. (208.71,198.87) ;  
\draw  [draw opacity=0][fill={rgb, 255:red, 245; green, 166; blue, 35 }  ,fill opacity=0.2 ] (209.88,197.74) .. controls (206.71,201.17) and (203.39,204.65) .. (199.93,208.15) .. controls (169.44,239.05) and (140.25,260.01) .. (134.4,255.26) -- (189.9,198.78) -- cycle ; \draw  [color={rgb, 255:red, 245; green, 166; blue, 35 }  ,draw opacity=1 ] (209.88,197.74) .. controls (206.71,201.17) and (203.39,204.65) .. (199.93,208.15) .. controls (169.44,239.05) and (140.25,260.01) .. (134.4,255.26) ;  
\draw  [draw opacity=0][fill={rgb, 255:red, 245; green, 166; blue, 35 }  ,fill opacity=0.2 ][dash pattern={on 4.5pt off 4.5pt}] (134.59,255.4) .. controls (134.48,255.32) and (134.37,255.24) .. (134.28,255.15) .. controls (129.27,250.48) and (145.42,226.05) .. (171.2,198.45) -- (189.9,198.78) -- cycle ; \draw  [color={rgb, 255:red, 245; green, 166; blue, 35 }  ,draw opacity=1 ][dash pattern={on 4.5pt off 4.5pt}] (134.59,255.4) .. controls (134.48,255.32) and (134.37,255.24) .. (134.28,255.15) .. controls (129.27,250.48) and (145.42,226.05) .. (171.2,198.45) ;  
\draw  [draw opacity=0][fill={rgb, 255:red, 245; green, 166; blue, 35 }  ,fill opacity=0.2 ] (208.64,198.15) .. controls (233.53,220.1) and (249.25,239.33) .. (245.57,244.34) -- (183.42,194.85) -- cycle ; \draw  [color={rgb, 255:red, 245; green, 166; blue, 35 }  ,draw opacity=1 ] (208.64,198.15) .. controls (233.53,220.1) and (249.25,239.33) .. (245.57,244.34) ;  
\draw  [draw opacity=0][fill={rgb, 255:red, 245; green, 166; blue, 35 }  ,fill opacity=0.2 ][dash pattern={on 4.5pt off 4.5pt}] (245.7,244.14) .. controls (245.64,244.26) and (245.56,244.37) .. (245.47,244.47) .. controls (240.59,250.24) and (208.86,232.7) .. (174.59,205.29) .. controls (173.43,204.37) and (172.29,203.45) .. (171.16,202.52) -- (183.42,194.85) -- cycle ; \draw  [color={rgb, 255:red, 245; green, 166; blue, 35 }  ,draw opacity=1 ][dash pattern={on 4.5pt off 4.5pt}] (245.7,244.14) .. controls (245.64,244.26) and (245.56,244.37) .. (245.47,244.47) .. controls (240.59,250.24) and (208.86,232.7) .. (174.59,205.29) .. controls (173.43,204.37) and (172.29,203.45) .. (171.16,202.52) ;  
\draw [color={rgb, 255:red, 245; green, 166; blue, 35 }  ,draw opacity=1 ] [dash pattern={on 0.84pt off 2.51pt}]  (171.14,202.51) -- (208.73,198.22) ;
\draw  [draw opacity=0] (104.65,177.22) .. controls (86.77,161.81) and (75.76,140.81) .. (75.76,117.68) .. controls (75.76,70.38) and (121.73,32.04) .. (178.44,32.04) .. controls (235.15,32.04) and (281.12,70.38) .. (281.12,117.68) .. controls (281.12,156.04) and (250.86,188.52) .. (209.16,199.41) -- (178.44,117.68) -- cycle ; \draw  [color={rgb, 255:red, 74; green, 144; blue, 226 }  ,draw opacity=1 ] (104.65,177.22) .. controls (86.77,161.81) and (75.76,140.81) .. (75.76,117.68) .. controls (75.76,70.38) and (121.73,32.04) .. (178.44,32.04) .. controls (235.15,32.04) and (281.12,70.38) .. (281.12,117.68) .. controls (281.12,156.04) and (250.86,188.52) .. (209.16,199.41) ;  
\draw  [draw opacity=0][dash pattern={on 0.84pt off 2.51pt}] (169.31,202.97) .. controls (146.2,201.28) and (125.29,193.19) .. (109.13,180.85) -- (178.44,117.68) -- cycle ; \draw  [color={rgb, 255:red, 74; green, 144; blue, 226 }  ,draw opacity=1 ][dash pattern={on 0.84pt off 2.51pt}] (169.31,202.97) .. controls (146.2,201.28) and (125.29,193.19) .. (109.13,180.85) ;  
\draw [color={rgb, 255:red, 245; green, 166; blue, 35 }  ,draw opacity=1 ]   (264.66,195.04) -- (277.26,194.35) ;
\draw [shift={(279.26,194.24)}, rotate = 176.86] [color={rgb, 255:red, 245; green, 166; blue, 35 }  ,draw opacity=1 ][line width=0.75]    (10.93,-3.29) .. controls (6.95,-1.4) and (3.31,-0.3) .. (0,0) .. controls (3.31,0.3) and (6.95,1.4) .. (10.93,3.29)   ;
\draw [color={rgb, 255:red, 245; green, 166; blue, 35 }  ,draw opacity=1 ]   (227.86,226.24) -- (235.94,216.29) ;
\draw [shift={(237.2,214.73)}, rotate = 129.07] [color={rgb, 255:red, 245; green, 166; blue, 35 }  ,draw opacity=1 ][line width=0.75]    (10.93,-3.29) .. controls (6.95,-1.4) and (3.31,-0.3) .. (0,0) .. controls (3.31,0.3) and (6.95,1.4) .. (10.93,3.29)   ;
\draw [color={rgb, 255:red, 245; green, 166; blue, 35 }  ,draw opacity=1 ]   (159.86,237.44) -- (170.08,246.23) ;
\draw [shift={(171.6,247.53)}, rotate = 220.66] [color={rgb, 255:red, 245; green, 166; blue, 35 }  ,draw opacity=1 ][line width=0.75]    (10.93,-3.29) .. controls (6.95,-1.4) and (3.31,-0.3) .. (0,0) .. controls (3.31,0.3) and (6.95,1.4) .. (10.93,3.29)   ;
\draw  [color={rgb, 255:red, 74; green, 144; blue, 226 }  ,draw opacity=1 ] (176.93,36.81) -- (166.4,32.53) -- (176.94,28.27) ;

\draw (276.5,53.33) node [anchor=north west][inner sep=0.75pt]  [color={rgb, 255:red, 74; green, 144; blue, 226 }  ,opacity=1 ] [align=left] {$\displaystyle \eta ( g_{1} ,g_{2} ,g_{3})$};
\draw (284.6,176) node [anchor=north west][inner sep=0.75pt]  [color={rgb, 255:red, 245; green, 166; blue, 35 }  ,opacity=1 ] [align=left] {$\displaystyle g_{3}$};
\draw (239,195.2) node [anchor=north west][inner sep=0.75pt]  [color={rgb, 255:red, 245; green, 166; blue, 35 }  ,opacity=1 ] [align=left] {$\displaystyle g_{2}$};
\draw (176.4,238.4) node [anchor=north west][inner sep=0.75pt]  [color={rgb, 255:red, 245; green, 166; blue, 35 }  ,opacity=1 ] [align=left] {$\displaystyle g_{1}$};

\end{tikzpicture}
    \caption{ The junction of topological defects in fig.~\ref{local-def-2grp}, associated with a higher-group, arranged in a configuration that can be used to wrap operators charged under $A^{(p+1)}$ and $G^{(p)}$.}
    \label{defects-2grp}
\end{figure}

We now move on to describe higher-groups and their representations, following eq.~\eqref{ses-higher}. We take, for definiteness, $q=p+1$, meaning that we have a $p$-form symmetry $G^{(p)}$ and a $(p+1)$-form symmetry $A^{(p+1)}$ that form a $(p+2)$-group defined by the exact sequence
\begin{equation}
    1\rightarrow A^{(p+1)}\rightarrow \underline{\Gamma} \rightarrow G^{(p)} \rightarrow 1 \ .
\end{equation}
As mentioned above, this extension is determined by a class $t_{p+3} \in H^{p+3}(B^{p+1}G,A)$, but we focus on the Postnikov class given by the image $\Phi^p[t_{p+3}]=\eta \in H^3(BG,A)$. At the level of topological defects, this implies that topological defects of the $A^{(p+1)}$ symmetry can end on a codimension-$(p+3)$ locus singled out by the intersection of four topological defects of $G^{(p)}$, as depicted in figure \ref{defects-2grp}.
We will argue that defect Hilbert spaces of the defects charged under $A^{(p+1)}$ support projective representations of $G^{(p)}$. 

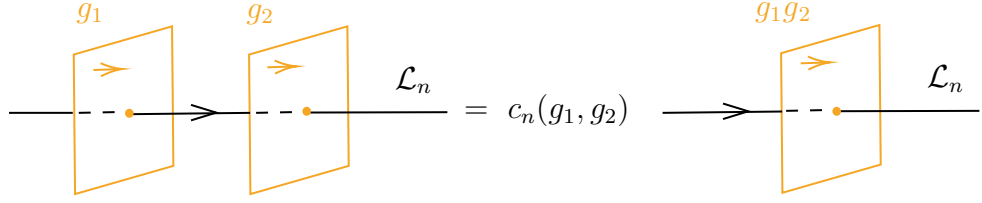
\begin{figure}
    \centering
   \tikzset{every picture/.style={line width=0.75pt}} 

\begin{tikzpicture}[x=0.75pt,y=0.75pt,yscale=-1,xscale=1]

\draw   (125.67,135.89) -- (138.33,139.78) -- (125.67,143.67) ;
\draw  [color={rgb, 255:red, 245; green, 166; blue, 35 }  ,draw opacity=1 ] (204.81,166.5) -- (155.47,180.65) -- (154.81,110.63) -- (204.15,96.48) -- cycle ;
\draw  [color={rgb, 255:red, 245; green, 166; blue, 35 }  ,draw opacity=1 ] (116.81,166.5) -- (67.47,180.65) -- (66.81,110.63) -- (116.15,96.48) -- cycle ;
\draw    (34.33,139.89) -- (66.33,139.89) ;
\draw    (95.33,140.56) -- (155.33,139.89) ;
\draw    (185.13,139.73) -- (254.33,139.56) ;
\draw   (392.67,135.22) -- (405.33,139.11) -- (392.67,143) ;
\draw  [color={rgb, 255:red, 245; green, 166; blue, 35 }  ,draw opacity=1 ] (471.81,165.84) -- (422.47,179.99) -- (421.81,109.96) -- (471.15,95.81) -- cycle ;
\draw    (362.33,139.89) -- (422.33,139.22) ;
\draw    (448.13,139.07) -- (521.33,138.89) ;
\draw [color={rgb, 255:red, 245; green, 166; blue, 35 }  ,draw opacity=1 ]   (76.66,118.38) -- (89.33,117.96) ;
\draw [shift={(91.33,117.89)}, rotate = 178.09] [color={rgb, 255:red, 245; green, 166; blue, 35 }  ,draw opacity=1 ][line width=0.75]    (10.93,-3.29) .. controls (6.95,-1.4) and (3.31,-0.3) .. (0,0) .. controls (3.31,0.3) and (6.95,1.4) .. (10.93,3.29)   ;
\draw [color={rgb, 255:red, 245; green, 166; blue, 35 }  ,draw opacity=1 ]   (431.66,115.04) -- (444.33,114.62) ;
\draw [shift={(446.33,114.56)}, rotate = 178.09] [color={rgb, 255:red, 245; green, 166; blue, 35 }  ,draw opacity=1 ][line width=0.75]    (10.93,-3.29) .. controls (6.95,-1.4) and (3.31,-0.3) .. (0,0) .. controls (3.31,0.3) and (6.95,1.4) .. (10.93,3.29)   ;
\draw [color={rgb, 255:red, 245; green, 166; blue, 35 }  ,draw opacity=1 ]   (163.99,117.04) -- (176.67,116.62) ;
\draw [shift={(178.67,116.56)}, rotate = 178.09] [color={rgb, 255:red, 245; green, 166; blue, 35 }  ,draw opacity=1 ][line width=0.75]    (10.93,-3.29) .. controls (6.95,-1.4) and (3.31,-0.3) .. (0,0) .. controls (3.31,0.3) and (6.95,1.4) .. (10.93,3.29)   ;
\draw  [dash pattern={on 4.5pt off 4.5pt}]  (69.33,139.89) -- (95.37,139.36) ;
\draw  [color={rgb, 255:red, 245; green, 166; blue, 35 }  ,draw opacity=1 ][fill={rgb, 255:red, 245; green, 166; blue, 35 }  ,fill opacity=1 ] (93.07,140.11) .. controls (93.07,139.15) and (93.81,138.36) .. (94.72,138.36) .. controls (95.63,138.36) and (96.37,139.15) .. (96.37,140.11) .. controls (96.37,141.08) and (95.63,141.86) .. (94.72,141.86) .. controls (93.81,141.86) and (93.07,141.08) .. (93.07,140.11) -- cycle ;
\draw  [dash pattern={on 4.5pt off 4.5pt}]  (158.33,139.89) -- (184.37,139.36) ;
\draw  [color={rgb, 255:red, 245; green, 166; blue, 35 }  ,draw opacity=1 ][fill={rgb, 255:red, 245; green, 166; blue, 35 }  ,fill opacity=1 ] (182.07,139.11) .. controls (182.07,138.15) and (182.81,137.36) .. (183.72,137.36) .. controls (184.63,137.36) and (185.37,138.15) .. (185.37,139.11) .. controls (185.37,140.08) and (184.63,140.86) .. (183.72,140.86) .. controls (182.81,140.86) and (182.07,140.08) .. (182.07,139.11) -- cycle ;
\draw  [dash pattern={on 4.5pt off 4.5pt}]  (424.33,138.89) -- (450.37,138.36) ;
\draw  [color={rgb, 255:red, 245; green, 166; blue, 35 }  ,draw opacity=1 ][fill={rgb, 255:red, 245; green, 166; blue, 35 }  ,fill opacity=1 ] (448.07,139.11) .. controls (448.07,138.15) and (448.81,137.36) .. (449.72,137.36) .. controls (450.63,137.36) and (451.37,138.15) .. (451.37,139.11) .. controls (451.37,140.08) and (450.63,140.86) .. (449.72,140.86) .. controls (448.81,140.86) and (448.07,140.08) .. (448.07,139.11) -- cycle ;

\draw (226.67,115.33) node [anchor=north west][inner sep=0.75pt]   [align=left] {$\displaystyle \mathcal{L}_{n}$};
\draw (66.67,83.33) node [anchor=north west][inner sep=0.75pt]   [align=left] {$\displaystyle \textcolor[rgb]{0.96,0.65,0.14}{g_{1}}$};
\draw (493.67,114.67) node [anchor=north west][inner sep=0.75pt]   [align=left] {$\displaystyle \mathcal{L}_{n}$};
\draw (407.33,83) node [anchor=north west][inner sep=0.75pt]   [align=left] {$\displaystyle \textcolor[rgb]{0.96,0.65,0.14}{g_{1} g}\textcolor[rgb]{0.96,0.65,0.14}{_{2}}$};
\draw (152,84) node [anchor=north west][inner sep=0.75pt]   [align=left] {$\displaystyle \textcolor[rgb]{0.96,0.65,0.14}{g_{2}}$};
\draw (261.33,129) node [anchor=north west][inner sep=0.75pt]   [align=left] {$\displaystyle =\ c_{n}( g_{1} ,g_{2})$};

\end{tikzpicture}
    \caption{Projective fusion of $G^{(p)}$ operators intersecting the defects $\mathcal{L}_n$, controlled by the phase $c_n$.}
    \label{fig:Gact}
\end{figure}

Let us consider a set of operators $\mathcal{L}_n(\gamma_{p+1})$ supporting a two-component linking action of $A^{(p+1)}$
\begin{figure}
    \centering
    \tikzset{every picture/.style={line width=0.75pt}} 

\begin{tikzpicture}[x=0.75pt,y=0.75pt,yscale=-1,xscale=1]

\draw    (60.13,120.73) -- (134.33,119.89) ;
\draw    (156.13,119.73) -- (230.33,119.89) ;
\draw [color={rgb, 255:red, 74; green, 144; blue, 226 }  ,draw opacity=1 ]   (144,59.22) -- (144.33,168.56) ;
\draw   (101.67,115.89) -- (114.33,119.78) -- (101.67,123.67) ;
\draw  [color={rgb, 255:red, 74; green, 144; blue, 226 }  ,draw opacity=1 ] (140.28,100.21) -- (143.83,87.45) -- (148.05,100.01) ;
\draw    (318.13,120.73) -- (488.33,119.89) ;
\draw   (373.67,116.89) -- (386.33,120.78) -- (373.67,124.67) ;
\draw [color={rgb, 255:red, 74; green, 144; blue, 226 }  ,draw opacity=1 ]   (406,59.22) -- (406.33,112.56) ;
\draw  [color={rgb, 255:red, 74; green, 144; blue, 226 }  ,draw opacity=1 ] (402.28,100.21) -- (405.83,87.45) -- (410.05,100.01) ;
\draw [color={rgb, 255:red, 74; green, 144; blue, 226 }  ,draw opacity=1 ]   (406,127.22) -- (406.33,167.56) ;

\draw (206.67,95.33) node [anchor=north west][inner sep=0.75pt]   [align=left] {$\displaystyle \mathcal{L}_{n}$};
\draw (130,64.67) node [anchor=north west][inner sep=0.75pt]   [align=left] {$\displaystyle \textcolor[rgb]{0.29,0.56,0.89}{a}$};
\draw (392,64.67) node [anchor=north west][inner sep=0.75pt]   [align=left] {$\displaystyle \textcolor[rgb]{0.29,0.56,0.89}{a}$};
\draw (462.67,95.33) node [anchor=north west][inner sep=0.75pt]   [align=left] {$\displaystyle \mathcal{L}_{n}$};
\draw (241.33,109) node [anchor=north west][inner sep=0.75pt]   [align=left] {$\displaystyle =\ \varphi _{n}( a)$};

\end{tikzpicture}
    \caption{Action of $A^{(p+1)}$ on the operators $\mathcal{L}_n$ via the phase $\varphi_n$.}
    \label{fig:Aact}
\end{figure}
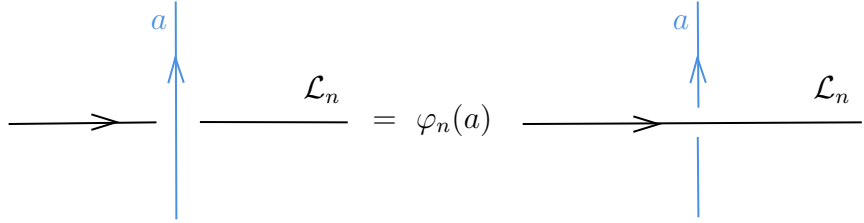
with charge $\varphi_n:A\rightarrow U(1)$, as shown in figure \ref{fig:Aact}. Moreover, we assume that the intersection between the defects $\mathcal{L}_n$ and the topological operators of $G^{(p)}$ is itself topological, i.e. the defects preserve the symmetry $G^{(p)}$. Then the fusion of the topological intersections produces a phase $c_n:G\times G \rightarrow U(1)$, as depicted in figure \ref{fig:Gact}.

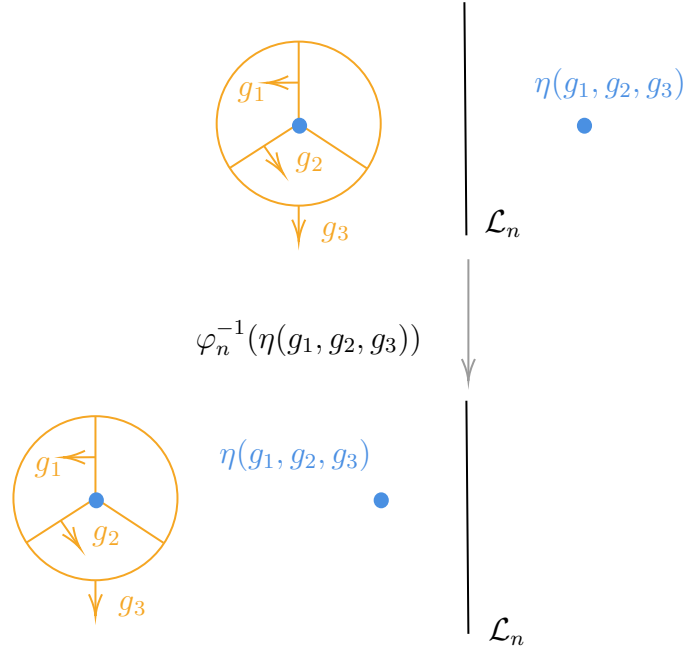
\begin{figure}
    \centering
    \tikzset{every picture/.style={line width=0.75pt}} 

\begin{tikzpicture}[x=0.75pt,y=0.75pt,yscale=-1,xscale=1]

\draw  [color={rgb, 255:red, 245; green, 166; blue, 35 }  ,draw opacity=1 ] (179,82.17) .. controls (179,59.43) and (197.43,41) .. (220.17,41) .. controls (242.9,41) and (261.33,59.43) .. (261.33,82.17) .. controls (261.33,104.9) and (242.9,123.33) .. (220.17,123.33) .. controls (197.43,123.33) and (179,104.9) .. (179,82.17) -- cycle ;
\draw    (303.33,21.24) -- (304.33,138.24) ;
\draw [color={rgb, 255:red, 245; green, 166; blue, 35 }  ,draw opacity=1 ]   (220.17,41) -- (220.17,82.17) ;
\draw [color={rgb, 255:red, 245; green, 166; blue, 35 }  ,draw opacity=1 ]   (220.17,82.17) -- (254.33,104.57) ;
\draw [color={rgb, 255:red, 245; green, 166; blue, 35 }  ,draw opacity=1 ]   (220.17,82.17) -- (185.33,104.57) ;
\draw  [color={rgb, 255:red, 74; green, 144; blue, 226 }  ,draw opacity=1 ][fill={rgb, 255:red, 74; green, 144; blue, 226 }  ,fill opacity=1 ] (360.29,83.22) .. controls (360.29,81.28) and (361.77,79.72) .. (363.59,79.72) .. controls (365.41,79.72) and (366.89,81.28) .. (366.89,83.22) .. controls (366.89,85.15) and (365.41,86.72) .. (363.59,86.72) .. controls (361.77,86.72) and (360.29,85.15) .. (360.29,83.22) -- cycle ;
\draw  [color={rgb, 255:red, 74; green, 144; blue, 226 }  ,draw opacity=1 ][fill={rgb, 255:red, 74; green, 144; blue, 226 }  ,fill opacity=1 ] (217.29,83.22) .. controls (217.29,81.28) and (218.77,79.72) .. (220.59,79.72) .. controls (222.41,79.72) and (223.89,81.28) .. (223.89,83.22) .. controls (223.89,85.15) and (222.41,86.72) .. (220.59,86.72) .. controls (218.77,86.72) and (217.29,85.15) .. (217.29,83.22) -- cycle ;
\draw [color={rgb, 255:red, 245; green, 166; blue, 35 }  ,draw opacity=1 ]   (220.17,123.33) -- (220.17,139.33) ;
\draw [shift={(220.17,141.33)}, rotate = 270] [color={rgb, 255:red, 245; green, 166; blue, 35 }  ,draw opacity=1 ][line width=0.75]    (10.93,-3.29) .. controls (6.95,-1.4) and (3.31,-0.3) .. (0,0) .. controls (3.31,0.3) and (6.95,1.4) .. (10.93,3.29)   ;
\draw [color={rgb, 255:red, 245; green, 166; blue, 35 }  ,draw opacity=1 ]   (202.75,93.37) -- (211.44,105.82) ;
\draw [shift={(212.58,107.46)}, rotate = 235.1] [color={rgb, 255:red, 245; green, 166; blue, 35 }  ,draw opacity=1 ][line width=0.75]    (10.93,-3.29) .. controls (6.95,-1.4) and (3.31,-0.3) .. (0,0) .. controls (3.31,0.3) and (6.95,1.4) .. (10.93,3.29)   ;
\draw [color={rgb, 255:red, 245; green, 166; blue, 35 }  ,draw opacity=1 ]   (220.17,61.58) -- (206,61.67) ;
\draw [shift={(204,61.68)}, rotate = 359.67] [color={rgb, 255:red, 245; green, 166; blue, 35 }  ,draw opacity=1 ][line width=0.75]    (10.93,-3.29) .. controls (6.95,-1.4) and (3.31,-0.3) .. (0,0) .. controls (3.31,0.3) and (6.95,1.4) .. (10.93,3.29)   ;
\draw  [color={rgb, 255:red, 245; green, 166; blue, 35 }  ,draw opacity=1 ] (77,270.17) .. controls (77,247.43) and (95.43,229) .. (118.17,229) .. controls (140.9,229) and (159.33,247.43) .. (159.33,270.17) .. controls (159.33,292.9) and (140.9,311.33) .. (118.17,311.33) .. controls (95.43,311.33) and (77,292.9) .. (77,270.17) -- cycle ;
\draw    (304.33,221.24) -- (305.33,338.24) ;
\draw [color={rgb, 255:red, 245; green, 166; blue, 35 }  ,draw opacity=1 ]   (118.17,229) -- (118.17,270.17) ;
\draw [color={rgb, 255:red, 245; green, 166; blue, 35 }  ,draw opacity=1 ]   (118.17,270.17) -- (152.33,292.57) ;
\draw [color={rgb, 255:red, 245; green, 166; blue, 35 }  ,draw opacity=1 ]   (118.17,270.17) -- (83.33,292.57) ;
\draw  [color={rgb, 255:red, 74; green, 144; blue, 226 }  ,draw opacity=1 ][fill={rgb, 255:red, 74; green, 144; blue, 226 }  ,fill opacity=1 ] (258.29,271.22) .. controls (258.29,269.28) and (259.77,267.72) .. (261.59,267.72) .. controls (263.41,267.72) and (264.89,269.28) .. (264.89,271.22) .. controls (264.89,273.15) and (263.41,274.72) .. (261.59,274.72) .. controls (259.77,274.72) and (258.29,273.15) .. (258.29,271.22) -- cycle ;
\draw  [color={rgb, 255:red, 74; green, 144; blue, 226 }  ,draw opacity=1 ][fill={rgb, 255:red, 74; green, 144; blue, 226 }  ,fill opacity=1 ] (115.29,271.22) .. controls (115.29,269.28) and (116.77,267.72) .. (118.59,267.72) .. controls (120.41,267.72) and (121.89,269.28) .. (121.89,271.22) .. controls (121.89,273.15) and (120.41,274.72) .. (118.59,274.72) .. controls (116.77,274.72) and (115.29,273.15) .. (115.29,271.22) -- cycle ;
\draw [color={rgb, 255:red, 245; green, 166; blue, 35 }  ,draw opacity=1 ]   (118.17,311.33) -- (118.17,327.33) ;
\draw [shift={(118.17,329.33)}, rotate = 270] [color={rgb, 255:red, 245; green, 166; blue, 35 }  ,draw opacity=1 ][line width=0.75]    (10.93,-3.29) .. controls (6.95,-1.4) and (3.31,-0.3) .. (0,0) .. controls (3.31,0.3) and (6.95,1.4) .. (10.93,3.29)   ;
\draw [color={rgb, 255:red, 245; green, 166; blue, 35 }  ,draw opacity=1 ]   (100.75,281.37) -- (109.44,293.82) ;
\draw [shift={(110.58,295.46)}, rotate = 235.1] [color={rgb, 255:red, 245; green, 166; blue, 35 }  ,draw opacity=1 ][line width=0.75]    (10.93,-3.29) .. controls (6.95,-1.4) and (3.31,-0.3) .. (0,0) .. controls (3.31,0.3) and (6.95,1.4) .. (10.93,3.29)   ;
\draw [color={rgb, 255:red, 245; green, 166; blue, 35 }  ,draw opacity=1 ]   (118.17,249.58) -- (104,249.67) ;
\draw [shift={(102,249.68)}, rotate = 359.67] [color={rgb, 255:red, 245; green, 166; blue, 35 }  ,draw opacity=1 ][line width=0.75]    (10.93,-3.29) .. controls (6.95,-1.4) and (3.31,-0.3) .. (0,0) .. controls (3.31,0.3) and (6.95,1.4) .. (10.93,3.29)   ;
\draw [color={rgb, 255:red, 155; green, 155; blue, 155 }  ,draw opacity=1 ]   (305.33,150.24) -- (305.33,209.24) ;
\draw [shift={(305.33,211.24)}, rotate = 270] [color={rgb, 255:red, 155; green, 155; blue, 155 }  ,draw opacity=1 ][line width=0.75]    (10.93,-3.29) .. controls (6.95,-1.4) and (3.31,-0.3) .. (0,0) .. controls (3.31,0.3) and (6.95,1.4) .. (10.93,3.29)   ;

\draw (230,130) node [anchor=north west][inner sep=0.75pt]   [align=left] {$\displaystyle \textcolor[rgb]{0.96,0.65,0.14}{g}\textcolor[rgb]{0.96,0.65,0.14}{_{3}}$};
\draw (217,95) node [anchor=north west][inner sep=0.75pt]   [align=left] {$\displaystyle \textcolor[rgb]{0.96,0.65,0.14}{g}\textcolor[rgb]{0.96,0.65,0.14}{_{2}}$};
\draw (188,60) node [anchor=north west][inner sep=0.75pt]   [align=left] {$\displaystyle \textcolor[rgb]{0.96,0.65,0.14}{g}\textcolor[rgb]{0.96,0.65,0.14}{_{1}}$};
\draw (337,53) node [anchor=north west][inner sep=0.75pt]   [align=left] {\textcolor[rgb]{0.29,0.56,0.89}{$\displaystyle \eta ( g_{1} ,g_{2} ,g_{3})$}};
\draw (128,318) node [anchor=north west][inner sep=0.75pt]   [align=left] {$\displaystyle \textcolor[rgb]{0.96,0.65,0.14}{g}\textcolor[rgb]{0.96,0.65,0.14}{_{3}}$};
\draw (115,283) node [anchor=north west][inner sep=0.75pt]   [align=left] {$\displaystyle \textcolor[rgb]{0.96,0.65,0.14}{g}\textcolor[rgb]{0.96,0.65,0.14}{_{2}}$};
\draw (86,248) node [anchor=north west][inner sep=0.75pt]   [align=left] {$\displaystyle \textcolor[rgb]{0.96,0.65,0.14}{g}\textcolor[rgb]{0.96,0.65,0.14}{_{1}}$};
\draw (179,239) node [anchor=north west][inner sep=0.75pt]   [align=left] {\textcolor[rgb]{0.29,0.56,0.89}{$\displaystyle \eta ( g_{1} ,g_{2} ,g_{3})$}};
\draw (167,179.57) node [anchor=north west][inner sep=0.75pt]   [align=left] {$\displaystyle \varphi _{n}^{-1}( \eta ( g_{1} ,g_{2} ,g_{3}))$};
\draw (312,124) node [anchor=north west][inner sep=0.75pt]   [align=left] {$\displaystyle \mathcal{L}_{n}$};
\draw (314,329) node [anchor=north west][inner sep=0.75pt]   [align=left] {$\displaystyle \mathcal{L}_{n}$};

\end{tikzpicture}
    \caption{Two-dimensional section of the operator $\mathcal{L}_n$ linking with the topological-defect configuration in fig.~\ref{defects-2grp}. The grey arrows describe the process of unlinking through the topological defect of $A^{(p+1)}$.}
    \label{fig:Aunlink}
\end{figure}

\begin{figure}
    \centering
    \input{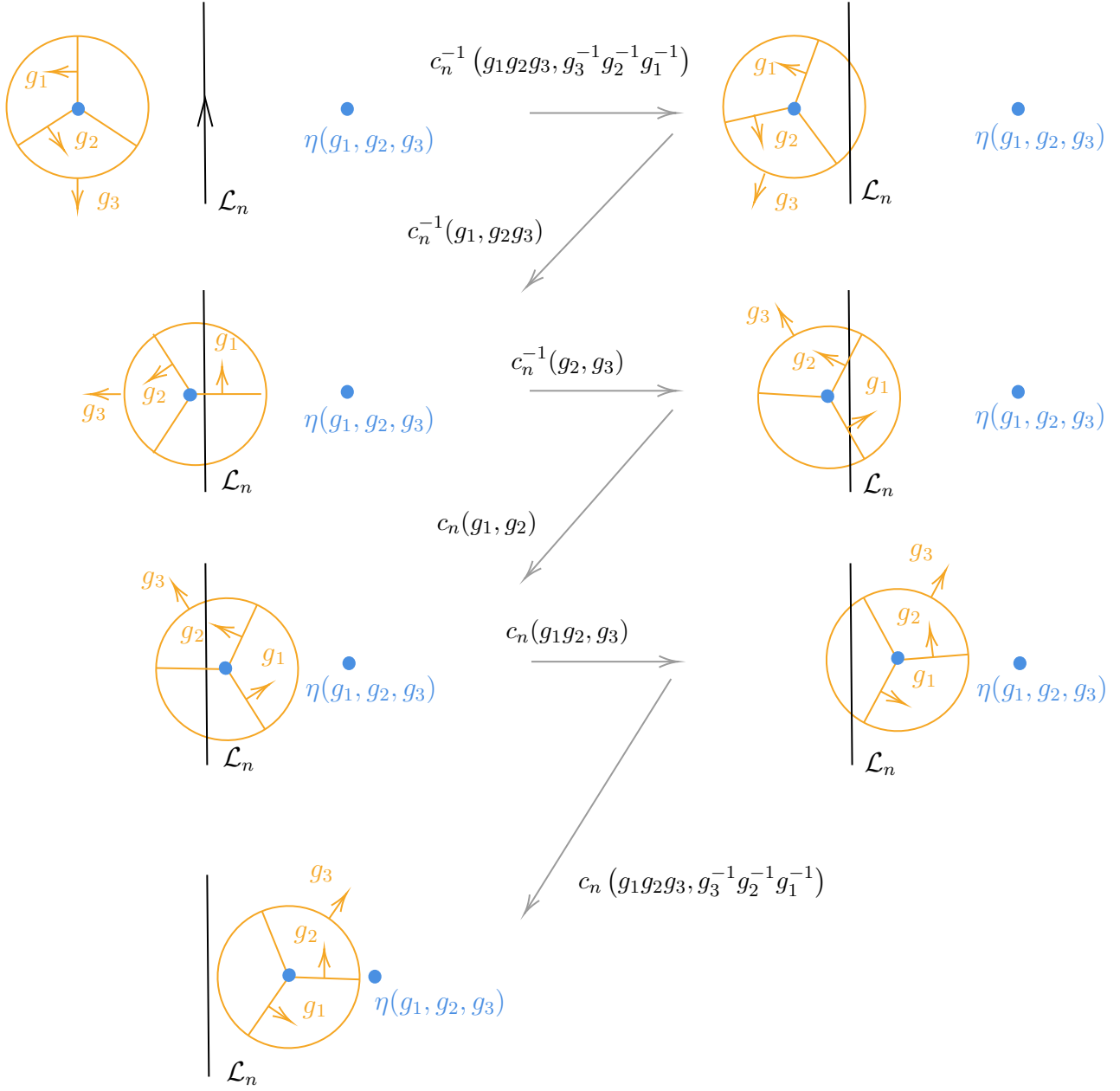}
    \caption{Two-dimensional section of the operator $\mathcal{L}_n$ linking with the topological-defect configuration in fig.~\ref{defects-2grp}. The grey arrows describe the process of unlinking through the topological defects of $G^{(p)}$.}
    \label{fig:Gunlink}
\end{figure}

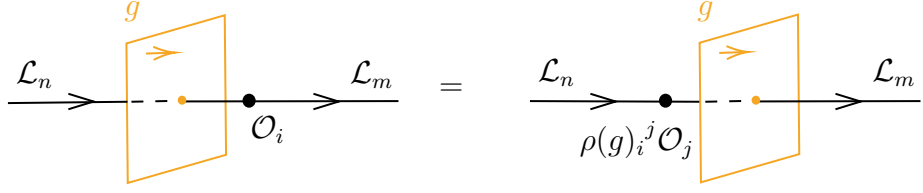
\begin{figure}
    \centering
    \tikzset{every picture/.style={line width=0.75pt}} 

\begin{tikzpicture}[x=0.75pt,y=0.75pt,yscale=-1,xscale=1]

\draw   (80.67,95.89) -- (93.33,99.78) -- (80.67,103.67) ;
\draw  [color={rgb, 255:red, 245; green, 166; blue, 35 }  ,draw opacity=1 ] (159.81,126.5) -- (110.47,140.65) -- (109.81,70.63) -- (159.15,56.48) -- cycle ;
\draw    (50.33,100.56) -- (110.33,99.89) ;
\draw    (139.13,99.73) -- (246.33,99.56) ;
\draw [color={rgb, 255:red, 245; green, 166; blue, 35 }  ,draw opacity=1 ]   (119.66,75.71) -- (132.33,75.29) ;
\draw [shift={(134.33,75.22)}, rotate = 178.09] [color={rgb, 255:red, 245; green, 166; blue, 35 }  ,draw opacity=1 ][line width=0.75]    (10.93,-3.29) .. controls (6.95,-1.4) and (3.31,-0.3) .. (0,0) .. controls (3.31,0.3) and (6.95,1.4) .. (10.93,3.29)   ;
\draw   (203.67,95.89) -- (216.33,99.78) -- (203.67,103.67) ;
\draw  [draw opacity=0][fill={rgb, 255:red, 0; green, 0; blue, 0 }  ,fill opacity=1 ] (168.07,99.36) .. controls (168.07,97.43) and (169.54,95.86) .. (171.37,95.86) .. controls (173.19,95.86) and (174.67,97.43) .. (174.67,99.36) .. controls (174.67,101.3) and (173.19,102.86) .. (171.37,102.86) .. controls (169.54,102.86) and (168.07,101.3) .. (168.07,99.36) -- cycle ;
\draw   (342.67,95.89) -- (355.33,99.78) -- (342.67,103.67) ;
\draw  [color={rgb, 255:red, 245; green, 166; blue, 35 }  ,draw opacity=1 ] (447.48,126.17) -- (398.14,140.32) -- (397.48,70.29) -- (446.82,56.14) -- cycle ;
\draw    (312.33,99.56) -- (397.33,99.89) ;
\draw    (427.13,99.73) -- (508.33,99.56) ;
\draw [color={rgb, 255:red, 245; green, 166; blue, 35 }  ,draw opacity=1 ]   (407.32,75.38) -- (420,74.96) ;
\draw [shift={(422,74.89)}, rotate = 178.09] [color={rgb, 255:red, 245; green, 166; blue, 35 }  ,draw opacity=1 ][line width=0.75]    (10.93,-3.29) .. controls (6.95,-1.4) and (3.31,-0.3) .. (0,0) .. controls (3.31,0.3) and (6.95,1.4) .. (10.93,3.29)   ;
\draw   (465.67,95.89) -- (478.33,99.78) -- (465.67,103.67) ;
\draw  [draw opacity=0][fill={rgb, 255:red, 0; green, 0; blue, 0 }  ,fill opacity=1 ] (377.07,99.36) .. controls (377.07,97.43) and (378.54,95.86) .. (380.37,95.86) .. controls (382.19,95.86) and (383.67,97.43) .. (383.67,99.36) .. controls (383.67,101.3) and (382.19,102.86) .. (380.37,102.86) .. controls (378.54,102.86) and (377.07,101.3) .. (377.07,99.36) -- cycle ;
\draw  [dash pattern={on 4.5pt off 4.5pt}]  (112.33,99.89) -- (138.37,99.36) ;
\draw  [color={rgb, 255:red, 245; green, 166; blue, 35 }  ,draw opacity=1 ][fill={rgb, 255:red, 245; green, 166; blue, 35 }  ,fill opacity=1 ] (136.07,99.11) .. controls (136.07,98.15) and (136.81,97.36) .. (137.72,97.36) .. controls (138.63,97.36) and (139.37,98.15) .. (139.37,99.11) .. controls (139.37,100.08) and (138.63,100.86) .. (137.72,100.86) .. controls (136.81,100.86) and (136.07,100.08) .. (136.07,99.11) -- cycle ;
\draw  [dash pattern={on 4.5pt off 4.5pt}]  (400.33,99.89) -- (426.37,99.36) ;
\draw  [color={rgb, 255:red, 245; green, 166; blue, 35 }  ,draw opacity=1 ][fill={rgb, 255:red, 245; green, 166; blue, 35 }  ,fill opacity=1 ] (424.07,99.11) .. controls (424.07,98.15) and (424.81,97.36) .. (425.72,97.36) .. controls (426.63,97.36) and (427.37,98.15) .. (427.37,99.11) .. controls (427.37,100.08) and (426.63,100.86) .. (425.72,100.86) .. controls (424.81,100.86) and (424.07,100.08) .. (424.07,99.11) -- cycle ;

\draw (53,78) node [anchor=north west][inner sep=0.75pt]   [align=left] {$\displaystyle \mathcal{L}_{n}$};
\draw (107.33,48) node [anchor=north west][inner sep=0.75pt]   [align=left] {$\displaystyle \textcolor[rgb]{0.96,0.65,0.14}{g}$};
\draw (220.67,78) node [anchor=north west][inner sep=0.75pt]   [align=left] {$\displaystyle \mathcal{L}_{m}$};
\draw (165.33,105.53) node [anchor=north west][inner sep=0.75pt]   [align=left] {$\displaystyle \ \mathcal{O}_{i}$};
\draw (315,77) node [anchor=north west][inner sep=0.75pt]   [align=left] {$\displaystyle \mathcal{L}_{n}$};
\draw (395,48) node [anchor=north west][inner sep=0.75pt]   [align=left] {$\displaystyle \textcolor[rgb]{0.96,0.65,0.14}{g}$};
\draw (482.67,77) node [anchor=north west][inner sep=0.75pt]   [align=left] {$\displaystyle \mathcal{L}_{m}$};
\draw (330.67,107.86) node [anchor=north west][inner sep=0.75pt]   [align=left] {$\displaystyle \ {\rho ( g)_{i}}^{j}\mathcal{O}_{j}$};
\draw (265.33,89) node [anchor=north west][inner sep=0.75pt]   [align=left] {$\displaystyle =$};

\end{tikzpicture}
    \caption{Action of $G^{(p)}$ on the junction operators $\mathcal{O}_i$ via the matrix $\rho$.}
    \label{fig:Gprojact}
\end{figure}

Now consider an operator $\mathcal{L}_n(\gamma_{p+1})$ linking with the topological-defect configuration shown in figure \ref{defects-2grp}. This configuration can be unlinked either through the topological defect of $A^{(p+1)}$ or through the topological defects of $G^{(p)}$. The first possibility produces a phase $\varphi_n^{-1}( \eta((g_1,g_2,g_3))$ as shown in figure \ref{fig:Aunlink}, while the second possibility produces a phase $(\delta c_n)^{-1}(g_1,g_2,g_3)$,\footnote{Recall that the definition of the differential in group cohomology is $(\delta c_n)(g_1,g_2,g_3)\equiv \frac{c_n(g_2,g_3)c_n(g_1,g_2g_3)}{c_n(g_1g_2,g_3)c_n(g_1,g_2)}$.} as shown in figure \ref{fig:Gunlink}. Hence, we derive the constraint
\begin{equation}\label{eq:2grpcharges}
 (\delta c_n)(g_1,g_2,g_3)=(\varphi_n \circ \eta)(g_1,g_2,g_3) \ .  
\end{equation}
Next we consider operators $\mathcal{O}_i(\gamma_p)$ living at junctions between two defects $\mathcal{L}_n(\gamma_{p+1})$ and $\mathcal{L}_m(\gamma'_{p+1})$, where $\gamma_p=\gamma'_{p+1}-\gamma_{p+1}$. 
\hspace{-.15cm}Consistency with the action of $A^{(p+1)}$ implies that $\varphi_n=\varphi_m$. The action of the intersection of topological operators of $G^{(p)}$ on the junctions is described by a unitary matrix $\rho$, see figure \ref{fig:Gprojact}.
Considering a pair of defects $g_1,g_2 \in G$, we can first fuse them in a single defect $g_1g_2$, and then act on $\mathcal{O}_i$,
or first act with each of them separately on $\mathcal{O}_i$, and only then fuse them in a single defect $g_1g_2$, see figure \ref{fig:Gprojunlink}.
\begin{figure}
  \centering
    \begin{subfigure}{0.79\textwidth}        \centering
    \tikzset{every picture/.style={line width=0.75pt}} 

\begin{tikzpicture}[x=0.75pt,y=0.75pt,yscale=-1,xscale=1]

\draw  [color={rgb, 255:red, 245; green, 166; blue, 35 }  ,draw opacity=1 ] (101.81,147.5) -- (52.47,161.65) -- (51.81,91.63) -- (101.15,77.48) -- cycle ;
\draw   (33.67,116.89) -- (46.33,120.78) -- (33.67,124.67) ;
\draw  [color={rgb, 255:red, 245; green, 166; blue, 35 }  ,draw opacity=1 ] (160.81,146.5) -- (111.47,160.65) -- (110.81,90.63) -- (160.15,76.48) -- cycle ;
\draw    (140.13,119.73) -- (222.33,119.56) ;
\draw [color={rgb, 255:red, 245; green, 166; blue, 35 }  ,draw opacity=1 ]   (120.66,95.71) -- (133.33,95.29) ;
\draw [shift={(135.33,95.22)}, rotate = 178.09] [color={rgb, 255:red, 245; green, 166; blue, 35 }  ,draw opacity=1 ][line width=0.75]    (10.93,-3.29) .. controls (6.95,-1.4) and (3.31,-0.3) .. (0,0) .. controls (3.31,0.3) and (6.95,1.4) .. (10.93,3.29)   ;
\draw   (182.67,115.89) -- (195.33,119.78) -- (182.67,123.67) ;
\draw  [draw opacity=0][fill={rgb, 255:red, 0; green, 0; blue, 0 }  ,fill opacity=1 ] (169.07,119.36) .. controls (169.07,117.43) and (170.54,115.86) .. (172.37,115.86) .. controls (174.19,115.86) and (175.67,117.43) .. (175.67,119.36) .. controls (175.67,121.3) and (174.19,122.86) .. (172.37,122.86) .. controls (170.54,122.86) and (169.07,121.3) .. (169.07,119.36) -- cycle ;
\draw   (349.67,218.89) -- (362.33,222.78) -- (349.67,226.67) ;
\draw  [color={rgb, 255:red, 245; green, 166; blue, 35 }  ,draw opacity=1 ] (440.48,249.17) -- (391.14,263.32) -- (390.48,193.29) -- (439.82,179.14) -- cycle ;
\draw    (323.33,222.56) -- (390.33,222.89) ;
\draw    (420.13,222.73) -- (471.33,222.56) ;
\draw [color={rgb, 255:red, 245; green, 166; blue, 35 }  ,draw opacity=1 ]   (400.32,198.38) -- (413,197.96) ;
\draw [shift={(415,197.89)}, rotate = 178.09] [color={rgb, 255:red, 245; green, 166; blue, 35 }  ,draw opacity=1 ][line width=0.75]    (10.93,-3.29) .. controls (6.95,-1.4) and (3.31,-0.3) .. (0,0) .. controls (3.31,0.3) and (6.95,1.4) .. (10.93,3.29)   ;
\draw   (443.67,218.89) -- (456.33,222.78) -- (443.67,226.67) ;
\draw  [draw opacity=0][fill={rgb, 255:red, 0; green, 0; blue, 0 }  ,fill opacity=1 ] (370.07,222.36) .. controls (370.07,220.43) and (371.54,218.86) .. (373.37,218.86) .. controls (375.19,218.86) and (376.67,220.43) .. (376.67,222.36) .. controls (376.67,224.3) and (375.19,225.86) .. (373.37,225.86) .. controls (371.54,225.86) and (370.07,224.3) .. (370.07,222.36) -- cycle ;
\draw  [dash pattern={on 4.5pt off 4.5pt}]  (113.33,119.89) -- (139.37,119.36) ;
\draw  [color={rgb, 255:red, 245; green, 166; blue, 35 }  ,draw opacity=1 ][fill={rgb, 255:red, 245; green, 166; blue, 35 }  ,fill opacity=1 ] (136.83,119.73) .. controls (136.83,118.77) and (137.57,117.98) .. (138.48,117.98) .. controls (139.39,117.98) and (140.13,118.77) .. (140.13,119.73) .. controls (140.13,120.7) and (139.39,121.48) .. (138.48,121.48) .. controls (137.57,121.48) and (136.83,120.7) .. (136.83,119.73) -- cycle ;
\draw  [dash pattern={on 4.5pt off 4.5pt}]  (393.33,222.89) -- (419.37,222.36) ;
\draw  [color={rgb, 255:red, 245; green, 166; blue, 35 }  ,draw opacity=1 ][fill={rgb, 255:red, 245; green, 166; blue, 35 }  ,fill opacity=1 ] (417.07,222.11) .. controls (417.07,221.15) and (417.81,220.36) .. (418.72,220.36) .. controls (419.63,220.36) and (420.37,221.15) .. (420.37,222.11) .. controls (420.37,223.08) and (419.63,223.86) .. (418.72,223.86) .. controls (417.81,223.86) and (417.07,223.08) .. (417.07,222.11) -- cycle ;
\draw [color={rgb, 255:red, 245; green, 166; blue, 35 }  ,draw opacity=1 ]   (61.66,96.71) -- (74.33,96.29) ;
\draw [shift={(76.33,96.22)}, rotate = 178.09] [color={rgb, 255:red, 245; green, 166; blue, 35 }  ,draw opacity=1 ][line width=0.75]    (10.93,-3.29) .. controls (6.95,-1.4) and (3.31,-0.3) .. (0,0) .. controls (3.31,0.3) and (6.95,1.4) .. (10.93,3.29)   ;
\draw  [dash pattern={on 4.5pt off 4.5pt}]  (57.33,120.89) -- (80.37,120.36) ;
\draw    (80.37,120.36) -- (111.33,119.89) ;
\draw    (51.33,120.89) -- (14.33,120.91) ;
\draw   (341.67,115.89) -- (354.33,119.78) -- (341.67,123.67) ;
\draw  [color={rgb, 255:red, 245; green, 166; blue, 35 }  ,draw opacity=1 ] (410.81,146.5) -- (361.47,160.65) -- (360.81,90.63) -- (410.15,76.48) -- cycle ;
\draw    (325.33,120.56) -- (361.33,119.89) ;
\draw    (390.13,119.73) -- (465.33,119.56) ;
\draw [color={rgb, 255:red, 245; green, 166; blue, 35 }  ,draw opacity=1 ]   (370.66,95.71) -- (383.33,95.29) ;
\draw [shift={(385.33,95.22)}, rotate = 178.09] [color={rgb, 255:red, 245; green, 166; blue, 35 }  ,draw opacity=1 ][line width=0.75]    (10.93,-3.29) .. controls (6.95,-1.4) and (3.31,-0.3) .. (0,0) .. controls (3.31,0.3) and (6.95,1.4) .. (10.93,3.29)   ;
\draw   (429.67,115.89) -- (442.33,119.78) -- (429.67,123.67) ;
\draw  [draw opacity=0][fill={rgb, 255:red, 0; green, 0; blue, 0 }  ,fill opacity=1 ] (419.07,119.36) .. controls (419.07,117.43) and (420.54,115.86) .. (422.37,115.86) .. controls (424.19,115.86) and (425.67,117.43) .. (425.67,119.36) .. controls (425.67,121.3) and (424.19,122.86) .. (422.37,122.86) .. controls (420.54,122.86) and (419.07,121.3) .. (419.07,119.36) -- cycle ;
\draw  [dash pattern={on 4.5pt off 4.5pt}]  (363.33,119.89) -- (389.37,119.36) ;
\draw  [color={rgb, 255:red, 245; green, 166; blue, 35 }  ,draw opacity=1 ][fill={rgb, 255:red, 245; green, 166; blue, 35 }  ,fill opacity=1 ] (387.07,119.11) .. controls (387.07,118.15) and (387.81,117.36) .. (388.72,117.36) .. controls (389.63,117.36) and (390.37,118.15) .. (390.37,119.11) .. controls (390.37,120.08) and (389.63,120.86) .. (388.72,120.86) .. controls (387.81,120.86) and (387.07,120.08) .. (387.07,119.11) -- cycle ;
\draw  [color={rgb, 255:red, 245; green, 166; blue, 35 }  ,draw opacity=1 ][fill={rgb, 255:red, 245; green, 166; blue, 35 }  ,fill opacity=1 ] (79.83,119.73) .. controls (79.83,118.77) and (80.57,117.98) .. (81.48,117.98) .. controls (82.39,117.98) and (83.13,118.77) .. (83.13,119.73) .. controls (83.13,120.7) and (82.39,121.48) .. (81.48,121.48) .. controls (80.57,121.48) and (79.83,120.7) .. (79.83,119.73) -- cycle ;

\draw (8,98) node [anchor=north west][inner sep=0.75pt]   [align=left] {$\displaystyle \mathcal{L}_{n}$};
\draw (108.33,63.67) node [anchor=north west][inner sep=0.75pt]   [align=left] {$\displaystyle \textcolor[rgb]{0.96,0.65,0.14}{g_{2}}$};
\draw (195.67,97) node [anchor=north west][inner sep=0.75pt]   [align=left] {$\displaystyle \mathcal{L}_{m}$};
\draw (166.33,128.53) node [anchor=north west][inner sep=0.75pt]   [align=left] {$\displaystyle \ \mathcal{O}_{i}$};
\draw (322,200) node [anchor=north west][inner sep=0.75pt]   [align=left] {$\displaystyle \mathcal{L}_{n}$};
\draw (370,167.33) node [anchor=north west][inner sep=0.75pt]   [align=left] {$\displaystyle \textcolor[rgb]{0.96,0.65,0.14}{g_{1} g_{2}}$};
\draw (452.67,200) node [anchor=north west][inner sep=0.75pt]   [align=left] {$\displaystyle \mathcal{L}_{m}$};
\draw (304.67,234.86) node [anchor=north west][inner sep=0.75pt]   [align=left] {$\displaystyle \ {\rho ( g_{1} g_{2})_{i}}^{j}\mathcal{O}_{j}$};
\draw (222.33,106) node [anchor=north west][inner sep=0.75pt]   [align=left] {$\displaystyle =\ c_{n}( g_{1} ,g_{2})$};
\draw (49.33,64.67) node [anchor=north west][inner sep=0.75pt]   [align=left] {$\displaystyle \textcolor[rgb]{0.96,0.65,0.14}{g_{1}}$};
\draw (319,98) node [anchor=north west][inner sep=0.75pt]   [align=left] {$\displaystyle \mathcal{L}_{n}$};
\draw (346.33,63.67) node [anchor=north west][inner sep=0.75pt]   [align=left] {$\displaystyle \textcolor[rgb]{0.96,0.65,0.14}{g_{1} g_{2}}$};
\draw (445.67,98) node [anchor=north west][inner sep=0.75pt]   [align=left] {$\displaystyle \mathcal{L}_{m}$};
\draw (416.33,127.53) node [anchor=north west][inner sep=0.75pt]   [align=left] {$\displaystyle \ \mathcal{O}_{i}$};
\draw (222.33,208) node [anchor=north west][inner sep=0.75pt]   [align=left] {$\displaystyle =\ c_{n}( g_{1} ,g_{2})$};

\end{tikzpicture}
    \caption{First fusing $G^{(p)}$ defects and then acting on the junction $\mathcal{O}_i$.}
    \end{subfigure}
    \medskip
    \begin{subfigure}{0.79\textwidth}
        \centering
        \input{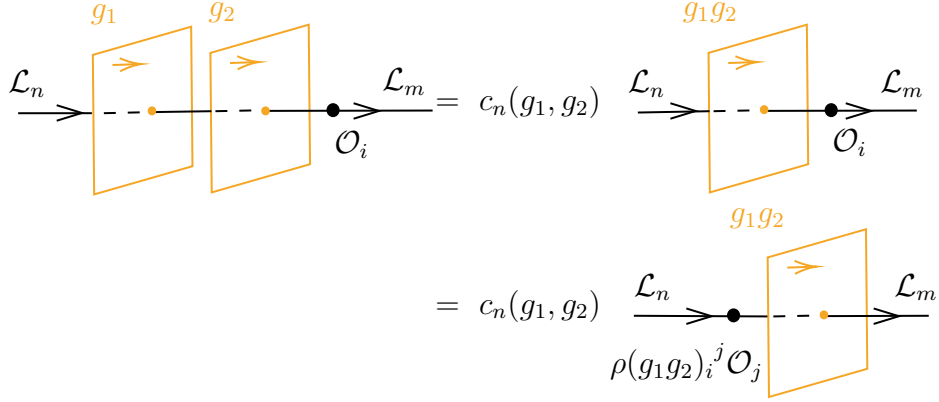}
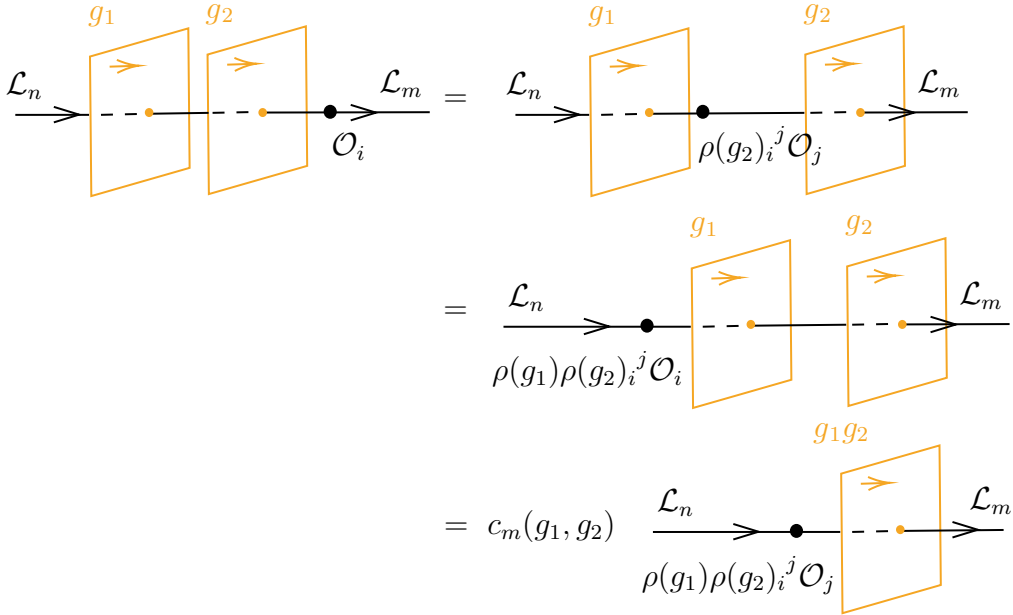
    \caption{First acting on the junction $\mathcal{O}_i$ and then fusing $G^{(p)}$ defects.}
    \end{subfigure}
    \caption{Manipulations that lead to the constraint \eqref{projGact}.}
    \label{fig:Gprojunlink}
\end{figure}
The compatibility of these two equivalent actions imposes the constraint
\begin{equation}\label{projGact}
    \rho(g_1)\rho(g_2)=\frac{c_n(g_1,g_2)}{c_m(g_1,g_2)}\rho(g_1g_2)\ .
\end{equation}
Using equation \eqref{eq:2grpcharges} and that $\varphi_n=\varphi_m$, we recognize that $(c_n/c_m)(g_1,g_2)$ is a closed $U(1)$-valued group 2-cocycle. This makes $\rho$ a projective representation. 

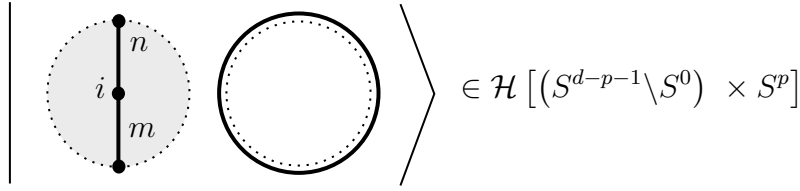
\begin{figure}[ht]
    \centering
    \tikzset{every picture/.style={line width=0.75pt}} 

\begin{tikzpicture}[x=0.75pt,y=0.75pt,yscale=-1,xscale=1]

\draw  [color={rgb, 255:red, 0; green, 0; blue, 0 }  ,draw opacity=1 ][dash pattern={on 0.84pt off 2.51pt}] (187,90.22) .. controls (187,70.24) and (203.19,54.05) .. (223.17,54.05) .. controls (243.14,54.05) and (259.33,70.24) .. (259.33,90.22) .. controls (259.33,110.19) and (243.14,126.38) .. (223.17,126.38) .. controls (203.19,126.38) and (187,110.19) .. (187,90.22) -- cycle ;
\draw  [fill={rgb, 255:red, 155; green, 155; blue, 155 }  ,fill opacity=0.2 ][dash pattern={on 0.84pt off 2.51pt}] (97.13,90.22) .. controls (97.13,69.87) and (113.09,53.38) .. (132.79,53.38) .. controls (152.49,53.38) and (168.46,69.87) .. (168.46,90.22) .. controls (168.46,110.56) and (152.49,127.05) .. (132.79,127.05) .. controls (113.09,127.05) and (97.13,110.56) .. (97.13,90.22) -- cycle ;
\draw    (78.33,44.38) -- (78.33,135.38) ;
\draw    (274.33,45.38) -- (291.33,90.38) -- (274.33,136.38) ;
\draw  [draw opacity=0][fill={rgb, 255:red, 0; green, 0; blue, 0 }  ,fill opacity=1 ] (129.49,90.22) .. controls (129.49,88.28) and (130.97,86.72) .. (132.79,86.72) .. controls (134.61,86.72) and (136.09,88.28) .. (136.09,90.22) .. controls (136.09,92.15) and (134.61,93.72) .. (132.79,93.72) .. controls (130.97,93.72) and (129.49,92.15) .. (129.49,90.22) -- cycle ;
\draw  [color={rgb, 255:red, 0; green, 0; blue, 0 }  ,draw opacity=1 ][line width=1.5]  (183.33,90.05) .. controls (183.33,67.96) and (201.24,50.05) .. (223.33,50.05) .. controls (245.42,50.05) and (263.33,67.96) .. (263.33,90.05) .. controls (263.33,112.14) and (245.42,130.05) .. (223.33,130.05) .. controls (201.24,130.05) and (183.33,112.14) .. (183.33,90.05) -- cycle ;
\draw [line width=1.5]    (132.79,53.38) -- (132.79,127.05) ;
\draw  [draw opacity=0][fill={rgb, 255:red, 0; green, 0; blue, 0 }  ,fill opacity=1 ] (129.73,53.88) .. controls (129.73,51.95) and (131.21,50.38) .. (133.03,50.38) .. controls (134.86,50.38) and (136.33,51.95) .. (136.33,53.88) .. controls (136.33,55.81) and (134.86,57.38) .. (133.03,57.38) .. controls (131.21,57.38) and (129.73,55.81) .. (129.73,53.88) -- cycle ;
\draw  [draw opacity=0][fill={rgb, 255:red, 0; green, 0; blue, 0 }  ,fill opacity=1 ] (129.73,126.88) .. controls (129.73,124.95) and (131.21,123.38) .. (133.03,123.38) .. controls (134.86,123.38) and (136.33,124.95) .. (136.33,126.88) .. controls (136.33,128.81) and (134.86,130.38) .. (133.03,130.38) .. controls (131.21,130.38) and (129.73,128.81) .. (129.73,126.88) -- cycle ;

\draw (303,76.38) node [anchor=north west][inner sep=0.75pt]   [align=left] {$\displaystyle \in \mathcal{H}\left[\left( S^{d-p-1} \backslash S^{0}\right) \ \times S^{p}\right]$};
\draw (137.03,60.38) node [anchor=north west][inner sep=0.75pt]   [align=left] {$\displaystyle n$};
\draw (136.37,103.72) node [anchor=north west][inner sep=0.75pt]   [align=left] {$\displaystyle m$};
\draw (119.7,80.38) node [anchor=north west][inner sep=0.75pt]   [align=left] {$\displaystyle i$};

\end{tikzpicture}
    \caption{States charged under the $(p+2)$-group.}
    \label{states-2grp}
\end{figure}

We can prepare states that span the representation space of these projective representations. These states sit in the Hilbert space of a pierced codimension-1 surface. As an example, we pick $\Sigma_{d-1}=(S^{d-p-1}\backslash S^0)\times S^p$. The states can be prepared by filling $S^{d-p-1}$ and inserting the $(p+1)$-dimensional configuration of $\mathcal{L}_{n}$, $\mathcal{L}_{m}$ and $\mathcal{O}_i$ in such a way that the $\mathcal{L}_{n,m}$ defects wrap $S^p$ and, along their remaining direction, connect the north and south poles of $S^{d-p-1}$, while the operator $\mathcal{O}_i$ connecting them wraps $S^p$ and is inserted at the center of $S^{d-p-1}$, see figure \ref{states-2grp}.

\section{Normalizer of \texorpdfstring{$\mathbb{Z}_N \times \mathbb{Z}_N \subset PSU(N)$}{ZN x ZN in PSU(N)}}\label{app-normalizer}

Given a group $G$ and a subgroup $H\subset G$, we denote by $N_G(H)$ the normalizer of $H$ in $G$, i.e.~the maximal subgroup of $G$ that contains $H$ as a normal subgroup
\begin{equation}
    N_G(H) = \{n\in G \ | \ \forall h \in H , \ nhn^{-1} \in H\} \ .
\end{equation}
Consider $G=PSU(N)$ and $H=\mathbb{Z}_N\times\mathbb{Z}_N$ as defined around eq.~\eqref{ZNxZN}. In the following, we want to show that
\begin{equation}
    N_{PSU(N)}(\mathbb{Z}_N\times \mathbb{Z}_N)
    =
    (\mathbb{Z}_N\times \mathbb{Z}_N) \rtimes
    SL(2,\mathbb{Z}_N) \ .
\end{equation}
Let us represent $PSU(N)$ as $SU(N)$ matrices modulo elements of the center. To check whether (the equivalence class of) a matrix $n$ belongs to the normalizer, it is enough to check if it sends the generator pair $(\rho_A,\rho_B)$ of $\mathbb{Z}_N\times \mathbb{Z}_N$ into another element of $\mathbb{Z}_N\times \mathbb{Z}_N$
\begin{equation}
\begin{split}
    n \rho_A n^{-1} & = \rho_A^a \rho_B^b \\
    n \rho_B n^{-1} & = \rho_A^c \rho_B^d \ , \\
\end{split}
\end{equation}
where $a,b,c,d\in\mathbb{Z}_N$. However, since we should preserve the commutation relation $\rho_A\rho_B=\eta\rho_B\rho_A$, we get the constraint
\begin{equation}\label{eq:nrnnrn}
    (n \rho_A n^{-1}) \  (n \rho_B n^{-1}) =
    \eta  \, (n \rho_B n^{-1}) \ (n \rho_A n^{-1}) \ .
\end{equation}
An explicit calculation gives
\begin{equation}
\begin{split}
    (n \rho_A n^{-1}) \  (n \rho_B n^{-1}) & =
    \rho_A^a \rho_B^b \rho_A^c \rho_B^d 
    \\ \ & =
    \eta^{-bc} \rho_A^{a+c} \rho_B^{b+d} \ , \\
    (n \rho_B n^{-1}) \  (n \rho_A n^{-1}) & =
    \rho_A^c \rho_B^d \rho_A^a \rho_B^b 
    \\ \ & =
    \eta^{-ad} \rho_A^{a+c} \rho_B^{b+d}~. \\
\end{split}
\end{equation}
As a result, \eqref{eq:nrnnrn} translates into the following condition on $a,b,c,d$
\begin{equation}
    \eta^{ad-bc} = \eta \ ,
\end{equation}
which means that the matrix $\begin{pmatrix}
    a & b \\ c & d
\end{pmatrix}$ is in $SL(2,\mathbb{Z}_N)$.

We have shown above that the transformed version of the generating pair $(\rho_A,\rho_B)$ still furnishes a representation of the Heisenberg algebra, that is $\rho_A\rho_B=\eta\rho_B\rho_A$. It is always possible to find a matrix $n\in U(N)$ that conjugates the two representations, because all irreducible representations of the Heisenberg algebra are unitarily equivalent. The latter statement can be proved as follows. Take two matrices $U,V\in U(N)$ satisfying
\begin{equation}
    U^N=V^N=
    \mathds{1}_N \ ,
    \qquad UV=\eta VU \ .
\end{equation}
Consider an eigenvector $u_0$ of $U$ with eigenvalue $\lambda$ (note that $\lambda\neq 0$ since $U^N=\mathds{1}_N$). If we apply $V$ $j$ times on $u_0$, we get another eigenvector of $U$ with eigenvalue $\lambda \eta^j$
\begin{equation}
    U (V^j u_0) = \eta^j V^j U u_0 =
    \lambda \eta^j (V^j u_0)
    \ .
\end{equation}
In this way we can build $N$ eigenvectors with different eigenvalues, which define a basis. In this basis $U$ is diagonal and $V$ acts as the shift matrix. So, up to phases, $U$ and $V$ are equivalent to $P$ and $Q$ defined in \eqref{PQ-def}, meaning that there exists a matrix $n\in U(N)$ that conjugates $(U,V)$ to $(P,Q)$. This still holds in the $N$ even case, where the $N$-th power of the matrices is minus the identity, since we can multiply the matrices by $\zeta^{-1}$ to get $U^N=V^N=\mathds{1}_N$ as evident from \eqref{Gamma-N}.
Finally, it is clear that we can choose $n$ to be in $SU(N)$ and thus it defines an element of $PSU(N)$.

When we gauge $H\subset G$ we expect to be left with an invertible symmetry
\begin{equation}
    N_{G}(H)/H \ ,
\end{equation}
while the defects belonging to $PSU(N)$ but not to $N_{G}(H)$ form a non-invertible coset symmetry \cite{Damia:2023gtc,Hsin:2024aqb,Hsin:2025ria}. In the case considered here, we have
\begin{equation}
  \frac{N_{G}(H)}{H} =   \frac{(\mathbb{Z}_N\times \mathbb{Z}_N) \rtimes
    SL(2,\mathbb{Z}_N)}{\mathbb{Z}_N\times \mathbb{Z}_N} = SL(2,\mathbb{Z}_N) \ .
\end{equation}

\section{On the reduction of \texorpdfstring{$w_2(PSU(N))$}{w2(PSU(N))}} 
\label{appendix-obstruction}

In this appendix, we show that $w_2(PSU(N))$ reduces as in \eqref{eq:w2-N-reduction} when the $PSU(N)$ bundle is restricted to a $\mathbb{Z}_N \times \mathbb{Z}_N$ bundle, where the $\mathbb{Z}_N \times \mathbb{Z}_N$ subgroup of $PSU(N)$ is the one defined around \eqref{ZNxZN}. To this end, we begin by reviewing some basic facts about gauge bundles.

\subsection{Generalities about gauge bundles}

A gauge bundle for a group $G$ on a compact manifold $\mathcal{M}$ is defined by an open cover $\{\mathcal{U}_i\}$ of $\mathcal{M}$ such that every patch $\mathcal{U}_i$ is isomorphic to $\mathbb{R}^d$, where $d=\dim \mathcal{M}$,\footnote{We also take the cover to be a \textit{good cover}, i.e.~a cover such that any nonempty finite intersection $\mathcal{U}_{i_1 \dots i_k}\equiv \mathcal{U}_{i_1} \cap \dots \cap \mathcal{U}_{i_k}$ is isomorphic to $\mathbb{R}^d$ (see \cite[pag.~42]{MR658304}). Every manifold admits a good cover. Given that the manifold is compact, the cover can be taken to be finite and we can give an ordering to the patches, i.e.~assign labels such that for any pair of distinct patches $\mathcal{U}_i,\mathcal{U}_j$ we have either $i<j$ or $i>j$.} and by a collection of $1$-forms $A_i$ valued in the Lie algebra of $G$ representing the connection of the bundle in every patch. 

Whenever two patches $\mathcal{U}_i,\mathcal{U}_j$ intersect we can relate the corresponding two 1-forms $A_i,A_j$ via a transition function
\begin{equation}
    g_{ij}:\mathcal{U}_{ij}=\mathcal{U}_i\cap \mathcal{U}_j \to G
\end{equation}
as
\begin{equation}\label{double-int-1}
    A_j = g_{ij}^{-1} A_i g_{ij} 
    -i g_{ij}^{-1} dg_{ij} \ .
\end{equation}

Whenever three patches $\mathcal{U}_i,\mathcal{U}_j,\mathcal{U}_k$ intersect, consistency requires the transition functions to satisfy the cocycle condition
\begin{equation}\label{cocycle-cond-g}
    g_{ij} g_{jk} g_{ki} = 1 \ .
\end{equation}

All these data are subject to redundancy. Indeed, we have local gauge transformations (local in the sense that they act patch-wise) that act on $A_i$ and $g_{ij}$ as
\begin{equation}\label{local-gt}
    A_i \mapsto U_i^{-1} A_i U_i 
    -iU_i^{-1} d U_i
    \ , \quad
    g_{ij} \mapsto {U_i}^{-1} g_{ij} U_j \ ,
\end{equation}
where $U_i: \mathcal{U}_i \to G$ is a family of locally defined functions.

For example, to define a Wilson line
\begin{equation}
    W_{\mathcal{R}}(\gamma)=\Tr_{\mathcal{R}} P \exp \left(i\int_\gamma A \right) \ ,
\end{equation}
one has to carefully define the holonomy $P\exp \left(i\int_\gamma A \right)$. To do so, one should cut the path $\gamma$ in arcs $\gamma_i\in\mathcal{U}_i$ and use the local connection to compute the holonomy along them, $\text{hol}_{\gamma_i}(A_i) =P\exp\left(i\int_{\gamma_i}A_i\right)$. The full holonomy is constructed by multiplying the local holonomies and inserting transition functions when the patch changes, namely
\begin{equation}
    P\exp \left(i\int_\gamma A \right)
    =
    \prod_{k=1,\dots} \text{hol}_{\gamma_{i_k}}(A_{i_k})
    g_{i_k i_{k+1}} \ .
\end{equation}
This definition guarantees invariance under local gauge transformations \eqref{local-gt}.

\subsection{From a \texorpdfstring{$U(1)$}{U(1)} bundle to a 
\texorpdfstring{$\mathbb{Z}_N$}{ZN} bundle}

Let us consider a flat $U(1)$ bundle, meaning that the collection of $1$-forms $A_i\in \Omega^1(\mathcal{U}_i,\mathfrak{u}(1))$ are all closed. Given that $\mathcal{U}_i\simeq \mathbb{R}^d$, we can find functions $\Lambda_i$ such that patch-wise
\begin{equation}
    A_i = d\Lambda_i \ .
\end{equation}
This means that via a local gauge transformation $U_i=\exp(-i\Lambda_i)$ we can set to zero all $A_i$'s
\begin{equation}
    A_i \mapsto 
    e^{i\Lambda_i} A_i e^{-i\Lambda_i}
    -i
    e^{i\Lambda_i} d (e^{-i\Lambda_i})
    =
    A_i - d\Lambda_i = 0 \ .
\end{equation}
This also modifies the transition functions
\begin{equation}
    g_{ij} \mapsto e^{i\Lambda_i} g_{ij} e^{-i\Lambda_j}
    =
    e^{i(\Lambda_i-\Lambda_j)} g_{ij} \ .
\end{equation}
This shows that a flat $U(1)$ bundle can be entirely specified via its transition functions. For such bundles the holonomy along a curve $\gamma$ is simply given by the product of the transition functions.

Now let us consider $\mathbb{Z}_N\subset U(1)$. We can describe a $\mathbb{Z}_N$ bundle by considering a flat $U(1)$ bundle and imposing that all transition functions are valued in $\mathbb{Z}_N$, that is 
\begin{equation}
    g_{ij} = \exp \left(\frac{2\pi i}{N} a_{ij} \right) \ , 
    \quad
    a_{ij} \in \mathbb{Z}_N \ .\footnote{Here, and for all the quantities appearing inside exponents, we use additive notation. }
\end{equation}
The cocycle condition \eqref{cocycle-cond-g} in terms of $a_{ij}$ becomes
\begin{equation}
    0 = a_{ij} + a_{jk} + a_{ki}
    =
    a_{jk} - a_{ik} + a_{ij}
    \equiv (\delta a)_{ijk} \ , 
\end{equation}
where we used $g_{ki}=g_{ik}^{-1}$ which implies $a_{ki}=-a_{ik}$ and $\delta$ is the coboundary operator,
\begin{equation}
    (\delta c)_{i_0,i_1,\dots,i_{p+2}} \coloneqq
    c_{i_1,i_2,\dots,i_{p+2}}
    - c_{i_0,i_2,\dots,i_{p+2}}
    +(-1)^{p+2} c_{i_0,i_1,\dots,i_{p+1}}
    =
    \sum_{k=0}^{p+2} (-1)^{k} c_{i_0,\dots,\widehat{i}_k,\dots,i_{p+2}} \ .
\end{equation}
The vanishing of $\delta a$ means that $a$ is a cocycle in $Z^1(\mathcal{M},\mathbb{Z}_N)$. This is the quantity we usually refer to as a discrete gauge field.

\subsection{\texorpdfstring{$PSU(N)$}{PSU(N)} bundles and \texorpdfstring{$w_2$}{w2}}\label{Appendix-PSU(N)-w2}

Consider now a $PSU(N)$ bundle with transition functions $g_{ij}:\mathcal{U}_{ij}\to PSU(N)$. Since
\begin{equation}\label{SESstructure-D}
   1 \rightarrow  \mathbb{Z}_N \xrightarrow i SU(N) \xrightarrow \pi PSU(N) \rightarrow 1 \ ,
\end{equation}
we can pick lifts $\tilde{g}_{ij}:\mathcal{U}_{ij}\to SU(N)$ such that $\pi(\tilde{g}_{ij})=g_{ij}$ and check if the cocycle condition, which is satisfied by $g_{ij}$,
\begin{equation}\label{cocycle-cond-g-3}
    g_{ij} g_{jk} g_{ki} = \mathds{1} \ ,
\end{equation}
is satisfied by their lifts. In general this won't be the case and actually
\begin{equation}\label{cocycle-cond-g-3-tilde}
    \tilde{g}_{ij} \tilde{g}_{jk} \tilde{g}_{ki} = \exp \left(\frac{2\pi i}{N} w_{ijk}\right)\mathds{1}  \ ,
\end{equation}
where $w_{ijk}$ takes values in $\mathbb{Z}_N$, so that the RHS is an element of $Z(SU(N))$, and defines a \v Cech 2-cocycle. 

\subsubsection{From \texorpdfstring{$w_2$}{w2} to \texorpdfstring{$a\cup b$}{a u b} for \texorpdfstring{$N=2$}{N=2}}

Now we want to consider a $\mathbb{Z}_2\times \mathbb{Z}_2 \subset SO(3)$ bundle. Being a discrete bundle, all its data are contained in its transition functions
\begin{equation}\label{Z2xZ2}
    g_{ij}= (R_x(\pi))^{a_{ij}} (R_y(\pi))^{b_{ij}} 
\end{equation}
where
\begin{equation}
    R_{x}(\pi) = \begin{pmatrix}
        1 & 0 & 0 \\
        0 & -1 & 0  \\
        0 & 0 & -1 
    \end{pmatrix}
    \quad , \quad
    R_{y}(\pi) = \begin{pmatrix}
    -1 & 0 & 0 \\
    0 & 1 & 0  \\
    0 & 0 & -1 
    \end{pmatrix}
\end{equation}
and $a_{ij}$ and $b_{ij}$ are the $\mathbb{Z}_2$ valued discrete gauge fields so that
\begin{equation}\label{g-reduced}
    g_{ij} = 
    \begin{pmatrix}
    (-1)^{b_{ij}} & 0 & 0 \\
    0 & (-1)^{a_{ij}} & 0 \\
    0 & 0 & (-1)^{a_{ij}+b_{ij}} \\
    \end{pmatrix} \ .
\end{equation}
Plugging \eqref{g-reduced} inside the cocycle condition \eqref{cocycle-cond-g-3} we get the cocycle conditions for $a$ and $b$ separately
\begin{equation}
\begin{split}
    (0)_2 = a_{ij} + a_{jk} + a_{ki}
    & = (\delta a)_{ijk} \\
    (0)_2 = b_{ij} + b_{jk} + b_{ki}
    & = (\delta b)_{ijk} \ , 
\end{split}
\end{equation}
which is satisfied for $a,b\in Z^1(\mathcal{M},\mathbb{Z}_2)$.

Now we consider the following commutative diagram of short exact sequences
\begin{displaymath}
\begin{tikzcd}\label{Ext_restrict-N=2}
1 \arrow[r]  & \mathbb{Z}_{2} \arrow[r, "i"] & SU(2) \arrow[r, "\pi"]  & SO(3) \arrow[r]&1\\
1 \arrow[r] & \mathbb{Z}_{2}\arrow[r,"i"] \arrow[hookrightarrow,u] & Q_8 \arrow[r,"\pi"]\arrow[hookrightarrow,u] &\mathbb{Z}_{2} \times \mathbb{Z}_{2} \arrow[r]\arrow[hookrightarrow,u]&1
\end{tikzcd}
\end{displaymath}
where the group $Q_8$
\begin{equation}
    Q_8 = \langle A,B,C | A^2=B^2=C,C^2=1,AB=BAC,AC=CA,BC=CB\rangle
\end{equation}
is embedded in $SU(2)$ as
\begin{equation}
    A \rightarrow i\sigma_x 
    \quad
    B \rightarrow i\sigma_y
    \quad
    C \rightarrow -\mathds{1}_2 \ .
\end{equation}
Starting from an element $g\in R_x(\pi)^aR_y(\pi)^b\in \mathbb{Z}_2 \times \mathbb{Z}_2 \subset SO(3)$, since $\pi:SU(2)\to SO(3)$ has kernel $\mathbb{Z}_2$, there exist exactly two elements $\tilde{g}\in Q_8 \subset SU(2)$ such that $\pi(\tilde{g})=g$. Moreover, since $\ker(\pi)=\{\mathds{1}_2,-\mathds{1}_2\}$, they differ by an overall minus sign. We write either of the two as $\tilde{g}=(i\sigma_x)^{\tilde{a}}(i\sigma_y)^{\tilde{b}}(-1)^c$ with $\tilde{a},\tilde{b}\in \mathbb{Z}_4$, $c\in \mathbb{Z}_2$, $(\tilde{a})_2=a,(\tilde{b})_2=b $ and identifications $(\tilde{a},\tilde{b},c)\sim(\tilde{a}+2,\tilde{b},c+1)\sim(\tilde{a},\tilde{b}+2,c+1)$.

We define lifts in $Q_8$ of the $\mathbb{Z}_2\times\mathbb{Z}_{2}$ valued transition functions \eqref{g-reduced} as
\begin{equation}\label{Q8}
    \tilde{g}_{ij} = {(i\sigma_x)}^{\tilde{a}_{ij}} {(i\sigma_y)}^{\tilde{b}_{ij}} {(-1)}^{c_{ij}} \ , \quad
    \text{for } i<j \ ,
\end{equation}
where $\tilde{a}_{ij}$ and $\tilde{b}_{ij}$ are $\mathbb{Z}_4$ lifts of $a_{ij},b_{ij}$ which satisfy
\begin{equation}\label{lift-Z4-Z2}
\begin{aligned}
    &\tilde{a}_{ij} \hspace{-2mm}& = a_{ij} \text{ mod } 2 \ ,  \\
    &\tilde{b}_{ij} \hspace{-2mm}& \hspace{-0.5mm}=\hspace{0.5mm} b_{ij} \text{ mod } 2 \ ,  
\end{aligned}
\end{equation}
and $c_{ij}$ is a $\mathbb{Z}_2$ valued cochain (generically non-closed).\footnote{The parametrization \eqref{Q8} is actually redundant, the triple $(\tilde{a}_{ij},\tilde{b}_{ij},c_{ij})$ is subject to identifications
\begin{equation}
\begin{split}
    \tilde{a}_{ij} & \to \tilde{a}_{ij} +2u_{ij}\\
    \tilde{b}_{ij} & \to \tilde{b}_{ij}
    +2v_{ij} \\   
    c_{ij} & \to c_{ij} -u_{ij} - v_{ij}
    \ ,\\ 
\end{split}
\end{equation}
where $u,v$ are $\mathbb{Z}_2$ cochains.
} The definition of $\tilde{g}_{ji}$ is fixed, by asking $\tilde{g}_{ij}\tilde{g}_{ji}=\mathds{1}$, to be 
\begin{equation}\label{Q8-inverse}
\begin{split}
    \tilde{g}_{ji} = \left(\tilde{g}_{ij}\right)^{-1}
    & =
    {(i\sigma_y)}^{-\tilde{b}_{ij}}
    {(i\sigma_x)}^{-\tilde{a}_{ij}}  
    {(-1)}^{-c_{ij}} \\
    \ & =
    {(i\sigma_x)}^{\tilde{a}_{ji}} {(i\sigma_y)}^{\tilde{b}_{ji}} {(-1)}^{c_{ji}-\tilde{a}_{ij}\tilde{b}_{ij}} \\
    \ & =
    {(i\sigma_x)}^{\tilde{a}_{ji}} {(i\sigma_y)}^{\tilde{b}_{ji}} {(-1)}^{c_{ji}-{a}_{ij}{b}_{ij}} \\
    \ & =
    {(i\sigma_x)}^{\tilde{a}_{ji}} {(i\sigma_y)}^{\tilde{b}_{ji}} {(-1)}^{c_{ji}- (a\cup_1b)_{ij}} \ ,
\end{split}
\end{equation}
where in the second-to-last step we have used \eqref{lift-Z4-Z2} to remove the dependence on the lifts $\tilde{a},\tilde{b}$, and we defined the $\cup_1$ product as $(a\cup_1b)_{ij}=a_{ij}b_{ij}=(a\cup_1b)_{ji}$. 
Now if we compute the cocycle condition for $i<j<k$, we get
\begin{equation}
\begin{split}
    \tilde{g}_{ij} \tilde{g}_{jk}
    & =
    (i\sigma_x)^{\tilde{a}_{ij}}(i\sigma_y)^{\tilde{b}_{ij}}
    (i\sigma_x)^{\tilde{a}_{jk}}(i\sigma_y)^{\tilde{b}_{jk}}
    (-1)^{c_{ij}+c_{jk}}\\
    & =
    (i\sigma_x)^{\tilde{a}_{ij}+\tilde{a}_{jk}}
    (i\sigma_y)^{\tilde{b}_{ij}+\tilde{b}_{jk}}
    (-1)^{\tilde{b}_{ij} \tilde{a}_{jk}
    +c_{ij}+c_{jk}}
    \\
    & =
    (i\sigma_x)^{(\delta\tilde{a})_{ijk}-\tilde{a}_{ki}}
    (i\sigma_y)^{(\delta\tilde{b})_{ijk}-\tilde{b}_{ki}}
    (-1)^{(b\cup a)_{ijk} 
    +(\delta c)_{ijk}-c_{ki}} \ .
    \\
\end{split}
\end{equation}
Since $\tilde{a}$ is a $\mathbb{Z}_4$ lift of a closed $\mathbb{Z}_2$ cochain, its $\delta$ is a multiple of $2$ and by definition gives the Bockstein of $[a]$
\begin{equation}
    \delta \tilde{a} = 2 \beta(a) \ .
\end{equation}
The same holds for $b$. Then
\begin{equation}
\begin{split}
    \tilde{g}_{ij} \tilde{g}_{jk}
    & =
    \tilde{g}_{ik} \,
    (-1)^{w_{ijk}}
\end{split}
\end{equation}
where
\begin{equation}
    w_{ijk} = 
    \beta(a)_{ijk} + \beta(b)_{ijk} + 
    (b\cup a)_{ijk}
    +(\delta c)_{ijk}
    \ ,
\end{equation}
or rearranging 
\begin{equation}
    \tilde{g}_{ij} \tilde{g}_{jk} \tilde{g}_{ki}
    =
    (-\mathds{1})^{w_{ijk}} \in Z(SU(2)) \ .
\end{equation}
$w_{ijk}$ can be checked to be closed and the presence of the term $(\delta c)_{ijk}$ shows that it is defined up to exact terms, so it is the representative of a cohomology class $[w]$. Moreover, from \cite[eq.~(A.19)]{Benini:2018reh} we have that 
\begin{equation}
    b \cup a =
    - a\cup b - \delta (b\cup_1 a)
\end{equation}
so we can write 
\begin{equation}
    w = \beta(a) + \beta(b) + a\cup b \ ,
\end{equation}
modulo $2$ and modulo exact terms.

\subsubsection{\texorpdfstring{$N\geq2$}{N>=2}}
We now want to repeat the same analysis for $N\geq2$ and the discrete subgroup defined in \eqref{Gamma-N}.
As a subgroup of $SU(N)$, it is a central extension of $\mathbb{Z}_N\times\mathbb{Z}_N$ by $\mathbb{Z}_N$, that is 
\begin{equation}\label{ext-N-Gamma}
    1 \to \mathbb{Z}_N \to \Gamma \to \mathbb{Z}_N \times \mathbb{Z}_N \to 1 \ .
\end{equation}
We can write any such extension as
\begin{equation}
    \Gamma = \langle A,B,C | 
    A^N=C^{k_1} , B^N=C^{k_2} , C^N=1,
    AB=BAC^{k_3} , CA=AC , CB=BC
    \rangle \ ,
\end{equation}
where $(k_1,k_2,k_3) \in \mathbb{Z}_N^3$. The subgroup defined in \eqref{Gamma-N} corresponds to the case $(0,0,1)$ for $N$ odd and $(N/2,N/2,1)$ for $N$ even. In what follows we do not specify $(k_1,k_2,k_3)$ and we will fix them only at the end.

Let us take a $\mathbb{Z}_N\times\mathbb{Z}_N$ bundle and see what the obstruction to lift it to a $\Gamma$ bundle is.
 We start from a pair of $\mathbb{Z}_N$ valued closed transition functions
\begin{equation}
\begin{split}
    (0)_N = a_{ij} + a_{jk} + a_{ki}
    & = (\delta a)_{ijk} \ ,\\
    (0)_N = \, b_{ij} + b_{jk} + b_{ki} \,
    & = (\delta b)_{ijk} \ . 
\end{split}
\end{equation}
We can write the $\mathbb{Z}_N \times \mathbb{Z}_N$ transition functions as
\begin{equation}
    g_{ij} = \big(\eta^{a_{ij}},\eta^{b_{ij}}\big) \ ,
\end{equation}
where $\eta=\exp(\frac{2\pi i}{N})$. Then we take as $\Gamma$ valued lifts
\begin{equation}
    \tilde{g}_{ij} = A^{\tilde{a}_{ij}} B^{\tilde{b}_{ij}} C^{c_{ij}} \ , \quad
    \text{for } i<j \ ,
\end{equation}
where $\tilde{a}$ and $\tilde{b}$ are $\mathbb{Z}_{N^2}$ lifts of $a,b$, and $c_{ij}$ is a $\mathbb{Z}_N$ valued cochain (generically non closed). This fixes
\begin{equation}
    \tilde{g}_{ji} = (\tilde{g}_{ij})^{-1}
    =
    A^{\tilde{a}_{ji}} B^{\tilde{b}_{ji}} C^{c_{ji} - k_3 (a\cup_1b)_{ji}} \ , \quad
    \text{for } i<j \ .
\end{equation}
Now, if we compute the cocycle condition for $i<j<k$, we get
\begin{equation}
\begin{split}
    \tilde{g}_{ij} \tilde{g}_{jk} =
    C^{\omega_{ijk}} \tilde{g}_{ik}
\end{split}
\end{equation}
where
\begin{equation}
    \omega_{ijk}= k_1 \beta_N(a)_{ijk}
    + k_2 \beta_N(b)_{ijk}
    - k_3 (b\cup a)_{ijk}
    + (\delta c)_{ijk} \ .
\end{equation}
The term $\delta c$ simply states that $\omega$, which is closed, is defined up to exact terms, so it is the representative of a cohomology class $[\omega]$ in $H^2(\mathcal{M},\mathbb{Z}_N)$ labeled by $(k_1,k_2,k_3)$ 
\begin{equation}
    [\omega]= k_1 \,\beta_N(a)
    + k_2 \,\beta_N(b)
    + k_3 \,a\cup b \ .
\end{equation}
The extensions are classified by a class in $H^2(\mathbb{Z}_N\times\mathbb{Z}_N,\mathbb{Z}_N)$ that is also labeled by $(k_1,k_2,k_3)$ and they are in one-to-one correspondence.

If $k_1$ divides $N$, then $k_1\beta_N(a)=k_1\beta_{N/k_1}(a)$ and we actually only needed to lift $a$ to a $\mathbb{Z}_{N\times N/k_1}$ cochain. Indeed
\begin{equation}\label{gcd_not_1}
    A^{\delta \tilde{a}_{ijk}}
    =
    A^{N \beta_{N/k_1}(a)_{ijk}}
    =
    C^{k_1\beta_{N/k_1}(a)_{ijk}} \ .
\end{equation}
The same holds for $k_2$ and $b$.

To summarize, when we embed the $\mathbb{Z}_N\times\mathbb{Z}_N$ group in $PSU(N)$, the obstruction reduces to
\begin{equation}
    w_2(PSU(N)) = 
    \begin{cases}
    \mathcal{A}_1 \cup \mathcal{B}_1 & \text{if } N \text{ odd}   \\ 
    \tfrac{N}{2}\beta_2(\mathcal{A}_1) + 
    \tfrac{N}{2}\beta_2(\mathcal{B}_1) + \mathcal{A}_1 \cup \mathcal{B}_1 & \text{if } N \text{ even}  \ , \\ 
    \end{cases}
\end{equation}
where $\mathcal{A}_1,\mathcal{B}_1$ are $\mathbb{Z}_N$ background fields, i.e.~eq.~\eqref{eq:w2-N-reduction}. See \cite{Tanizaki:2018xto} for a different derivation in the $N=3$ case.

\bibliography{refs}
\bibliographystyle{JHEP}

\end{document}